\documentclass[nohyper,12pt,letterpaper]{JHEP3}
\usepackage{amsmath}
\usepackage{amsthm}
\usepackage{amsfonts}

\DeclareMathOperator{\rank}{rank} % rank
\DeclareMathOperator{\mult}{mult} % multiplicity of a root
\DeclareMathOperator{\ad}{ad} % adjoint action
\DeclareMathOperator{\hgt}{ht} % Height of a root
\DeclareMathOperator{\Ch}{Ch} % Character
\DeclareMathOperator{\Max}{Max} % maximum
\DeclareMathOperator{\Aut}{Aut} % Automorphism

\date{\today}

\title{M-theory and $E_{10}$: Billiards, Branes, and Imaginary Roots}

\author{
Jeffrey Brown${}^1$, Ori J. Ganor${}^2$ and Craig Helfgott${}^2$\\ 
${}^1$Department of Mathematics,
             University of California, Berkeley, CA 94720 \\
${}^2$Department of Physics,
             University of California, Berkeley, CA 94720 \\
  and\\
${}^2$Physics Division,
Ernest Orlando Lawrence Berkeley National Laboratory, \\ Berkeley, CA 94720\\
Emails:\\ \email{jbrown@Math.Berkeley.edu},
       \email{origa@socrates.berkeley.edu},
       \email{helfgott@socrates.berkeley.edu}}

\abstract{
Eleven dimensional supergravity compactified on $T^{10}$ admits
classical solutions describing
what is known as {\it billiard cosmology} -- a dynamics
expressible as an abstract (billiard) ball moving in the 10-dimensional
root space of the infinite dimensional Lie algebra $E_{10},$ occasionally
bouncing off walls in that space.
Unlike finite dimensional Lie algebras, $E_{10}$ has negative and zero norm roots,
in addition to the positive norm roots.
The walls above are related to physical fluxes that, in turn,
are related to positive norm roots (called {\it real roots}) of $E_{10}.$
We propose that zero and negative norm roots, called {\it imaginary roots},
are related to physical branes.
Adding ``matter'' to the billiard cosmology corresponds
to adding potential terms associated to imaginary roots.
The, as yet, mysterious relation between $E_{10}$ and M-theory on $T^{10}$
can now be expanded as follows: real roots correspond to fluxes or instantons,
and imaginary roots correspond to particles and branes (in the cases we checked).
Interactions between fluxes and branes and between branes and branes are classified
according to the inner product of the corresponding roots (again in the cases we checked).
We conclude with a discussion of an effective Hamiltonian description 
that captures some features of M-theory on $T^{10}.$
}


\received{July 14, 2004}

\preprint{\hepth{0401053}\\ UCB-PTH-04-01\\ LBNL-54228}
\begin{document}

\newtheorem{thm}{Theorem}[section]
\newtheorem{lem}[thm]{Lemma}
\newtheorem{prop}[thm]{Proposition}
\newtheorem{claim}[thm]{Claim}
\newtheorem{conj}[thm]{Conjecture}

\theoremstyle{definition}
\newtheorem{defn}{Definition}[section]

\numberwithin{equation}{section}

\newcommand{\thmref}[1]{Theorem~\ref{#1}}
\newcommand{\secref}[1]{\S\ref{#1}}
\newcommand{\lemref}[1]{Lemma~\ref{#1}}
\newcommand{\clmref}[1]{Claim~\ref{#1}}

\newcommand{\propref}[1]{Proposition~\ref{#1}}
\newcommand{\figref}[1]{Figure~\ref{#1}}

% ========================================================================== 

\def\be{\begin{equation}} 
\def\ee{\end{equation}} 
\def\bear{\begin{eqnarray}} 
\def\eear{\end{eqnarray}} 
\def\nn{\nonumber} 
 
\newcommand\bra[1]{{\langle {#1}|}}  % QM bran
\newcommand\ket[1]{{|{#1}\rangle}}  % QM ket
 
\def\defineas{{\,\stackrel{{\mbox{\tiny def}}}{=}\,}} 
 
\def\a{\alpha} 
\def\b{\beta} 
\def\g{\gamma} 
\def\u{\mu} 
\def\v{\nu} 
\def\r{\rho} 
\def\th{{\theta}} 
\def\lam{{\lambda}} 

\def\bth{{\overline{\theta}}} 
\def\blam{{\overline{\lambda}}} 
\def\bpsi{{\overline{\psi}}} 
\def\bsig{{\overline{\sigma}}} 
\def\Dslash{{\relax{\not\kern-.18em \partial}}} % Dirac operator
\def\Spin{{{\mbox{\rm Spin}}}} % spin group
\def\SL{{{\mbox{\rm SL}}}} % SL() group
\def\rt{{\rightarrow}}  
\def\cc{{\mbox{c.c.}}} 

\newcommand\SUSY[1]{{${\cal N}={#1}$}}  % supersymmetry
\newcommand\px[1]{{\partial_{#1}}} 
\newcommand\qx[1]{{\partial^{#1}}} 

\newcommand\ppx[1]{{\frac{\partial}{\partial {#1}}}} 
\newcommand\pspxs[1]{{\frac{\partial^2}{\partial {#1}^2}}} 
\newcommand\pspxpx[2]{{\frac{\partial^2}{\partial {#1}\partial {#2}}}} 

% Fields and rings etc.
\newcommand{\field}[1]{\mathbb{#1}}
\newcommand{\ring}[1]{\mathbb{#1}}
\newcommand{\C}{\field{C}}
\newcommand{\R}{\field{R}}
\newcommand{\Z}{\ring{Z}}
\newcommand{\N}{\ring{N}}

%%%%%%%%%%%%%%%%%%%%%%%%%%%%%%%%%%%%%%%%%%%%%%%%%%%%%%%%%%%%%%%%%%%%%%%%
% Vertical lines
\providecommand{\abs}[1]{{\lvert#1\rvert}}
\providecommand{\norm}[1]{{\lVert#1\rVert}}
\providecommand{\divides}{{\vert}}
\providecommand{\suchthat}{{\vert\quad}}

\def\bz{{\overline{z}}}
\def\gYM{{g_{\mbox{\rm\tiny YM}}}} % YM coupling constant
\def\gst{{g_s}} % string coupling constant
\def\Mst{{M_s}} % string scale
\newcommand\rep[1]{{\bf {#1}}} % representation
\def\Brane{{\cal B}} % brane
\def\Ac{{S}} % Action
\def\MinkAc{{\tilde{S}}} % Minkowski Action
\def\Lag{{L}} % Lagrangian

\newcommand\inner[2]{{\langle {#1}, {#2} \rangle}} % inner product
\newcommand\kform[2]{{({#1}|{#2})}} % Killing form

\def\rtB{{\Theta}} % Brane name
\def\ct{{\tilde{\tau}}} % conformal time

\def\CarEX{{\hat{\frak h}_\R}} % Cartan subalgebra of E10
\def\CarEIX{{{\frak h}_\R}} % Cartan subalgebra of E9
\def\CarEVIII{{\dot{\frak h}_\R}} % Cartan subalgebra of E8
\def\algEX{{{\bf \hat{g}}}} % algebra E10
\def\algEIX{{{\bf g}}} % algebra E9
\def\algEVIII{{{\bf \dot{g}}}} % algebra E8
\def\rtEX{{\hat{\Delta}}} % roots of E10
\def\rtEIX{{\Delta}} % roots of E9
\def\rtEVIII{{\dot{\Delta}}} % roots of E8
\def\latEX{{\hat{Q}}} % root lattice of E10
\def\latEIX{{Q}} % root lattice of E9
\def\latEVIII{{\dot{Q}}} % root lattice of E8
\def\weylEX{{\hat{W}}} % Weyl group of E10
\def\weylEIX{{W}} % Weyl group of E9
\def\weylEVIII{{\dot{W}}} % Weyl group of E8
\def\wgtEX{{\hat{P}}} % weights of E10
\def\wgtEIX{{P}} % weights of E9
\def\wgtEVIII{{\dot{P}}} % weights of E8
\newcommand\fwEX{{\hat{\Lambda}}} % a particular weight of E10
\newcommand\fwEIX{{\Lambda}} % a particular weight of E9
\newcommand\fwEVIII{{\dot{\Lambda}}} % a particular weight of E8
\def\komEX{{{\bf \hat{k}}}} % Chevalley invariant sub algebra E10
\def\komEIX{{{\bf k}}} % Chevalley invariant subalgebra E9
\def\komEVIII{{{\bf \dot{k}}}} % Chevalley invariant subalgebra E8

\def\ReRootsEX{{\hat{\Delta}_{\mbox{\rm\tiny Re}}}} % Real roots of E10
\def\ImRootsEX{{\hat{\Delta}_{\mbox{\rm\tiny Im}}}} % imaginary roots of E10
\def\NegImRootsEX{{\hat{\Delta}_{\mbox{\rm\tiny Im}}^{-}}} % negative imaginary roots of E10
\def\PosImRootsEX{{\hat{\Delta}_{\mbox{\rm\tiny Im}}^{+}}} % positive imaginary roots of E10
\def\PosReRootsEX{{\hat{\Delta}_{\mbox{\rm\tiny Re}}^{+}}} % positive real roots of E10

\newcommand\RootsEXatLevel[1]{{\hat{\Delta}_{[{#1}]}}} % roots at a given level

\newcommand\pcoeff[1]{{p^{({#1})}}} % coefficient in partition generating function

\def\Inst{{\Phi}} % Instanton action
\def\Fl{{\cal C}} % flux
\def\Kom{{K}} % maximal compact subgroup
\def\ModSp{{\cal M}} % moduli space
\def\vp{{\vec{p}}} % vector of Kasner powers
\def\vh{{\vec{h}}} % vector of \log R's
\def\veps{{\vec{\epsilon}}} % infinitesimal change in \vh
\def\EffMet{{g}} % effective metric
\def\Monod{{M}} % monodromy
\def\wMonod{{\widetilde{M}}} % another monodromy

\def\TubSig{{\cal N}} % Tubular neighborhood
\def\Disc{{\bf  D}} % Disc

\def\Ptl{{{\cal V}}} % potential term
\def\PtlInst{{{\cal V}_{\mbox\rm inst}}} % instanton related potential term
\def\HamMass{{{\cal H}_{\mbox\rm mass}}} % mass related potential term
\def\Ham{{{\cal H}}} % Hamiltonian
\def\InstOp{{{\cal O}}} % Instanton creation operator
\def\Npairs{{N_{\mbox{\rm\tiny pairs}}}} % number of pairs

%\newenvironment{definition}[1]{\vskip 10pt\noindent{\bf Definition}\newline
%{\it {#1}} -- }{\vskip 10pt}
%\newenvironment{theorem}[1]{\vskip 0.5cm\noindent{\bf Theorem:} {\it {#1}}\newline
%\noindent}{\vskip 0.5cm}
%\newenvironment{notation}{\vskip 0.5cm\noindent{\bf Notation:}\newline
%\noindent}{\vskip 0.5cm}
%\newenvironment{claim}{\vskip 0.5cm\noindent{\bf Claim:}\newline
%\noindent}{\vskip 0.5cm}
%\newcounter{PropositionNumber}
%\setcounter{PropositionNumber}{0}
%\newenvironment{proposition}[1]{\vskip 0.5cm\noindent
%   \addtocounter{PropositionNumber}{1}
%   {\bf Proposition {\arabic{PropositionNumber}}{#1}:}\newline\noindent}{\vskip 0.5cm}

%\newtheorem{definition}{Definition}
%\newtheorem{theorem}{Theorem}
%\newtheorem{notation}{Notation}
%\newtheorem{proposition}{Proposition}
%\newtheorem{claim}{Claim}
%\newtheorem{conjecture}{Conjecture}

% ---------------------------------------------------------------------------
\centerline{\bf DISCLAIMER}
This document was prepared as an account of work sponsored by the United States Government. While this document is believed to contain correct information, neither the United States Government nor any agency thereof, nor The Regents of the University of California, nor any of their employees, makes any warranty, express or implied, or assumes any legal responsibility for the accuracy, completeness, or usefulness of any information, apparatus, product, or process disclosed, or represents that its use would not infringe privately owned rights. Reference herein to any specific commercial product, process, or service by its trade name, trademark, manufacturer, or otherwise, does not necessarily constitute or imply its endorsement, recommendation, or favoring by the United States Government or any agency thereof, or The Regents of the University of California. The views and opinions of authors expressed herein do not necessarily state or reflect those of the United States Government or any agency thereof or The Regents of the University of California.
% ---------------------------------------------------------------------------

\tableofcontents

\section{Introduction}\label{sec:intro}
% ========================================================================== 
% Setting: M-theory on $T^{10}$
Our setting is M-theory with all of space compactified by periodic boundary
conditions.
When more than $d=8$ dimensions are compact, there is no notion of
{\it moduli space of vacua}; the metric and even the topology
of the compact directions should
be allowed to fluctuate and should be treated quantum mechanically.
But a complete quantum mechanical formulation of this setting is, of course,
at the moment unknown.

It has been suggested over two decades ago that the infinite dimensional
Kac-Moody Lie algebra $E_{10}$ is relevant to the formulation of this theory
\cite{Julia}.
Since then, the possible connection between M-theory and $E_{10}$ has
been discussed in various
settings (see \cite{Marcus:1983td}-\cite{Keurentjes:2003hc} for a sample). 
There are also recent conjectures about a formulation of {\it uncompactified}
M-theory in terms of $E_{10}$ 
\cite{Damour:2002cu}
and about a description of
the behavior of M-theory near spacelike singularities in terms of $E_{10}$ 
\cite{Damour:2000wm}-\cite{deBuyl:2003ub}.
$E_{10}$ and other Kac-Moody and Generalized Kac-Moody algebras
also appeared in other contexts in string theory (see for instance
\cite{Dijkgraaf:1996it}-\cite{DeWolfe:1998pr}) which
we will not discuss here.
Although a lot of progress has been made
%\cite{Barwald:1997gm}-\cite{Olive:2002yf},
\cite{Bauer:1996ca}-\cite{Olive:2002yf},
a full understanding
of the connection of M-theory to $E_{10}$ is still an open problem.

One of the features that distinguish infinite dimensional Kac-Moody Lie algebras,
such as $E_{10},$ from the finite dimensional ones is the existence of 
{\it imaginary roots} \cite{KacBook} in the root space.
{}From the physical point of view these roots are mysterious, and to the best
of our knowledge their physical interpretation has not been explored.

In this paper, we will study these imaginary roots from a physical perspective.
We will propose that they can be matched with actual branes.

% cosmology
We find it convenient to work with periodic boundary conditions,
although our proposal about the relation of imaginary roots and branes
can be readily adapted to the noncompact setting of
\cite{Damour:2000wm}-\cite{deBuyl:2003ub}.
The simplest way to set periodic boundary conditions on all 10 spatial
directions is to pick a topology of $T^{10}.$
Classically, a homogeneous Kasner metric on $T^{10}$ of the form
\be\label{KasnerMetricI}
ds^2 = -dt^2 + \sum_{i=1}^{10} R_i(t)^2 dx_i^2,\qquad
0\le x_i\le 2\pi,\qquad i=1\dots 10,
\ee
can be a solution to Einstein's equations if all $\log R_i$'s
are linear in $\log t.$
We set the slope to be a constant $p_i$ so that
\be\label{logRlogp}
\log \frac{R_i(t)}{R_i(t_0)} = p_i\log \frac{t}{t_0},\qquad i=1\dots 10.
\ee
Without matter, the Kasner metric \eqref{KasnerMetricI} is a solution provided 
the constants $p_1,\dots,p_{10}$
satisfy $\sum p_i = \sum p_i^2 = 1.$ This metric describes a universe
that is contracting in some directions (where $p_i<0$) and expanding
in other directions (where $p_i>0$).
This metric was extensively studied in \cite{Banks:1998vs}, where it was
shown that a classical treatment of a Kasner metric \eqref{KasnerMetricI}
is asymptotically trustworthy in the far future
if the vector of powers $\vp\equiv (p_1,\dots, p_{10})$
describes a timelike vector
in $\R^{9,1}$ (unrelated to the geometrical spacetime) with a suitably chosen
metric
\be\label{EtenCartanMetric}
\norm{\vp}^2 \equiv \sum p_i^2 -\Bigl(\sum p_i\Bigr)^2.
\ee
\textit{$\norm{\vp}^2=0$} if 
\textit{$\sum p_i = \sum p_i^2 = 1$}, and thus $\vp$ can never
be timelike unless we also include matter.
But before we add matter in the form of Kaluza-Klein particles and
branes, let us discuss the dynamics in the presence of fluxes.
A {\it flux} in this context could be either a constant $G = dC$
(where $C$ is the $3$-form of 11D supergravity) or a U-dual field.
A U-dual field could describe, for example, a nontrivial fibration of
one of the ten spatial directions over the remaining nine.
The fluxes are quantized and have discrete values, and
there are instanton effects that change the fluxes by integer amounts.
Explicit constructions of such instanton terms appear in
\cite{Instantons}.
%\cite{Green:1997as}-\cite{Pioline:2001jn}.

With fluxes, the classical dynamics of the scale factors $\log R_i$ is no longer
linear in $\log t.$ It was argued in 
\cite{Ivashchuk:1999rm}\cite{Damour:2000wm}-\cite{deBuyl:2003ub}
that the evolution of the $\log$s of the
scale factors can be approximated by a piecewise linear function that describes
Kasner epochs separated by sharp changes in the vector $\vp.$ 
The changes correspond to
reflections off (abstract) walls in $(\log R_i)$-space. The
walls correspond to the various fluxes that are present.
The orientation of each wall is determined by the type of flux, and its
position is determined by the amount of flux.
This evolution is called {\it billiard cosmology} since the dynamics is analogous
to that of a billiard ball in an abstract 10-dimensional space with coordinates
$\log R_i$, and the reflections are analogous to the ball bouncing off the walls.
Even in the absence of fluxes the walls above are present quantum mechanically.
They represent the necessary U-duality transformations that can be used to convert
small dimensions to large dimensions \cite{Banks:1998vs}.
Without matter, these reflections lead to a chaotic evolution \cite{Damour:2000wm}.

$E_{10}$ makes its appearance when we identify the
(``billiard table'') 10-dimensional
space with the Cartan subalgebra of
the infinite dimensional hyperbolic Kac-Moody Lie algebra, 
and identify each reflection off
a wall with a fundamental reflection
generator of the Weyl group.
The metric \eqref{EtenCartanMetric} can be identified
with the Cartan metric of $E_{10}$ [which has signature $(9,1)$].

The infinite dimensional
noncompact group $G_{10}$ that is defined as the exponential of a certain
real form of the Lie algebra $E_{10}$
is a natural extension of the finite dimensional
noncompact groups $G_d = \exp E_d$
with $d\le 8$ that appear as classical symmetry groups
of the low energy limit of M-theory compactified on $T^d.$ 
On the classical level, these symmetry groups are spontaneously broken,
and $G_d$ acts transitively on the moduli space of vacua whose metric
and topology can be summarized by writing the moduli space as 
$\Gamma'_d\backslash G_d/K_d.$ Here $\Kom_d$ is the maximal compact subgroup of
$G_d$ and $\Gamma'_d = SL(d,\Z)\subset G_d.$
For $d=8$ we have $G_d\equiv E_{8(8)}(\R),$
and $\Kom_d = \Spin(16)/\Z_2$ \cite{Cremmer:1979up}\cite{Marcus:1983hb}.
On the quantum level, these groups are explicitly broken
by loop and instanton effects,
and are {\it not} good symmetries. 
This point is demonstrated in explicit formulas for low-energy 
effective scattering amplitudes (presented as terms in the low-energy effective
action that contain, say, products of 4 curvature tensors) that
appear in \cite{Instantons}.
The quantum moduli space also contains extra identifications which
extend $\Gamma_d'$ to the full U-duality group $\Gamma_d$ \cite{Hull:1994ys}.
It is a discrete subgroup of $G_d$ that preserves a lattice in
an appropriate representation of $G_d$ \cite{Witten:1995ex}.
The extension of $\Gamma_d'$ to $\Gamma_d$ makes the volume of
the moduli space finite.
For $d=8$ we have $\Gamma_d \equiv E_{8(8)}(\Z).$
It is therefore also clear that $G_{10} \equiv\exp E_{10}$ cannot
be an unbroken symmetry group of any formulation of M-theory on $T^{10}$ 
that includes instanton effects.
It has to be either explicitly or spontaneously broken.

% instantons and roots
Nevertheless, $E_{10}$ provides a nice
characterization of the instanton effects.
It is well known that a positive root $+\a$ of the Lie algebra $E_d$ corresponds
to an instanton $\Brane_\a$ of M-theory compactified on $T^d$
(see \cite{Obers:1997kk}\cite{Obers:1998fb}
and section \secref{subsec:instantons} for a review).
For example, for $d=8$, if the metric on $T^8$ is diagonal and there are no fluxes,
the instantons are Kaluza-Klein particles, M2-branes,
M5-branes, and Kaluza-Klein  monopoles with a Euclidean world-volume.
We will review this correspondence between positive roots and instantons
in \secref{subsec:instantons}.

% this paper
In this paper we will study the case $d=10.$
This case is unique in that the Lie algebra
$E_{10}$ is the first $E_d$ with a Cartan form
that is not semi-positive definite. It is a hyperbolic Kac-Moody algebra
with a Cartan form of signature $(9,1).$
We recall that a Kac-Moody algebra with a simply-laced
connected Dynkin diagram is said to be {\it hyperbolic}
if its Cartan form is of indefinite type and every connected
subdiagram of the Dynkin diagram is of affine or finite type \cite{KacBook}.
Hyperbolic Kac-Moody algebras have rank $\le 10,$ and in this sense
the case $d=10$ is also maximal.

If the Cartan form is of indefinite type,
as is the case for $E_{10},$ the roots $\a$ do not necessarily square
to $2$. In fact the roots of an infinite dimensional Kac-Moody Lie algebra
can be classified as {\it real} and {\it imaginary} \cite{KacBook}.
Real roots satisfy $\a^2 = 2$, and all other roots 
are called {\it imaginary} and satisfy $\a^2 \le 0$.
The Weyl group acts transitively on the real roots.
We will review these facts in more detail in \secref{subsec:KM}.

The familiar instantons such as Kaluza-Klein particles, M2-branes,
M5-branes, and Kaluza-Klein  monopoles
all correspond to {\it real} positive roots of $E_{10}.$
In fact, as will be reviewed in \secref{subsec:instantons},
the Weyl group of $E_{10}$ formally acts as U-duality on the instanton
\cite{Obers:1997kk}\cite{Obers:1998fb}.
Hence, every object that can be obtained by U-duality 
from the above list of objects is also related to a positive real root,
and, vice versa, every positive real root is related to an object
that can be obtained by a formal U-duality transformation on, say, a
Euclidean M2-brane.

The question arises: {\it what is the physical interpretation of the
imaginary roots?}

The purpose of this paper is to study the roots with $\a^2 \le 0$ and to 
relate them to physical objects. 
We begin in \secref{sec:comb} by associating
a formal ``action'' to the root, and we study the ``combinatorial''
properties of this action as a function of radii $R_1,\dots R_{10}.$
In this section we explore a ``naive'' interpretation of imaginary roots
simply as new types of instantons with very large actions.

In \secref{sec:imroots} we propose a different interpretation,
which is one of the main points of this paper.
We propose that certain imaginary roots correspond to Minkowski objects.
To support this claim, we construct the Minkowski objects -- say branes --
via a creation process by pushing one instanton through another.
For example, one can construct an M2-brane by pushing an M5-brane
through another M5-brane \cite{Hanany:1996ie}.
We use a Wick rotated version of that process where one instanton is translated
in time until it crosses over another.

Once we accept the connection between imaginary roots and physical branes,
we can study the interactions of branes with branes and the interactions of
branes with fluxes from the Lie algebraic point of view.
We characterize various interactions according to the inner product of the 
participating roots.

Finally, we attempt to collect all the information 
together and construct an effective Hamiltonian that
describes the masses of the branes.
The model is a $\sigma$-model on a coset \textit{$G_{10}/\Kom_{10}$} of \textit{$G_{10}$}.
The Hamiltonian is, up to a sign, simply the 
\textit{$G_{10}(\equiv\exp E_{10})$} left-invariant
Laplacian $\Ham = -\triangle$ and the wave-function satisfies 
a Wheeler-DeWitt equation $\Ham\Psi=0.$
\textit{$G_{10}$} is spontaneously broken
to the U-duality subgroup 
\textit{$E_{10}(\Z)$} by requiring $\Psi$ to be \textit{$E_{10}(\Z)$}
invariant.
This suggestion is rather old, but the new point is 
to try to analyze the modes that correspond to imaginary roots quantum mechanically.
Doing that, we discover a piece in the Hamiltonian that is analogous
to a particle in a magnetic field. We compare the $n^{th}$ excited Landau level
to a state with $n$ branes (or Kaluza-Klein particles).
The energy separation between the Landau levels almost matches the energy of a brane,
but unfortunately there is a mismatch by a factor of $2\pi.$
There are also a few other puzzles, related to charge neutrality and zero-point energies.

The paper is organized as follows.
In \secref{sec:prelim},
we review the construction of infinite dimensional
Lie algebras as presented in \cite{KacBook}.
We also review billiard cosmology and
the connection between M-theory on $T^d$ and the Lie algebra $E_d.$
In particular we discuss real and imaginary roots
of $E_{10}$ and their multiplicities.
In \secref{sec:comb}, we explore the combinatorial properties of branes
that correspond to imaginary roots.
In \secref{sec:imroots}, we argue that certain imaginary roots
correspond to Minkowski branes and we study the various constructions
of such branes via a brane creation process involving two instantons.
As an application,
in \secref{subsec:withmatter}, we add matter
in the form of Kaluza-Klein particles and branes to billiard cosmology.
The matter component corresponds to potentials in $(\log R_i)$-space
oriented in directions corresponding to imaginary roots.
In \secref{sec:interactions}, we study how 
interactions of pairs of branes and
the interaction of a brane with a flux are encoded in the product
of the corresponding roots.
In \secref{sec:fermions}, we show that each instanton
defines a subgroup of the maximally compact subgroup
$K_{10}\subset \exp E_{10}.$ 
This is an extension of the statement for $d=8$ that a BPS instanton
preserves half of the supersymmetry generators, and therefore defines
a subgroup of the R-symmetry group $\Spin(16),$
which is the double cover of the compact subgroup
$\Spin(16)/\Z_2\subset E_{8(8)}(\R).$
% We explore possible generalizations for imaginary roots.
In \secref{sec:hamiltonian} we explore a possible Hamiltonian
formulation and compare 
our proposal to the ``small tension expansion''
of \cite{Damour:2002cu}.
We conclude with some open questions and a few conjectures.

\section{Preliminaries}\label{sec:prelim}
% ========================================================================== 

% Math
\subsection{Infinite dimensional Kac-Moody Lie algebras}\label{subsec:KM}
% --------------------------------------------------------------------------
In this subsection we will review the salient features of infinite
dimensional Kac-Moody Lie algebras.
Our discussion is taken from
\cite{KacLT}\cite{KMW}\cite{KacBook}.

Readers who are familiar with this subject and readers who are
not interested in the mathematical details are (reluctantly) advised to 
read \secref{subsubsec:construction}
and then skip to \secref{subsubsec:physicalbasis}.
In \secref{subsubsec:construction},
we review the construction of Kac-Moody algebras,
and demonstrate it for the hyperbolic Kac-Moody algebra
of interest $E_{10}$ and also for its subalgebra
$E_9$ which is an example of an affine Lie algebra \cite{ALA}\cite{KacBook}.
In \secref{subsubsec:multiplicity},
we explain the multiplicity formula for level $0,1$ roots obtained in \cite{KMW}.

\subsubsection{Review of Kac Moody Algebras and Root Spaces}
\label{subsubsec:construction}
% - - - - - - - - - - - - - - - - - - - - - - - - - - - - - - - - - - - - -
We recall the definition
of the Kac-Moody algebra $E_{10}$ and distinguished subalgebras 
$E_8, E_9.$  The construction of $E_{10}$
is a special case of the general construction of Kac Moody algebras in \cite{KacLT}.
We start with the Dynkin diagram of $E_{10}$:
% and $E_{10}$
\vskip 10pt
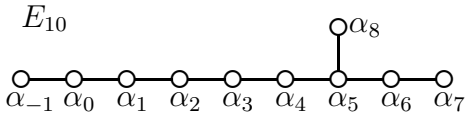
\begin{picture}(370,80)
%
% $E_{10}$
%
\thicklines
\multiput(20,20)(20,0){9}{\circle{6}}
\multiput(23,20)(20,0){8}{\line(1,0){14}}

\put(140,23){\line(0,1){14}}
\put(140,40){\circle{6}}
%%% axes
\put( 14,10){$\a_{-1}$}
\put( 36,10){$\a_0$}
\put( 56,10){$\a_1$}
\put( 76,10){$\a_2$}
\put( 96,10){$\a_3$}
\put(116,10){$\a_4$}
\put(136,10){$\a_5$}
\put(156,10){$\a_6$}
\put(176,10){$\a_7$}

\put(144,38){$\a_8$}
\put(20,40){$E_{10}$}
\end{picture}
\vskip 10pt
\noindent
We then associate to the diagram a 
Cartan matrix $A_{10}=(a_{ij}), (i,j=-1,0\cdots 8),$
by defining 
$$
a_{ij}\defineas\left\{\begin{array}{rl}
2 & {\mbox{if $i=j$}} \\
-1 & {\mbox{if nodes $i,j$ are connected by a line}} \\
0 & {\mbox{otherwise}}
\end{array}\right.
$$
The matrix $A_{10}$ is symmetric and $\det(A_{10})=-1$;
therefore ${\mbox{rank}}(A_{10})=10$. 
Choose a real vector space $\CarEX$ of dimension 10,
and linearly independent sets
\textit{$\Pi\defineas\{\a_{-1},\cdots,\a_{8}\}\subset \CarEX^*$}
(where $*$ denotes the dual space) and 
\textit{$\Pi^\vee\defineas\{\a_{-1}^\vee,\dots,\a_8^\vee\}\subset \CarEX$}
and define 
\textit{$\a_{j}(\a_i^\vee)\defineas a_{ij}$}.  We note that for general
Kac-Moody algebras
\be\label{DimCartan}
\dim\CarEX=2n-\rank(A)
\ee
where $n$ is the number of nodes
in the Dynkin diagram and
$A$ is the matrix associated to the diagram. 

The Kac-Moody algebra $E_{10}$
is the Lie algebra over $\C$ with the set of generators
\textit{$\CarEX\cup \{e_i, f_i\}_{i=-1}^8$},
and relations
$$
[h,h']=0,\quad
[e_i, f_j]=\delta_{ij}\a_i^\vee,\quad
[h,e_i]=\a_i(h)e_i,\quad
[h,f_i]=-\a_i(h)f_i,\qquad
h,h'\in\CarEX,
$$
\be\label{AdRule}
\ad(e_i)^{1-a_{ij}}e_j=0,\quad
\ad(f_i)^{1-a_{ij}}f_j=0,\qquad
i\neq j,
\ee
where \textit{$\ad(x)y\defineas [x,y]$},
\textit{$\ad(x)^2 y\equiv [x,[x,y]]$}, and so on.
Since $\CarEX$ has a basis of dimension 10, there are $30$ linearly independent
generators. These are called {\it Chevalley generators.}
$\CarEX$ is called the {\it Cartan subalgebra} of $E_{10}$
and is an
abelian subalgebra of maximal dimension
under which $E_{10}$ is completely reducible.
\footnote{
There are abelian subalgebras that are bigger than $\CarEX,$
but $E_{10}$ is not completely reducible with respect to those subalgebras.
Examples can be deduced from the constructions of \cite{Cremmer:1997ct}, and
we are grateful to the anonymous referee for pointing this out.
} 

We next identify an $E_9$ subalgebra of $E_{10}$ as
the Kac Moody algebra obtained from
the subdiagram of the $E_{10}$ diagram by deleting the $(-1)$-node
and the line connecting it to the $0$-node.
Similarly we identify an $E_8$ subalgebra by deleting the $-1,0$
nodes and the lines
connecting nodes $-1,0$ and nodes $0,1$.
\vskip 10pt
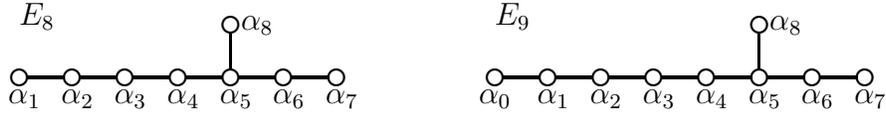
\begin{picture}(370,80)
%
% $E_8$
%
\thicklines
\multiput(40,20)(20,0){7}{\circle{6}}
\multiput(43,20)(20,0){6}{\line(1,0){14}}

\put(120,23){\line(0,1){14}}
\put(120,40){\circle{6}}
%%% axes
\put(36,10){$\a_1$}
\put(56,10){$\a_2$}
\put(76,10){$\a_3$}
\put(96,10){$\a_4$}
\put(116,10){$\a_5$}
\put(136,10){$\a_6$}
\put(156,10){$\a_7$}

\put(124,38){$\a_8$}
\put(40,40){$E_{8}$}
%
% $E_9$
%
\thicklines
\multiput(220,20)(20,0){8}{\circle{6}}
\multiput(223,20)(20,0){7}{\line(1,0){14}}

\put(320,23){\line(0,1){14}}
\put(320,40){\circle{6}}
%%% axes
\put(214,10){$\a_0$}
\put(236,10){$\a_1$}
\put(256,10){$\a_2$}
\put(276,10){$\a_3$}
\put(296,10){$\a_4$}
\put(316,10){$\a_5$}
\put(336,10){$\a_6$}
\put(356,10){$\a_7$}

\put(324,38){$\a_8$}
\put(220,40){$E_{9}$}
\end{picture}
\vskip 10pt

We then construct corresponding Cartan matrices
$A_{8}$ and $A_{9}$ following the procedure outlined above,
and view these matrices as minors of $A_{10}$.
The defining relations for $E_8$ and $E_9$
are thus inherited from the relations for $E_{10}.$  

We let $\CarEVIII$ denote the 
Cartan subalgebra (CSA) of $E_8$ and $\CarEIX$ the CSA of $E_9$.
We note that $\det(A_{8})=1$ and $\det(A_{9})=0$.  
A basis for the kernel
of $A_{9}$ is \textit{$\{(0,1,2,3,4,5,6,4,2,3)^{t}\}$}.
We then see from the adaptation of
formula \eqref{DimCartan} to $E_9$ that $\dim \CarEIX=10,$ and thus 
$\CarEIX = \CarEX.$
In keeping with the notation of \cite{KMW},
we define $\algEVIII\defineas E_8$ with CSA $\CarEVIII,$
$\algEIX\defineas E_9$ with CSA $\CarEIX$,
and $\algEX\defineas E_{10}$ with CSA $\CarEX$.
We have the root space decompositions of each algebra with respect to its CSA.
For example, \textit{$\algEX =\bigoplus_{\a\in \CarEX^*}\algEX_\a$} where
$$
\algEX_\a\defineas\{x\in \algEX: [h,x]=\a(h)x, \quad\forall h\in \CarEX\},
$$
and we define the {\it root space}
$$
\rtEX\defineas \{\a\in \CarEX^*: \algEX_\a\neq 0, \a\neq 0\}.
$$
We let $\latEX\defineas\sum_{i=-1}^{8}\Z\a_i$, and
$\latEX_{+}\defineas\sum_{i=-1}^{8}\N\a_i.$
($\N$ will denote the non-negative integers.)
Finally, define $\rtEX_{+}=\rtEX\cap\latEX_{+}$,
the set of positive roots of $\algEX$.
We then have $\rtEX=\rtEX_{+}\cup\rtEX_{-}$
where $\rtEX_{-}=-\rtEX_{+}$ \cite{KacBook}.
We define $\latEVIII\subset \latEX$, $\rtEVIII\subset \rtEX$, etc.,
analogously for the algebras $\algEVIII$, and similarly for
$\algEIX.$  

The signature of the inner product on the root lattice of a finite-dimensional 
simple Lie algebra is well known to be positive definite \cite{KacBook},
so from the $E_8$ subalgebra of $E_{10}$ and the fact that
$\det(A_{10})=-1$, we see that the inner product on $\latEX$ must have signature
$(9,1)$.

We partial-order $\CarEX^*$ by $\a\succeq\b$ if
$\a-\b\in\latEX_{+}$.  For $\a=\sum_{i=-1}^{8}k_i\a_i\in \latEX$
we define the {\it height} as $\hgt(\a)\defineas\sum_{i=-1}^8 k_i$.
Finally, we introduce the {\it Weyl group} $\weylEX$ of $\algEX$
as the subgroup of \textit{$\Aut\CarEX^*$} 
(the group of metric preserving linear transformations of
$\CarEX$) generated by {\it simple reflections}
$$
r_{i}(\lambda)=\lambda-\lambda(\a_i^\vee)\a_i,\qquad i=-1,\dots, 8,
\qquad
\lambda\in\CarEX^*.
$$
A root $\a\in \rtEX$ is called a {\it real root} if there exist 
$w\in {\hat W}$ such that $w(\a)=\a_i$ for some $-1\le i\le 8$; otherwise
$\a$ is an {\it imaginary root.}
As $\algEVIII$ is a finite dimensional Lie algebra, all of its roots are real.
In general, a root $\a$ is real if and only if $\kform{\a}{\a}>0$.
For $E_9\equiv\algEIX$, all the imaginary roots are integer multiples
of the root
\be\label{DeltaRoot}
\delta \defineas \a_0 +2\a_1 +3\a_2 +4\a_3 +5\a_4 +6\a_5 +4\a_6 +2\a_7 +3\a_8
\in\rtEIX.
\ee
It satisfies
$$
\kform{\delta}{\a_i} = 0,\qquad i=0,\dots,8,
$$
We denote the set of imaginary roots of $E_{10}$ as
$$
\ImRootsEX\defineas\{\a\in\rtEX: \kform{\a}{\a} \le 0\},
$$
and we define the set of positive (negative) imaginary roots as
$\PosImRootsEX \defineas\ImRootsEX\cap\rtEX_{+}$ 
($\NegImRootsEX \defineas\ImRootsEX\cap\rtEX_{-}$).

%({\bf We need to define ``integrable'' representation somewhere!})
%
The adjoint action of $\algEX$ on itself
is an integrable representation, which means that
$$
\forall x\in\algEX\qquad
\exists n\in\Z_{+}:\quad
\ad(e_{\a_i})^n(x) = 0,\quad
\ad(f_{\a_i})^n(x) = 0,\qquad
i=-1\dots 8
$$
Among other things, it implies that the Lie group
$\exp \algEX$ can be defined.
It also implies that $\weylEX$ preserves multiplicities of roots.
Therefore, all real roots have multiplicity 1.
However, imaginary roots can have multiplicities greater than 1.
The multiplicities of the imaginary roots of $\algEIX$ are given by
$$
\mult(n\delta)=8, \qquad 0\neq n\in \Z.
$$
There is no known closed formula for the multiplicities of the imaginary roots
of $\algEX\equiv E_{10}$.
However, a closed formula
has been derived in \cite{KMW} for roots of ``affine levels'' 1 and 2
(this term will be explained below).
We outline the derivation of these multiplicities in the next subsection.
See also \cite{Bauer:1996ca}\cite{Gebert:1997hx}
for a list of many roots and their multiplicities.

We will make a few extra observations before we continue.
\begin{prop}[Lemma 5.3 and Theorem 5.4 of \cite{KacBook}]
\label{prop:weylimaginary}
Every imaginary root can be uniquely written as $\g = w(\a)$ for a Weyl-group element
$w\in\weylEX$ and $\a\equiv\sum_{i=-1}^8 k_i\a_i\in\latEX$ satisfying: (i)
$\kform{\a}{\a_i}\le 0$ for all simple roots ($i=-1\dots 8$), and (ii)
the subdiagram of the Dynkin diagram consisting of all vertices such that
$k_i\neq 0$ is connected.
\end{prop}

\begin{prop}\label{prop:isotropic}
Every imaginary root $\a\in\ImRootsEX$ that satisfies $\kform{\a}{\a}=0$
is $\weylEX$-equivalent to $n\delta$ for some $0\neq n\in\Z.$
Its multiplicity is therefore exactly $8.$
\end{prop}
\begin{proof}
This follows immediately from proposition 5.7 of \cite{KacBook},
which uses \propref{prop:weylimaginary}.
\end{proof}

\begin{prop}\label{prop:sqmintwo}
Every positive imaginary root $\a\in\ImRootsEX$ that satisfies $\kform{\a}{\a}=-2$
is $\weylEX$-equivalent to $\a_{-1}+2\a_0+4\a_1+6\a_2+8\a_3+10\a_4+12\a_5+8\a_6+4\a_7+6\a_8.$
\end{prop}
\begin{proof}
We use the same technique as in the proof of proposition 5.7 of \cite{KacBook}.
We set
$\a=\sum_{i=-1}^8 k_i\a_i$ with $k_i\ge 1$ (otherwise $\a^2\ge 0$).
Using \propref{prop:weylimaginary}, we may assume that
$\kform{\a}{\a_i}\le 0$ for all $i=-1\dots 8$.
Then
$-2 = \kform{\a}{\a} = \sum_{i=-1}^8 k_i\kform{\a}{\a_i}.$
But every term on the righthand side is negative or zero.
Since all $k_i$'s are positive we are left with three options:
(i) $\kform{\a}{\a_s} = \kform{\a}{\a_t} = -1$ for some $-1\le s<t\le 8,$
and $\kform{\a}{\a_i}=0$ for all $i\neq s,t$;
(ii) $\kform{\a}{\a_s} = -2$ and $k_s = 1$ for some $-1\le s\le 8,$
and $\kform{\a}{\a_i}=0$ for all $i\neq s$;
(iii) $\kform{\a}{\a_s} = -1$ and $k_s = 2$ for some $-1\le s\le 8,$
and $\kform{\a}{\a_i}=0$ for all $i\neq s.$

Using the inverse of the Cartan matrix given in \eqref{AInverse}
below, we can solve all $k_i$'s in each case above, and check whether
$\a^2 = -2.$ It turns out that there
is a unique solution, and only for case (iii) with $s=2,$ which is the root
given above.
\end{proof}
As we shall see in \secref{subsubsec:multiplicity}, the multiplicity of
the root is $44.$ Therefore, all roots $\a$ with $\a^2=-2$ have multiplicity $44.$

\begin{defn}
We will say that a root $\a$ is {\it prime} if it cannot be written
as $\a = n\b$ for some integer $n>1$ and a root $\b.$
\end{defn}
All real roots are prime, but imaginary roots are not necessarily prime.
Since all roots with $\kform{\a}{\a}=0$ are Weyl-equivalent to 
a multiple of the root $\delta,$
it follows that all positive
prime roots with $\kform{\a}{\a}=0$ are Weyl equivalent
to the root $\delta.$

Let us  summarize the various terms in the following table:
\vskip 5pt
\begin{tabular}{ll}
$\CarEX$ & Cartan subalgebra of $E_{10}$\\
$\rtEX$ & Set of all roots of $E_{10}$\\
$\rtEX_{+}$ & Set of positive roots of $E_{10}$\\
$\ReRootsEX$ & Set of real roots ($\a^2 = 2$) of $E_{10}$\\
$\ImRootsEX$ & Set of imaginary roots ($\a^2 \le 0$) of $E_{10}$\\
$\latEX$ & Root lattice of $E_{10}$\\
$\weylEX$ & Weyl group of $E_{10}$\\
$\succeq$ & partial order on $\CarEX^*$ \\
$\hgt$ & height of a root\\
$\CarEIX$ & Cartan subalgebra of $E_9$ \\
$\rtEIX$ & Set of all roots of $E_9$\\
$\vdots$ & \\
$\delta$ & minimal positive imaginary root of $E_9$ \\
$\CarEVIII$ & Cartan subalgebra of $E_8$ \\
$\rtEVIII$ & Set of all roots of $E_8$\\
$\vdots$ & \\
\end{tabular} 

\subsubsection{Dimensions of Level-1 Root Spaces}
\label{subsubsec:multiplicity}
% - - - - - - - - - - - - - - - - - - - - - - - - - - - - - - - - - - - - -
For an element \textit{$\a=\sum_{i=-1}^{8}k_{i}\a_{i}\in \latEX$},
$-k_0=\kform{\a}{\delta}$ is called the {\it affine level} of $\a$.
Here $\delta\in\rtEIX\subset\rtEX$ was defined in \eqref{DeltaRoot}.
We denote the set of all roots of $E_{10}$ at affine level $l$ by
$\RootsEXatLevel{l}.$

The formula
$$
\mult(\a)=\pcoeff{8}\bigl(1-\frac{\kform{\a}{\a}}{2}\bigr),\qquad
-k_0 = 0,1.
$$
is derived in \cite{KMW} for $\a$
a level 0 or level 1 root of $E_{10}$.
By definition, $\pcoeff{8}(k)$ is the coefficient of $q^k$ in
$1/\prod_{n=1}^{\infty}(1-q^n)^8.$
Up to numerical prefactors, the generating function of bosonic objects
$\prod_{n=1}^{\infty}[1/(1-q^n)]$ is ubiquitous in string theory,
and its appearance in this new context is very intriguing.

The derivation in \cite{KMW} makes reference to Chapter 12 in \cite{KacBook},
and we briefly fill in those details here.  
Define the {\it weights} of $E_{10}$ as
$$
\wgtEX \defineas
\{\lambda\in \CarEX^*:\kform{\lambda}{\a_i}\in \Z,\quad i=-1,0,\cdots, 8\},
$$
and define the {\it dominant weights} as
\be\label{DefDominantWeightsI}
\wgtEX_{+}\defineas
\{\lambda\in \wgtEX: \kform{\lambda}{\a_i}\geq 0,\quad i=-1,0,\cdots, 8\}
\ee
In \cite{KMW}, the {\it dominant weights} in the weight lattice $\latEX$ 
are defined in a different way:
$$
\wgtEX_{+}=\{\sum_{i=-1}^{8} k_i\fwEX_i: k_i\in \N\}
$$
where $\fwEX_i$ ($i=-1,0,\dots, 8$)
are the {\it fundamental weights},
\be\label{DefDominantWeightsII}
\kform{\fwEX_i}{\a_j}=\delta_{i,j},\qquad i,j=-1,0,\dots,8.
\ee
The two definitions are equivalent.
It is obvious that $\wgtEX\supseteq\latEX$,
which is true for any Kac-Moody algebra.
For $E_{10}$, since $\det(A_{10})=-1$, it follows that $\wgtEX=\latEX$.

The fundamental weights are calculated as follows \cite{KMW}.
Expand $\fwEX_i=\sum_{k=-1}^{8}c_{ik}\a_k$.
Then we solve 
$$
\sum_{k=-1}^{8}c_{ik}\kform{\a_k}{\a_j}
=\sum_{k=-1}^{8}c_{ik} a_{kj}=\delta_{i,j}.
$$
Thus, the coefficients $c_i$ are the rows
of the inverse of the Cartan matrix $A_{10}=(a_{ij})$:
\be\label{AInverse}
-(A_{10})^{-1}=
 \left( \begin{array}{cccccccccc}
       0 & 1 & 2 & 3 & 4 & 5 & 6 & 4 & 2 & 3 \\
       1 & 2 & 4 & 6 & 8 & 10 & 12 & 8 & 4 & 6 \\
       2 & 4 & 6 & 9 & 12 & 15 & 18 & 12 & 6 & 9 \\
       3 & 6 & 9 & 12 & 16 & 20 & 24 & 16 & 8 & 12 \\
       4 & 8 & 12 & 16 & 20 & 25 & 30 & 20 & 10 & 15 \\
       5 & 10 & 15 & 20 & 25 & 30 & 36 & 24 & 12 & 18 \\
       6 & 12 & 18 & 24 & 30 & 36 & 42 & 28 & 14 & 21 \\
       4 & 8 & 12 & 16 & 20 & 24 & 28 & 18 & 9 & 14 \\
       2 & 4 & 6 & 8 & 10 & 12 & 14 & 9 & 4 & 7 \\ 
       3 & 6 & 9 & 12 & 15 & 18 & 21 & 14 & 7 & 10 \end{array}\right)
\ee
We have
$$
\fwEX_{-1}=-\delta,\qquad
\fwEX_{0}=-\a_{-1}-2\delta.
$$
{}From \cite{KacBook} (chapter 5) we know that the set of negative imaginary roots
$\NegImRootsEX$ is $\weylEX$-invariant.
The orbit of $\weylEX$ on an imaginary root of $\algEX=E_{10}$ intersects
$\wgtEX_{+}$ exactly once; the intersection
root $\mu$ is the one that maximizes $\hgt(\mu)$ \cite{KacBook} (chapter 5).
Since $\algEX$ is integrable as
an adjoint representation of itself, 
the Weyl group $\weylEX$ preserves root multiplicities, so it suffices to find the multiplicities of $\NegImRootsEX\cap\wgtEX_{+}$. 
It is easily checked from the second definition of
$\wgtEX_{+},$ given above \eqref{DefDominantWeightsII}, that dominant weights
that are also roots at level-1 are of the form
$$
\RootsEXatLevel{1}\cap \wgtEX_{+}
=\{\fwEX_{0}+k_{-1}\fwEX_{-1}
=-\a_{-1}-(k_{-1}+2)\delta: k_{-1}\in \N\}.
$$
The idea in \cite{KMW} is to determine the multiplicities of these level-1 roots.

Given $\fwEX\in\CarEX$, denote by $L(\fwEX)$ the irreducible representation
of $E_9$ with highest weight $\fwEX$ (Chapter 9 of \cite{KacBook}).
$L(\fwEX)$ has weight space decomposition
$L(\fwEX)=\bigoplus_{\lambda\leq\fwEX}V_{\lambda}$, where $\dim(V_{\fwEX})=1$.
$L(\fwEX)$ is integrable if and only if $\fwEX\in \wgtEX_{+}$
(Chapter 10 of \cite{KacBook}).

Note that 
$\RootsEXatLevel{1}$, defined at the beginning of this subsection,
is a representation of $\algEIX\equiv E_9.$
{}From now till the rest of this subsection we restrict attention to this
representation.
%{}From now on we restrict attention to the representation
%$\algEX_{[1]}=\RootsEXatLevel{1}\cap\algEX$ of $\algEIX=E_9$.

We note that $-\a_{-1}=\fwEX_{0}+2\delta$,
and $L(\fwEX_{0}+2\delta)$ is an integrable
highest weight representation of $E_9$.
The level of the representation, as a representation of
an affine Lie algebra, is $\kform{-\a_{-1}}{\delta}=1$.   
In general,
let $P(\fwEIX)$ be the set of weights of a representation $L(\fwEIX)$ of
$\algEIX\equiv E_9$,
with $\fwEIX\in \wgtEIX_{+}$, 
where $\wgtEIX, \wgtEIX_{+}$ are defined as in
\eqref{DefDominantWeightsI} but for $E_9$:
$$
\wgtEIX\defineas\{\lambda\in \CarEIX^{*}:
\kform{\lambda}{\a_i}\in \Z,\quad i=0,\dots, 8\},
$$
$$
\wgtEIX_{+}\defineas
\{\lambda\in \wgtEIX: \kform{\lambda}{\a_i}\geq 0,\quad i=0,\dots, 8\}.
$$
Then $\lambda\in P(\fwEIX)$ is called {\it maximal} if
$\lambda+\delta \notin P(\fwEIX)$.
We denote the set of maximal weights of $L(\fwEIX)$ by
$$
\Max(\fwEIX)\defineas \{\lambda\in P(\fwEIX): \lambda+\delta\notin P(\fwEIX)\}
$$
\begin{claim}
$\Max(\Lambda)$ is preserved by the Weyl group $\weylEIX$ of $\algEIX\equiv E_9$.
\end{claim}
\begin{proof}
Suppose $w(\lambda)+\delta\in P(\fwEIX)$ for some $\lambda\in \Max(\fwEIX)$
and $w\in \weylEIX$.  Then $w^{-1}(w(\lambda)+\delta)\in P(\fwEIX)$.
But $\lambda+w^{-1}(\delta)=\lambda+\delta$, so we have a contradiction.
\end{proof}

Any orbit of $\weylEIX$ on $P(\fwEIX)$ intersects $\wgtEIX_{+}$ once;
the intersection weight $\mu$ being the weight such that $\hgt(\fwEIX-\mu)$
is minimal in its $\weylEIX$ orbit.
In particular, any maximal weight is
$\weylEIX$-equivalent to a maximal weight in $\wgtEIX_{+}$.
Since $L(\fwEX_{0}+2\delta=-\a_{-1})$
is highest weight, $-\a_{-1}+\delta$ is not in $P(\fwEX_{0}+2\delta)$;
therefore $\fwEX_{0}+2\delta$ is a maximal weight 
in $P(\fwEX_{0}+2\delta)\cap \wgtEIX_{+}$.
It is the unique such weight \cite{KMW}.
{}From previous remarks it then follows that any weight of $\Max(\fwEIX_{0}+2\delta)$
is $\weylEIX$-equivalent to $\fwEX_{0}+2\delta$.
We now state
\begin{prop}[$12.5 (e)$ of \cite{KacBook}]
For any $\mu\in P(\fwEIX)$, there exists a unique $\lambda\in \Max(\fwEIX)$
and unique $n\geq 0$ such that $\mu=\lambda-n\delta$.
Furthermore, for $\lambda\in P(\fwEIX)$, the set
$\{n\in \Z : \lambda-n\delta\in P(\fwEIX)\}$
is an interval $[-p,\infty)$ with $p\geq 0$, and the function
$t\mapsto\mult_{L(\fwEIX)}(\lambda-t\delta)$
is non-decreasing on the interval.
Moreover, if $0\neq x\in \algEIX_{-\delta},$ (where $\algEIX_{-\delta}\subset\algEIX$
is the $8$-dimensional subspace of the Lie algebra $E_9$
of all the elements $x\in\algEIX$ with weight $-\delta$) the map 
$\ad (x):L(\fwEIX)_{\lambda-t\delta}\rightarrow L(\fwEIX)_{\lambda-(t+1)\delta}$
given by $y\mapsto [x,y]$ is injective.
\end{prop}
These observations imply
\be\label{DecompositionPLambda}
P(\fwEIX)=\bigsqcup_{\lambda\in \Max(\fwEIX)}\{\lambda-n\delta: n\geq 0\}.
\ee
(The union is disjoint.)

We will now define a few characters.
The expressions below are {\it formal} series in the formal
variables $e^\mu$ where $\mu$ runs over all possible weights.
They are of the form $\sum_\mu k_\mu e^\mu$ where $k_\mu$
are integers. Two such series can be multiplied to yield a series of
a similar form, and the integer multiplicities $k_\mu$ can be read off
the coefficient of $e^\mu.$ (There are actually
some restrictions on multiplying two series -- it is required that
each resulting
$k_\mu$ will have a finite number of contributions, but we do not 
need to worry about that here.)
We will also use the convention that 
$(1-e^{-\mu})^{-1}=1+e^{-\mu}+e^{-2\mu}+\cdots$.

First, for $\lambda\in \Max(\fwEIX),$ define
$$
a_{\lambda}^{\fwEIX}
\defineas\sum_{n=0}^{\infty}\mult_{L(\fwEIX)}(\lambda-n\delta)e^{-n\delta}
$$
Also, define the character  
$\Ch L(\fwEIX)$ of $P(\fwEIX)$ as
$$
\Ch L(\fwEIX)\defineas
\sum_{\lambda\in P(\fwEIX)}(\dim_{L(\fwEIX)}\lambda)e^{\lambda}.
$$ 
The above decomposition \eqref{DecompositionPLambda} of $P(\fwEIX)$ implies that 
$$
\Ch L(\fwEIX)=\sum_{\lambda\in \Max(\fwEIX)}e^{\lambda}a_{\lambda}^{\fwEIX}.
$$ 
We now return to the level-1 representation of interest, $L(\fwEX_{0}+2\delta)$.
We proved above that any $\weylEIX$-orbit in
$\Max(\fwEIX)$ intersects $\wgtEIX_{+}$ exactly once.
Since $\wgtEIX_{+}\cap\RootsEXatLevel{1}=\fwEX_{0}+2\delta$,
there is therefore only one $\weylEIX$ orbit.  
The character $\Ch L(\fwEX_{0}+2\delta)$ therefore contains the term
$$
e^{\fwEX_{0}+2\delta}a_{\fwEX_{0}+2\delta}^{\fwEX_{0}+2\delta},
$$
and a term with the same
root multiplicities and the same values of $\kform{\a}{\a}$
for each maximal weight that is $\weylEIX$-equivalent to $\fwEX_{0}+2\delta$. 
To proceed, we quote
\begin{prop}[12.13 of \cite{KacBook}]
 Let $\fwEIX\in\wgtEIX_{+}^{1}\defineas \wgtEIX_{+}\cap\RootsEXatLevel{1}.$
Then (and, by the way,
this is true in general for affine algebras of type $X_{N}^{(r)}$, 
where $X=A,D$ or $E$),
$$
a_{\fwEIX}^{\fwEIX}=\prod_{n=1}^{\infty}(1-e^{-n\delta})^{-\mult(n\delta)}.
$$
\end{prop}

Recalling the realization of the affine algebra $\algEIX\equiv E_9$
as a Lie algebra of regular polynomial maps from $\C^{*}$ to $\algEVIII\equiv E_8$,
we know that $\dim(\algEIX_{n\delta})=\dim(\CarEVIII\otimes t^n)=8.$  
Since $\fwEX_{0}+2\delta\in \wgtEIX_{+}^{1}$,
the above observation implies a term
$$
e^{\fwEX_{0}+2\delta}\prod_{n=1}^{\infty}(1-e^{-n\delta})^{-8}
$$  
in the character $\Ch L(\fwEX_{0}+2\delta).$
Let $\a=\fwEX_{0}+2\delta-k\delta=-\a_{-1}-k\delta.$
We have
$$
\kform{\a}{\a}=2(1-k)\Rightarrow k=1-\frac{\kform{\a}{\a}}{2}.
$$
Define 
$\pcoeff{8}(k)$ to be the coefficient of $e^{-k\delta}$ in 
$\prod_{n=1}^{\infty}(1-e^{-n\delta})^{-8}$, and we have that $\mult(\a)=\pcoeff{8}(k)$.
Putting this together gives Kac's result
$$
\mult(\a)=\pcoeff{8}\bigl(1-\frac{\kform{\a}{\a}}{2}\bigr)
$$
for $\a$ a level-0 or level-1 root. 

\subsubsection{``Physical'' basis for the Cartan subalgebra of $E_{10}$}
\label{subsubsec:physicalbasis}
% - - - - - - - - - - - - - - - - - - - - - - - - - - - - - - - - - - - - -
It is convenient to pick a basis for the Cartan subalgebra of $E_{10}$
that exhibits the $sl(10)\subset E_{10}$ subalgebra manifestly.
In this basis a vector $\vh\in\CarEX$ has components
\be\label{vhComponents}
\vh = (h_1, h_2, \dots, h_{10}).
\ee
The relation to the basis $\a_{-1},\dots,\a_8$
of \secref{subsubsec:construction} is given by
\be\label{vhSimple}
\vh = 
\sum_{i=-1}^5 \Bigl(\sum_{j=1}^{i+2} h_j\Bigr)\a_i^\vee
+\frac{1}{3}\Bigl(2\sum_{j=1}^8 h_j -h_9 -h_{10}\Bigr)\a_6^\vee
+\frac{1}{3}\Bigl(\sum_{j=1}^9 h_j -2h_{10}\Bigr)\a_7^\vee
+\frac{1}{3}\Bigl(\sum_{j=1}^{10}h_j\Bigr)\a_8^\vee.
\ee
Acting as a subgroup of the Weyl group $\weylEX$ of $E_{10},$
the Weyl group of $sl(10)$, which is the permutation group $S_{10}$
simply permutes the components $h_1,\dots, h_{10}.$
The Cartan metric can be written in this basis as
\be\label{CartanMetricH}
\norm{\vh}^2 = \sum_{i=1}^{10} h_i^2-\Bigl(\sum_{i=1}^{10} h_i\Bigr)^2.
\ee
Similarly, we define a ``physical'' basis for $\CarEX^*$ as follows:
\bear
\a_{-1} &=& (1,-1,0,0,0,0,0,0,0,0),\nn\\
\a_0 &=& (0,1,-1,0,0,0,0,0,0,0),\nn\\
&\vdots&\nn\\
\a_7 &=& (0,0,0,0,0,0,0,0,1,-1),\nn\\
\a_8 &=& (0,0,0,0,0,0,0,1,1,1),\nn
\eear
For any root $\a,$ we will frequently use the notation
$$
\a^2 \defineas \kform{\a}{\a}.
$$

% Physics
\subsection{Billiard cosmology}\label{subsec:billiard}
% --------------------------------------------------------------------------
We will now review the classical evolution of a universe constructed
from M-theory on $T^{10}.$ We used the term ''constructed from'' because,
as we shall see, inclusion of generalized fluxes can change the topology.
Our review is based in part on \cite{Banks:1998vs} and \cite{Damour:2002et}.

The initial ansatz is a Kasner-like metric
\be\label{KasnerMetric}
ds^2 = -dt^2 + \sum_{i=1}^{10} R_i(t)^2 dx_i^2,\qquad
0\le x_i< 2\pi,\qquad i=1\dots 10.
\ee
Einstein's equations are solved by
$$
\log \frac{R_i(t)}{R_i(t_0)}= p_i\log \frac{t}{t_0},\qquad i=1\dots 10,
$$
(for some fixed arbitrary $t_0$) provided that
\be\label{ConditionPs}
\sum_{i=1}^{10} p_i = \sum_{i=1}^{10} p_i^2 = 1.
\ee
For fixed
$$
\tau\equiv\log \frac{t}{t_0},
$$
define the ten-dimensional vector 
\be\label{vhDef}
\vh \equiv (\log [M_p R_1], \dots,\log [M_p R_{10}]).
\ee
It is convenient to interpret this vector as a point in the
Cartan subalgebra $\CarEX\subset E_{10}$
according to \eqref{vhComponents}.
The classical evolution of the universe is now mapped to an abstract
mechanical system of a single particle moving
on a straight line in $\CarEX.$ If we identify $\tau=\log (t/t_0)$ as
the time variable then the particle has constant velocity.
Note that, with the Cartan metric \eqref{CartanMetricH}, the configuration
space $\CarEX$ is identified with $\R^{9,1}.$

Excluding the very special case that one $p_i$ is $1$ and the rest are $0,$
\eqref{ConditionPs} implies that at least one $p_i$ has to be
negative and at least one other $p_j$ has to be positive.
This means that in the far past and in the far future at least one dimension
shrinks to zero, according to the classical solution. This observation 
invalidates the assumptions of classical 10+1D geometry both in the far past and
in the far future. As shown in \cite{Banks:1998vs}, it is still possible
to have a weakly coupled description after dimensional reduction, provided
that $d\vh/d\ct$ is timelike in the Cartan metric \eqref{CartanMetricH}.
This will not be the case if equation \eqref{ConditionPs} is satisfied,
but it could be true if we add matter. But first we include fluxes.

Denote the 4-form field strength of 10+1D supergravity by $G = dC.$
To start, suppose we turn on only the component $G_{1234}.$
Flux quantization requires it to be an integer.
It then contributes a potential term to the classical supergravity action
proportional to
$$
\sqrt{g} |G|^2 = \frac{G_{1234}^2}{(R_1 R_2 R_3 R_4)^2}V_{10}
 = \frac{G_{1234}^2}{V_{10}}(R_5 \cdots R_{10})^2,\qquad
V_{10} \equiv R_1 \cdots R_{10}.
$$
Note that in the absence of fluxes, the condition \eqref{ConditionPs}
implies that
$$
d\tau = \frac{dt}{t} = C\frac{dt}{M_p^{10} V_{10}},\qquad (G = 0).
$$
where $C$ is a constant.
In the presence of fluxes, it is more convenient to define 
{\it conformal time} as
\be\label{DefConformalTime}
%\ct \equiv \frac{V_{10}(t_0)}{t_0}\int^t_{t_0}\frac{dt'}{V_{10}(t')}
\ct \equiv \int^t_{t_0}\frac{dt'}{2\pi M_p^9 V_{10}(t')}
\ee
for some initial time $t_0.$
It then turns out that the classical equations of motion are encoded
by the Lagrangian 
\be\label{LagrangianWithFlux}
\Lag = 2\pi\Bigl\lVert\frac{d\vh}{d\ct}\Bigr\rVert^2
-\pi [(2\pi)^3 G_{1234}]^2 e^{2 (h_5 + h_6 + h_7 + h_8 + h_9 + h_{10})}
\ee
with the extra constraint that only trajectories with total energy zero
(defined with respect to the conformal time) are allowed.
The potential term that is proportional to the square of the flux $G_{1234}$
can be modeled as a sharp wall at position
$$
h_5 + \cdots + h_{10} \sim -\log G_{1234}.
$$
The mechanical system is now described by a particle moving at constant velocity
(with respect to the conformal time $\ct$) until it hits the wall.
After the collision the particle reflects off the wall, conserving energy
and momentum parallel to the wall, and continues at a constant velocity on
its new trajectory.
It turns out that the reflection off the wall can be interpreted as a Weyl reflection
in $\CarEX.$ 
That is, the reflection off the wall defines
a linear transformation on the velocity vector $d\vh/d\ct$,
which is precisely a Weyl reflection.
The position of the particle is therefore confined to lie within
a fundamental Weyl chamber of $E_{10}$
\cite{Damour:2000wm}\cite{Damour:2001sa}\cite{Damour:2002et}.
We will return to this point in \secref{subsec:withmatter}.

U-duality \cite{Hull:1994ys} acts on the vector $\vh.$
In fact the U-duality group has a subgroup that is generated
by permutations of the indices
$h_1,\dots, h_{10}$  and by the transformation
$$
\vh\longrightarrow
\bigl(
h_1 -\tfrac{2}{3}h_{123},\,
h_2 -\tfrac{2}{3}h_{123},\,
h_3 -\tfrac{2}{3}h_{123},\,
h_4 +\tfrac{1}{3}h_{123},\,
\cdots
h_{10} +\tfrac{1}{3}h_{123}
\bigr),\quad
h_{123}\equiv h_1 + h_2 + h_3.
$$
(The remaining U-duality transformation generators are transformations
that enforce periodicity of gauge fluxes such as $C_{123}$.)
It turns out that this subgroup is the Weyl group of $E_{10}$
\cite{Obers:1997kk}\cite{Obers:1998fb}.
These linear transformations preserve
the kinetic term of \eqref{LagrangianWithFlux},
since the Weyl group preserves the Cartan metric.
But they can act nontrivially on the potential term.
As we have seen above,
each potential term corresponds to a Weyl reflection in $\CarEX$.
The Weyl group acts on these reflections by conjugation, and hence changes
the position of the walls.

Some of the new walls obtained this way correspond to other fluxes, while other
walls correspond to a topology change, because
U-duality can turn the components $G_{1234}$ into components of the metric.
For example, one can get a wall that corresponds to the potential term
\be\label{GrWall}
\pi k^2 \exp\{2 (h_1 + h_2 + h_3 + h_4 + h_5 + h_6 + h_7 + 2 h_{10})\}.
\ee
This wall describes a topology change from $T^{10}$ to a circle fibration
of the $10^{th}$ direction over the $8^{th}$ and $9^{th}$ with first Chern
class $c_1 = k$. The metric is given by
$$
ds^2 = -dt^2 + \sum_{i=1}^9 R_i(t)^2 dx_i^2 
+ R_{10}(t)^2 (dx_{10} - \tfrac{k}{2\pi} x_9 dx_8)^2,
$$
and the boundary conditions are such that $x_9\rightarrow x_9+2\pi$
must be accompanied by $x_{10}\rightarrow x_{10} + k x_8.$

Finally, consider two fluxes in transverse directions, say $G_{1234}$ and
$G_{5678}$. The term \textit{$\int C\wedge G\wedge G$} of 10+1D supergravity 
implies that $G\wedge G$ is a source of 3-form flux. Since all 10 spatial
dimensions are compact an anti M2-brane must be present to absorb the flux
\cite{Sethi:1996es}. The Kasner cosmology must now also contain matter
in addition to fluxes.

% Math-Physics connection
\subsection{Instantons and positive roots}\label{subsec:instantons}
% --------------------------------------------------------------------------
We have mentioned in \secref{subsec:billiard} that fluxes
such as $G_{1234}$ correspond to real positive roots of $E_{10}.$
We will now discuss this correspondence in more detail.
Instead of discussing the fluxes themselves, it is convenient to discuss 
processes that change the flux by one unit.
These are the {\it instantons} of M-theory.
For example, the flux $G_{1234}$ can be changed by one unit via an instanton
that can be interpreted as an M5-brane with Euclidean world-volume,
wrapping the $5^{th},\dots,10^{th}$ directions
\cite{Becker:1995kb}\cite{Witten:1996bn}.
Analogous Euclidean branes can also be constructed in string theory
and supergravity. In this context they are
known as S-branes \cite{Gutperle:2002ai}-\cite{Chen:2002yq}.

Let us list the various possible Euclidean objects present for M-theory
on $T^d$ with $d\le 8.$
These are: Kaluza-Klein particles, M2-branes, M5-branes,
and Kaluza-Klein monopoles.
Let $R_1,\dots, R_d$ be the radii
of $T^d$ and $M_p$ be the Planck mass.
In the absence of fluxes, the
actions of these objects are, up to permutations of the indices,
\be\label{InstActionsExplicit}
2\pi R_1 R_2^{-1},\quad
2\pi M_p^3 R_1 R_2 R_3,\quad
2\pi M_p^6 R_1 R_2 R_3 R_4 R_5 R_6,\quad
2\pi M_p^9 R_1 R_2 R_3 R_4 R_5 R_6 R_7 R_8^2.
\ee
The correspondence with positive roots of $E_8$
allows us to write down a simple formula for such actions.
Up to a $2\pi$ factor,
the $\log$ of the action $\Ac_\a$ of instanton $\Brane_\a$ is given by 
$\inner{\a}{\vh}$
where $\vh$ is the vector in the Cartan subalgebra $\CarEX$ that is related 
to $R_1,\dots,R_d$ by  $\vh=(\log [M_p R_1], \dots,\log [M_p R_d]),$
similarly to equation \eqref{vhDef}.
$$
\Ac_\a = 2\pi e^{\inner{\a}{\vh}}.
$$
If $\vh$ is in a region of $\CarEX$ such that $\inner{\a_i}{\vh}\gg 1$
for all simple roots $i=-1,\dots,8$ then the Euclidean objects
can be safely interpreted as instantons. Generically, if the $R_i$'s
are given by \eqref{logRlogp} with $\vp$ timelike in the metric 
\eqref{EtenCartanMetric} then there is some choice of simple roots for $E_{10}$
for which all the instanton actions above are large 
at very late times \cite{Banks:1998vs}.

The Euclidean objects contribute instanton terms to amplitudes.
These instanton terms could, for example, be corrections to $R^4$ terms
(contractions of 4 curvature tensors) or
$\lambda^{16}$ terms
(contractions of 16 fermions) in the low-energy effective
action in the $(11-d)$ noncompact dimensions \cite{Obers:1998fb}.
%\cite{Green:1997as}-\cite{Pioline:2001jn}.
The instanton terms behave as $\Inst = \exp (-\Ac_\a + i \Fl_\a),$
where $\Fl_\a$ is the flux that couples to the object.
For example: for the Kaluza-Klein particle with action 
\textit{$\Ac_\a = 2\pi R_1/R_2$},
this flux is the ratio of metric components 
\textit{$\Fl_\a = 2\pi g_{12}/g_{22}$},
for the M2-brane with action 
\textit{$\Ac_\a = 2\pi M_p^3 R_1 R_2 R_3$}, the flux is the M-theory
3-form component \textit{$\Fl_\a = (2\pi)^3 C_{123}$}.

Strictly speaking, the instanton actions
in \eqref{InstActionsExplicit} are in the absence of off-diagonal
metric components such as $g_{12},\dots,$ and in the absence of
3-form fluxes such as $C_{123},$ etc.
In order to avoid confusion with the 4-form flux $G=dC$
we will refer to all the former collectively as {\it $\theta$-angles.}
In the presence of $\theta$-angles, the action $\Ac_\a$ is, in general, modified.
All the $\theta$-angles, together with the radii $R_1,\dots,R_d$ parameterize
the moduli space $\ModSp_d = E_d(\Z)\backslash E_d(\R)/\Kom_d$ 
\cite{Cremmer:1979up}\cite{Marcus:1983hb} where $\Kom_d$ is the
maximal compact subgroup of $E_d(\R).$
It turns out that $\Inst = \exp(-\Ac_\a + i \Fl_\a)$ is 
a harmonic function on $E_d(\R)/\Kom_d$ with respect to
the $E_d(\R)$-left invariant
metric \cite{Instantons}\cite{Ganor:1999ui}.

Furthermore, actions of simple combinations of instantons, corresponding
to Wick rotated bound states, are also given by harmonic functions. 
This observation allows us to algebraically relate 
bound states of Euclidean branes to the Lie algebra roots.
For example, an M2-brane that is wrapping the diagonal of the $R_3-R_4$
torus has action 
\textit{$\Ac = 2\pi M_p^3 R_1 R_2 \sqrt{R_3^2 +R_4^2}$}, in the absence of
$\theta$-angles.
In the presence of $\theta$-angles, this action is, in general, modified to
$$
\Ac' = \tfrac{1}{(2\pi)^2}M_p^3\int_{M2}\sqrt{\EffMet} d^3 x + i\int_{M2} C,
$$
where $\EffMet$ is the induced metric on the M2-brane and
the integrals are performed on the M2-brane worldvolume.
Let us assume that the only nonzero $\theta$-angles are
\textit{$\Fl_\a \equiv C_{123}$} and
\textit{$\Fl_\b\equiv C_{124}$}, where we have introduced the two Lie algebra roots
$\a$ and $\b$ with
$$
\Ac_\a = 2\pi M_p^3 R_1 R_2 R_3, \qquad 
\Ac_\b = 2\pi M_p^3 R_1 R_2 R_4.
$$
The harmonic function then reduces to
$$
\Phi'\Longrightarrow e^{-\Ac'+i (\Fl_\a + \Fl_\b)},\qquad
\Ac' = \sqrt{\Ac_\a^2 +\Ac_\b^2}.
$$
The point is that we can determine $\Ac'$ by calculating
the absolute value of a harmonic function
whose phase behaves as $\exp\{i\Fl_\a +i\Fl_\b\}.$
(See \cite{Ganor:1999ui} for more details.)

\section{Combinatorics}\label{sec:comb}
% ========================================================================== 
As we have reviewed in \secref{subsec:instantons}, each positive
real root $\a$ of $E_{10}$ corresponds to a unique Euclidean brane, and the action
of the brane, in the absence of fluxes,
is given by \textit{$2\pi\exp\inner{\a}{\vh}$},
where $\vh$ is the vector of $\log$s of radii
given by \eqref{vhDef}.
The actions are of the form \textit{$2\pi\prod_{i=1}^{10} (M_p R_i)^{n_i}$}
where $n_i$ are positive integers, except for Kaluza-Klein instantons
in which case one $n_i$ is $-1.$
{}From this action we can read off the dimension of the brane by counting
the number of powers for which $n_i=1$. For example for an M2-brane
the action could be \textit{$2\pi M_p^3 R_1 R_2 R_3$} and the dimension is $3.$

Formally, we can define an action corresponding to imaginary roots  of $E_{10}$
in exactly the same manner, \textit{$2\pi\exp\inner{\a}{\vh}$}.
We can then ask similar questions, such as how many $n_i$'s are $1$, about
the imaginary roots as well.
The purpose of this section is to study such ``combinatorial'' properties
of the real as well as the imaginary branes.
In this section we will naively interpret the imaginary roots as Euclidean branes.
However, in \secref{sec:imroots},
we will propose another interpretation that we believe is better.

\subsection{Root properties}\label{subsec:properties}
% --------------------------------------------------------------------------
% singleton count
% doubleton count
% tripleton count
% real roots with positive singleton count
% ...
We will now work with the Lie algebra $E_{10}.$
As we have seen in \secref{subsec:KM},
our convenient basis for the weight space $\R^{10}$ is
such that the root lattice $\Gamma\subset \R^{10}$ 
is spanned by vectors
$$
\a = (n_1, n_2, n_3, n_4, n_5, n_6, n_7, n_8, n_9, n_{10}),
\quad
n_i\in \Z,\quad (i=1\dots 10),
\quad 
\sum_{i=1}^{10} n_i \equiv 0 \mod 3,
$$
and such that the Cartan product is given by
$$
\a^2 = \sum_{i=1}^{10} n_i^2 -\frac{1}{9}\Bigl(\sum_{i=1}^{10}n_i\Bigr)^2
=\frac{1}{9}\sum_{1\le i<j\le 10} (n_i - n_j)^2
-\frac{1}{9}\sum_{i=1}^{10} n_i^2.
$$
The action of the corresponding {\it formal} brane $\Brane_\a$ 
is \textit{$2\pi\prod_i (M_p R_i)^{n_i}$} where $R_i$ are here best thought of
as abstract variables (formally, the
radii of the $10$ directions of $T^{10}$).
Strictly speaking, this is the action of an instanton in the absence of 
{$\theta$-angles,} i.e. off-diagonal
metric terms such as $g_{12},\dots,$ and in the absence of 3-form fluxes such
as $C_{123},$ etc.

The inner product of two roots $\a$ and $\a'$ is given by
$$
\inner{\a}{\a'} \equiv 
\sum_{i=1}^{10} n_i n_i' -\frac{1}{9}\Bigl(\sum_{i=1}^{10}n_i\Bigr)
\Bigl(\sum_{i=1}^{10}n_i'\Bigr)
$$
We can translate the actions to type-IIA by defining
\be\label{typeIIAred}
m_0\equiv -n_{10}+\frac{1}{3}\sum_{i=1}^{10} n_i,\quad
m_i\equiv n_i,\qquad
i=1\dots 9,
\ee
The action can then {\it formally} 
be written as \textit{$2\pi\gst^{-m_0}\prod_{i=1}^9 (\Mst l_i)^{m_i}$}
where $\gst$ is the string coupling constant, $\Mst$ is the string scale,
and $l_i$ are the {\it formal} compactification radii.
(These formulas again assume that all $\theta$-angles are zero.)
The inner product can then be written as
$$
\inner{\a}{\a'} =
2 m_0 m_0' +\sum_{i=1}^9 m_i m_i'
-\frac{1}{2}m_0\sum_{i=1}^9 m_i'-\frac{1}{2}m_0'\sum_{i=1}^9 m_i
$$

For future use we need to define
\begin{defn}
We say that a root $\a$ with indices $n_i$ is {\it thicker} ({\it thinner}) than
a root $\a'$ with indices $n_i'$ if $n_i \ge n_i'$ ($n_i\le n_i'$) for $i=1\dots 10.$
%We denote this by $\a\flat\a'$ ($\a'\flat\a$).
\end{defn}

The roots given by
$$
\rtB_{1\dots 9} \equiv (1,1,1,1,1,1,1,1,1,0)
$$
and all its permutations are the thinnest among
all the imaginary roots.

\begin{defn}
we define the {\it void count} of the root $\a$ to be the number of $i$'s for
which $n_i=0.$
\end{defn}

\begin{defn}
we define the {\it singleton count} of the root $\a$ to be the number of $i$'s for
which $n_i=1.$
\end{defn}

\begin{defn}
we define the {\it doubleton count} of the root $\a$ to be the number of $i$'s for
which $n_i=2.$
\end{defn}

\begin{claim}
\label{thm:voidimag}
The only positive imaginary roots with void count $> 0$
are permutations of the following:
$$
(0,n,n,n,n,n,n,n,n,n),\qquad n>0.
$$
\end{claim}
\begin{proof}
Without loss of generality we may assume that $n_1 = 0.$
The root is then an element of the $\algEIX=E_9$ subalgebra.
The claim immediately follows from the characterization of the imaginary roots
of $E_9$ as $n\delta.$
\end{proof}

\begin{claim}
\label{thm:nonegni}
A positive imaginary root has no negative $n_i$'s.
A positive real root has negative $n_i$'s only if it is a permutation
of $(1,-1,0,\dots,0).$
\end{claim}
The proof is given in the appendix.

Let us now describe the imaginary roots $\a$ up to the action
of the Weyl group.
\begin{prop}\label{prop:imaguptow}
Every positive imaginary root $\g\in\PosImRootsEX$ 
of $\algEX=E_{10}$ can be uniquely written
as $\g=w(\a)$ with $w\in\weylEX$ an element of the Weyl group, and
$\a\in\latEX$ given by
$$
\a = (n_1, n_2, \dots, n_{10}),
$$
and satisfying
$$
0< n_1 \le n_2 \le \cdots\le n_{10},\qquad
2(n_8 + n_9 + n_{10}) \le n_1 + n_2 + \cdots + n_7.
$$
\end{prop}
\begin{proof}
This follows immediately from
\propref{prop:weylimaginary} and \clmref{thm:nonegni}.
\end{proof}

\begin{thm}
\label{thm:imaginaryroots}
The only imaginary roots with a singleton count $s\ge 2$ are permutations of
the roots given in the table of \figref{table:ImRoots}.
In that table, we have indicated the square of the root, the singleton count $s$,
the doubleton count $d$, and the multiplicity of the root $m.$
There is an infinite number of imaginary roots of singleton count $s=1.$
\end{thm}
\noindent The proofs are given in the appendix.
\vskip 10pt
\begin{figure}[ht]
\begin{tabular}{||l|l|l|l|l||}\hline\hline
Root & Square & $s$ & $d$ & multiplicity  \\ \hline
$\rtB_{2\dots 10}          \equiv (0,1,1,1,1,1,1,1,1,1)$ &
 $(\rtB_{2\dots 10})^2 = 0$ & 9 & 0 & $m=8$\\ \hline
$\rtB_{1\dots 8;9\,10}      \equiv (1,1,1,1,1,1,1,1,2,2)$ &
 $(\rtB_{1\dots 8;9\,10})^2 = 0$ & 8 & 2 & $m=8$\\ \hline
$\rtB_{1\dots 5;6\dots 10} \equiv (1,1,1,1,1,2,2,2,2,2)$ &
 $(\rtB_{1\dots 5;6\dots 10})^2 = 0$ & 5 & 5 & $m=8$\\ \hline
$\rtB_{123;4\dots 9;10}    \equiv (1,1,1,2,2,2,2,2,2,3)$ &
 $(\rtB_{123;4\dots 9;10})^2 = 0$ & 3 & 6 & $m=8$\\ \hline
$\rtB_{12;3\dots 7;89\,10}  \equiv (1,1,2,2,2,2,2,3,3,3)$ &
 $(\rtB_{12;3\dots 7;89\,10})^2 = 0$ & 2 & 5 & $m=8$\\ \hline
$\rtB_{12;34;5 \dots 10}   \equiv (1,1,2,2,3,3,3,3,3,3)$ &
 $(\rtB_{12;34;5\dots 9\,10})^2 = 0$ & 2 & 2 & $m=8$\\ \hline
$\rtB_{12;;3\dots 9;10}    \equiv (1,1,3,3,3,3,3,3,3,4)$ &
 $(\rtB_{12;;3\dots 9;10})^2 = 0$ & 2 & 0 & $m=8$\\ \hline
$\rtB_{12;3\dots 10}       \equiv (1,1,2,2,2,2,2,2,2,2)$ &
 $(\rtB_{12;3\dots 10})^2 = -2$ & 2 & 8 & $m=44$\\ \hline
\end{tabular}
\caption{Imaginary roots of $E_{10}$ with singleton count$\ge 2.$}
\label{table:ImRoots}
\end{figure}
\vskip 10pt

Our notation $\rtB_{i_1 i_2 \dots i_s; j_1 j_2 \dots j_d; \cdots}$
indicates the indices
$i_1,\dots,i_s$ that have $n_{i_1} = \cdots = n_{i_s} = 1,$
then the indices $j_1, j_2,\dots, j_d$ that have 
$n_{j_1} = \cdots = n_{j_d} = 2,$ and so on.
By analogy with instantonic branes associated to real roots,
we will say that {\it the imaginary 
root is extended in directions $i_1,\dots,i_s.$}
Ironically, in \secref{sec:imroots} we will argue that
such a root corresponds to a brane that is {\it not} extended
in these directions. But, for the naive interpretation of
the present section the terminology above is natural.

The formal actions associated with the imaginary roots are
\bear
\Ac_{2\dots 10} &=& 2\pi M_p^9 R_2 \cdots  R_{10},\nn\\
\Ac_{1\dots 8;9\,10} &=& 2\pi M_p^{12} R_1 \cdots  R_8 (R_9 R_{10})^2,\nn\\
\Ac_{1\dots 5;6\dots 10} &=& 2\pi M_p^{15} R_1 \cdots  R_5 (R_6 \cdots R_{10})^2,\nn\\
\Ac_{123;4\dots 9;10} &=& 2\pi M_p^{18} R_1 R_2 R_3 (R_4 \cdots R_9)^2 R_{10}^3,\nn\\
\Ac_{12;3\dots 7;89\,10} &=& 2\pi M_p^{21} R_1 R_2 (R_3 \cdots R_7)^2 (R_8 R_9 R_{10})^3,\nn\\
\Ac_{12;34;5\dots 10} &=& 2\pi M_p^{24} R_1 R_2 (R_3 R_4)^2 (R_5 \cdots R_{10})^3,\nn\\
\Ac_{12;;3\dots 9;10} &=& 2\pi M_p^{27} R_1 R_2 (R_3 \cdots R_9)^3 R_{10}^4,\nn\\
\Ac_{12;3\dots 10} &=& 2\pi M_p^{18} R_1 R_2 (R_3 \cdots R_{10})^2.\nn\\
&& \label{ImInstantonActions}
\eear
For completeness we present:
\newcounter{theorem:realroots}
\begin{thm}
\label{thm:realroots}
The only real roots with a singleton count $s\ge 2$ are permutations of
the roots given in the table of \figref{table:ReRoots}.
There is an infinite number of real roots of singleton count $s=1.$
\end{thm}
\noindent
The proof is also outlined in the appendix,
and see also \cite{Elitzur:1997zn}\cite{Obers:1997kk}.
\vskip 10pt
\begin{figure}[ht]
\begin{tabular}{||l|l|l|l|l||}\hline\hline
Root & Square & $s$ & $d$ & multiplicity  \\ \hline
$\rtB_{89\,10}   \equiv (0,0,0,0,0,0,0,1,1,1)$ &
 $(\rtB_{89\,10})^2 = 2$ & 3 & 0 & $m=1$\\ \hline
$\rtB_{5\dots 10}   \equiv (0,0,0,0,1,1,1,1,1,1)$ &
 $(\rtB_{5\dots 10})^2 = 2$ & 6 & 0 & $m=1$\\ \hline
$\rtB_{3\dots 9;10}   \equiv (0,0,1,1,1,1,1,1,1,2)$ &
 $(\rtB_{3\dots 9;10})^2 = 2$ & 7 & 1 & $m=1$\\ \hline
$\rtB_{2\dots 7;89\,10}   \equiv (0,1,1,1,1,1,1,2,2,2)$ &
 $(\rtB_{2\dots 7;89\,10})^2 = 2$ & 6 & 3 & $m=1$\\ \hline
$\rtB_{234;5\dots 10}   \equiv (0,1,1,1,2,2,2,2,2,2)$ &
 $(\rtB_{234;5\dots 10})^2 = 2$ & 3 & 6 & $m=1$\\ \hline
$\rtB_{1\dots 9;;10}   \equiv (1,1,1,1,1,1,1,1,1,3)$ &
 $(\rtB_{1\dots 9;;10})^2 = 2$ & 9 & 0 & $m=1$\\ \hline
$\rtB_{1\dots 6;789;10}   \equiv (1,1,1,1,1,1,2,2,2,3)$ &
 $(\rtB_{1\dots 6;789;10})^2 = 2$ & 6 & 3 & $m=1$\\ \hline
$\rtB_{1\dots 4;5\dots 8;9\,10}  \equiv (1,1,1,1,2,2,2,2,3,3)$ &
 $(\rtB_{1\dots 4;5\dots 8;9\,10})^2 = 2$ & 4 & 4 & $m=1$\\ \hline
$\rtB_{123;456;7\dots 10}  \equiv (1,1,1,2,2,2,3,3,3,3)$ &
 $(\rtB_{123;456;7\dots 10})^2 = 2$ & 3 & 3 & $m=1$\\ \hline
$\rtB_{123;;4\dots 10}  \equiv (1,1,1,3,3,3,3,3,3,3)$ &
 $(\rtB_{123;;4\dots 10})^2 = 2$ & 3 & 0 & $m=1$\\ \hline
$\rtB_{12;3\dots 8;9;10}  \equiv (1,1,2,2,2,2,2,2,3,4)$ &
 $(\rtB_{12;3\dots 8;9;10})^2 = 2$ & 2 & 6 & $m=1$\\ \hline
$\rtB_{12;345;6\dots 9;10}  \equiv (1,1,2,2,2,3,3,3,3,4)$ &
 $(\rtB_{12;345;6\dots 9;10})^2 = 2$ & 2 & 3 & $m=1$\\ \hline
$\rtB_{12;3;4\dots 8;9\,10} \equiv (1,1,2,3,3,3,3,3,4,4)$ &
 $(\rtB_{12;3;4\dots 8;9\,10})^2 = 2$ & 2 & 1 & $m=1$\\ \hline
$\rtB_{12;;3\dots 6;7\dots 10} \equiv (1,1,3,3,3,3,4,4,4,4)$ &
 $(\rtB_{12;;3\dots 6;7\dots 10})^2 = 2$ & 2 & 0 & $m=1$\\ \hline
$\rtB_{12;;3;4\dots 10} \equiv (1,1,3,4,4,4,4,4,4,4)$ &
 $(\rtB_{12;;3;4\dots 10})^2 = 2$ & 2 & 0 & $m=1$\\ \hline
\end{tabular}
\caption{Real roots of $E_{10}$ with singleton count$\ge 2.$}
\label{table:ReRoots}
\end{figure}
\vskip 10pt
\noindent
Note that $\rtB_{89\,10}$ corresponds to an M2-brane,
$\rtB_{5\dots 10}$ to an M5-brane,
$\rtB_{3\dots 9;10}$ to a Kaluza-Klein monopole, and
$\rtB_{1\dots 9;;10}$ becomes a D8-brane after reduction to type-IIA
on the $10^{th}$ direction, as in equation \eqref{typeIIAred}.

\begin{defn}
we define the {\it hyperplane of the root}
to be the subspace of $\R^{10}$ generated by unit vectors in all directions
$i$ for which $n_i=1.$
\end{defn}
Obviously,
the hyperplane of the root has a dimension equal to the singleton count.

In the notation and terminology of \secref{subsubsec:construction},
the imaginary roots listed above can be constructed as follows.
We first write down all the positive {\it real} roots that can be obtained from
the simple root $\a_{-1}$ (with corresponding action \textit{$2\pi R_1 R_2^{-1}$})
by Weyl reflections in the Weyl group $\weylEIX$ of $E_9.$ These reflections
are generated by the simple reflections $r_0,\dots,r_8.$
The simple reflections $r_0,\dots, r_7$ act simply as permutations
of the indices of $R_2,\dots, R_9$ and we can ignore them.
Successive application of the simple reflection $r_8$ on $\a_{-1},$
with suitable permutations of the indices in between,
produces the following list of real roots:
\bear
&& 
\Ac_{\a_{-1}} = 2\pi R_1 R_2^{-1},
\nn\\
r_8\cdots r_j(\a_{-1}) &=& 
\Bigl(\sum_{i=-1}^5\a_i\Bigr) +\a_8
\Rightarrow
\Ac = 2\pi M_p^3 R_1 R_9 R_{10};
\nn\\
r_8\cdots r_j r_8\cdots r_k (\a_{-1}) &=& 
\Bigl(\sum_{i=-1}^8\a_i\Bigr) +
\a_4 + 2\a_5 + \a_6 +\a_8
\Rightarrow
\Ac = 2\pi M_p^6 R_1 R_6 R_7 R_8 R_9 R_{10};
\nn\\
r_8\cdots r_j r_8\cdots r_k r_8\cdots r_l (\a_{-1}) &=& 
\a_{-1} +2\a_0 +3\a_1 + 4(\a_2 +\a_3 +\a_4) + 5\a_5 + 3\a_6 + \a_7 + 3\a_8
\nn\\ &&
\Rightarrow
\Ac = 2\pi M_p^9 R_1 R_2 R_3 R_4 R_7 R_8 R_9 R_{10}^2;\qquad
(j,k,l=0\dots 7).
\nn
\eear
We know from \secref{subsubsec:multiplicity} that all of the remaining
level-1 roots are simply translations of 
the maximal weights of \textit{$L(-\a_{-1})$} by $-n\delta$
for $n\ge 1.$ 
[Recall that 
\textit{$\Ac_{n\delta} = 2\pi(M_p^9 R_2 R_3 R_4 R_5 R_6 R_7 R_8 R_9 R_{10})^n$}.]

For $n=1$ these translations give us the following imaginary roots
with multiplicity $m=8$:
$$
\rtB_{13456789\,10}\, (s=9),\quad
\rtB_{12345678;9\,10}\, (s=8),\quad
\rtB_{12345;6789\,10}\, (s=5),\quad
\rtB_{156;234789;10}\, (s=3),
$$
For $n=2$ these translations give us the following imaginary roots
with multiplicity $m=44$:
$$
\rtB_{12;3456789\,10}\, (s=2),\quad
\rtB_{1;2345678;9\,10}\, (s=1),\quad
\rtB_{1;2345;6789\,10}\, (s=1),\quad
\rtB_{1;56;234789;10}\, (s=1),
$$
The last three roots did not appear in our table
in \figref{table:ImRoots} above since their
singleton count is smaller than $2.$

%\subsection{Translation to type-IIA variables}\label{subsec:typeIIA}
%% --------------------------------------------------------------------------
Translating the root actions above to type-IIA notation we obtain
\bear
\Ac_{1\dots 9} \rightarrow \left\{\begin{array}{l}
\frac{2\pi}{\gst^2}\Mst^8 l_1 \cdots l_8\\
\frac{2\pi}{\gst^3}\Mst^9 l_1 \cdots l_9\\
\end{array}\right.
&\qquad&
\Ac_{1\dots 8;9\,10} \rightarrow \left\{\begin{array}{l}
\frac{2\pi}{\gst^2}\Mst^{10} l_1 \cdots l_8 l_9^2\\
\frac{2\pi}{\gst^3}\Mst^{11} l_1 \cdots l_7(l_8 l_9)^2\\
\end{array}\right.
\nn\\
\Ac_{1\dots 5;6789\,10} \rightarrow \left\{\begin{array}{l}
\frac{2\pi}{\gst^3}\Mst^{13} l_1 \cdots l_5 (l_6\cdots l_9)^2\\
\frac{2\pi}{\gst^4}\Mst^{14} l_1 \cdots l_4 (l_5\cdots l_9)^2\\
\end{array}\right.
&\qquad&
\Ac_{123;456789;10} \rightarrow \left\{\begin{array}{l}
\frac{2\pi}{\gst^3}\Mst^{15} l_1 \cdots l_3 (l_4\cdots l_9)^2\\
\frac{2\pi}{\gst^4}\Mst^{16} l_1 \cdots l_3 (l_4\cdots l_8)^2 l_9^3\\
\frac{2\pi}{\gst^5}\Mst^{17} l_1 l_2 (l_3\cdots l_8)^2 l_9^3\\
\end{array}\right.\nn\\
\Ac_{12;3456789\,10} \rightarrow \left\{\begin{array}{l}
\frac{2\pi}{\gst^4}\Mst^{16} l_1 l_2 (l_3\cdots l_9)^2\\
\frac{2\pi}{\gst^5}\Mst^{17} l_1(l_2\cdots l_9)^2\\
\end{array}\right.
&\qquad&
\Ac_{12;34567;89\,10} \rightarrow \left\{\begin{array}{l}
\frac{2\pi}{\gst^4}\Mst^{18} l_1 l_2 (l_3\cdots l_7)^2 (l_8 l_9)^3\\
\frac{2\pi}{\gst^5}\Mst^{19} l_1 l_2 (l_3\cdots l_6)^2 (l_7 l_8 l_9)^3\\
\frac{2\pi}{\gst^6}\Mst^{20} l_1 (l_2\cdots l_6)^2 (l_7 l_8 l_9)^3\\
\end{array}\right.\nn\\
\Ac_{12;34;56789\,10} &\rightarrow& \left\{\begin{array}{l}
\frac{2\pi}{\gst^5}\Mst^{21} l_1 l_2 (l_3 l_4)^2 (l_5\cdots l_9)^3\\
\frac{2\pi}{\gst^6}\Mst^{22} l_1 l_2 l_3^2 (l_4\cdots l_9)^3\\
\frac{2\pi}{\gst^7}\Mst^{23} l_1 (l_2 l_3)^2 (l_4\cdots l_9)^3\\
\end{array}\right.\nn
\eear

\subsection{Orthogonal roots}\label{subsec:orthogonality}
% --------------------------------------------------------------------------
For two real roots $\a,\b,$ the condition $\kform{\a}{\b}=0$
has a physical interpretation in terms of the corresponding
instantons 
\cite{Ivashchuk:1996ru}-\cite{Englert:2003pd}\cite{Ganor:1999ui}\cite{Ganor:1999uy}.
It means that the two instantons can ``bind at threshold,'' 
\cite{Sen:1995vr}\cite{Douglas:1995bn} so that
the bound instanton has only one time-translation zero-mode and its action is
the sum of the actions of the two individual instantons (at least when all the
$\theta$-angles are set to zero).
For example, an M2-brane with action \textit{$2\pi M_p^3 R_1 R_4 R_5$} can bind
at threshold to an M2-brane with action \textit{$2\pi M_p^3 R_1 R_2 R_3$}.
It can also bind at threshold to a Kaluza-Klein instanton
with action \textit{$2\pi R_2 R_4^{-1}$}, and so on.
We will now calculate which imaginary roots from the lists above
are orthogonal to various real roots.

An imaginary root that is extended in directions $1,\dots, s$ 
(see the definitions above) is orthogonal
to all the real roots corresponding to Kaluza-Klein instantons that have actions
\textit{$2\pi R_k R_l^{-1}$} for $1\le k<l\le s.$

As another example, let $\a$ be a real root that corresponds to an M2-brane
instanton.
It is orthogonal to $\rtB_{1\dots 9}$ if the
M2-brane's hyperplane is a subset of the $\rtB_{1\dots 9}$'s hyperplane.

The real root $\a$ is orthogonal to $\rtB_{1\dots 8;9\,10}$ if their hyperplanes
intersect on a dimension-$2$ plane. In this case, the intersection
has co-dimension $1$ inside the M2-brane's hyperplane. It is therefore
tempting to say that the M2-brane {\it can end} on the
$\rtB_{1\dots 8;9\,10}$-instanton,
just like an M2-brane can end on an M5-brane
\cite{Strominger:1995ac}.

The real root $\a$ is orthogonal to $\rtB_{1\dots 5;6789\,10}$ if their hyperplanes
intersect on a dimension-$1$ hyperplane (a line).

There are two distinct possibilities for $\a$ to be orthogonal to
$\rtB_{123;456789;10}$.
In one, the corresponding hyperplanes intersect along a line,
and in the other the hyperplanes intersect only at the origin.

Similarly, an M2-brane root $\a$ is orthogonal to
$\rtB_{12;34567;89\,10}$ in two distinct cases. In one,
the intersection of their hyperplanes is exactly the origin, and
in the other the intersection is a dimension-$1$ hyperplane (a line).

An M2-brane root $\a$ is orthogonal to
$\rtB_{12;34;56789\,10}$ or $\rtB_{12;3456789\,10}$ only if
the intersection of their hyperplanes is exactly the origin.

We can perform a similar analysis for a root $\a$ that corresponds to
an M5-brane, but we will not present it here.

\section{Physical interpretation of imaginary roots}\label{sec:imroots}
% ========================================================================== 

The discussion in \secref{sec:comb} assumed that imaginary
roots correspond to Euclidean branes.
We can always {\it define} the action corresponding to an imaginary
root as \textit{$2\pi\exp\inner{\a}{\vh}$}, as we did in \secref{sec:comb},
and study its combinatorial properties.
But this definition lacks sufficient physical motivation.

In this section we would like to propose an alternative interpretation
that, we believe, is more physical.
We propose that a prime positive imaginary root $\g$ with $\g^2=0$
corresponds to a Minkowski brane, and \textit{$2\pi\exp\inner{\vh}{\g}$}
describes its mass in units inverse to conformal time.
We will begin to study the imaginary roots $\g$ by
looking for two real positive roots $\a$ and $\b$ such that $\g=\a+\b.$

\subsection{Prime roots with $\g^2 = 0$}\label{subsec:primesqzero}
% --------------------------------------------------------------------------
Let $\rtB_{2\dots 10}$ be the imaginary root that corresponds to the action
$$
\Ac_{\rtB_{2\dots 10}}
 = 2\pi e^{\inner{\delta}{\vh}} 
 = 2\pi M_p^9 R_2 R_3 R_4 R_5 R_6 R_7 R_8 R_9 R_{10}.
$$
It satisfies $(\rtB_{2\dots 10})^2 = 0$,
and it is minimal in the sense that no other imaginary root
is thinner (see \secref{sec:comb}).

We will start with the roots 
that can be obtained from $\rtB_{2\dots 10}$ by a Weyl reflection.
These are all the prime roots that square to zero.
For a specific example, take $\a, \b$ corresponding to M5-branes with actions
\be\label{abRoots}
\Ac_\a =
2\pi e^{\inner{\a}{\vh}} = 2\pi M_p^6 R_1 R_2 R_3 R_4 R_5 R_6,\qquad
\Ac_\b =
2\pi e^{\inner{\b}{\vh}} = 2\pi M_p^6 R_1 R_2 R_7 R_8 R_9 R_{10},
\ee
Then $\g=\a+\b$ is an imaginary root with $\g^2 = 0$ and multiplicity $m=8$ and
\be\label{gCreated}
\Ac_\g = 2\pi M_p^{12} (R_1 R_2)^2 R_3 R_4 R_5 R_6 R_7 R_8 R_9 R_{10}.
\ee
Let \textit{$J^{+\a}$} and \textit{$J^{+\b}$}
be elements of $E_{10}$ that correspond to the real roots. 
Then the commutator \textit{$[J^{+\a}, J^{+\b}]$} is in the weight
space corresponding to $\g.$

\vskip 10pt
\begin{figure}[t]
\begin{picture}(380,160)
\thinlines

% =============
% $t_\a < t_\b$
% =============
\put(10,0){\begin{picture}(120,160)
\put(-5,0){\vector(0,1){150}}
\put(-8,155){$t$}

\newcounter{Offset}
\setcounter{Offset}{10}
\put(45,170){(a)}

% Instanton $\a$
\put(48,53){$\a$}
\put(50,50){\circle*{4}}
\multiput(-3,50)(\value{Offset},0){11}{\line(1,0){6}}
\put(45,88){$\b$}
\put(50,100){\circle*{4}}
\multiput(-3,100)(\value{Offset},0){11}{\line(1,0){6}}

\put(107,47){$t_\a$}
\put(107,97){$t_\b$}

% ++++++++++++++++++
% The hatch patterns
% ++++++++++++++++++

\thinlines

\multiput(\value{Offset},50)(\value{Offset},0){4}{\line(1,-1){50}}
\multiput(\value{Offset},100)(\value{Offset},0){4}{\line(1,1){50}}

% variables
% ---------
\newcounter{Xco}
\newcounter{Yco}
\newcounter{TLen}
\setcounter{TLen}{50}
\loop

% lower left $t_\a < t_\b$
% ------------------------
\setcounter{Xco}{0}
\setcounter{Yco}{0}
\addtocounter{Yco}{\value{TLen}}
\put(\value{Xco},\value{Yco}){\line(1,-1){\value{TLen}}}

% lower right  $t_\a < t_\b$
% --------------------------
\setcounter{Xco}{100}
\setcounter{Yco}{50}
\addtocounter{Xco}{-\value{TLen}}
\put(\value{Xco},\value{Yco}){\line(1,-1){\value{TLen}}}

% Upper left $t_\a < t_\b$
% ------------------------
\setcounter{Xco}{0}
\setcounter{Yco}{150}
\addtocounter{Yco}{-\value{TLen}}
\put(\value{Xco},\value{Yco}){\line(1,1){\value{TLen}}}

% Upper right  $t_\a < t_\b$
% --------------------------
\setcounter{Xco}{100}
\setcounter{Yco}{100}
\addtocounter{Xco}{-\value{TLen}}
\put(\value{Xco},\value{Yco}){\line(1,1){\value{TLen}}}
\addtocounter{TLen}{-\value{Offset}}
\ifnum \value{TLen}>0
\repeat
\end{picture}}

% =============
% $t_\b < t_\a$
% =============
\put(260,0){\begin{picture}(120,160)
\thicklines
\put(50,50){\line(0,1){50}}

\thinlines
\put(-5,0){\vector(0,1){150}}
\put(-8,155){$t$}

%\newcounter{Offset}
\setcounter{Offset}{10}
\put(45,170){(b)}

% Instanton $\a$
\put(48,105){$\a$}
\put(50,100){\circle*{4}}
\multiput(-3,100)(\value{Offset},0){11}{\line(1,0){6}}
\put(48,38){$\b$}
\put(50,50){\circle*{4}}
\multiput(-3,50)(\value{Offset},0){11}{\line(1,0){6}}

\put(107,47){$t_\b$}
\put(107,97){$t_\a$}

% ++++++++++++++++++
% The hatch patterns
% ++++++++++++++++++

\thinlines

% variables
% ---------
%\newcounter{Xco}
%\newcounter{Yco}
%\newcounter{TLen}
\setcounter{TLen}{80}
\loop

% lower left $t_\b < t_\a$
% ------------------------
\setcounter{Xco}{0}
\setcounter{Yco}{0}
\addtocounter{Yco}{\value{TLen}}
\put(\value{Xco},\value{Yco}){\line(1,-1){\value{TLen}}}

% lower right  $t_\b < t_\a$
% --------------------------
\setcounter{Xco}{100}
\setcounter{Yco}{100}
\addtocounter{Xco}{-\value{TLen}}
\put(\value{Xco},\value{Yco}){\line(1,-1){\value{TLen}}}

% Upper left $t_\b < t_\a$
% ------------------------
\setcounter{Xco}{0}
\setcounter{Yco}{150}
\addtocounter{Yco}{-\value{TLen}}
\put(\value{Xco},\value{Yco}){\line(1,1){\value{TLen}}}

% Upper right  $t_\b < t_\a$
% --------------------------
\setcounter{Xco}{100}
\setcounter{Yco}{50}
\addtocounter{Xco}{-\value{TLen}}
\put(\value{Xco},\value{Yco}){\line(1,1){\value{TLen}}}
\addtocounter{TLen}{-\value{Offset}}
\ifnum \value{TLen}>0
\repeat

% lower left $t_\b < t_\a$ extra part through $\b$
% ------------------------------------------------
\put(0,100){\line(1,-1){51}}
\put(60,40){\line(1,-1){40}}

\put(0,90){\line(1,-1){45}}
\put(55,35){\line(1,-1){35}}

% lower right $t_\b < t_\a$ extra
% -------------------------------
\put(10,100){\line(1,-1){90}}

% upper left $t_\b < t_\a$ extra part through $\a$
% ------------------------------------------------
\put(0,60){\line(1,1){45}}
\put(53,113){\line(1,1){37}}

\put(0,50){\line(1,1){53}}
\put(58,108){\line(1,1){42}}

% upper right $t_\b < t_\a$ extra
% -------------------------------
\put(10,50){\line(1,1){90}}

\end{picture}}

\end{picture}
\caption{
Two instantons associated with the real roots $\a,\b.$
Each instanton creates a jump in the associated flux.
The fluxes are depicted by the diagonal pattern. 
Instanton $\a$ creates a jump from a nonzero value to $0,$ while instanton
$\b$ changes another flux from $0$ to a nonzero value.
(a)~Instanton $\a$ occurs
before instanton $\b,$ and the different fluxes do not overlap;
(b)~Instanton $\a$ occurs
after instanton $\b,$ and the fluxes overlap between times $t_\b$ and $t_\a.$
In addition, a particle (the thick vertical line) associated to $\g=\a+\b$ 
is created between the two instantons.
}
\label{fig:BraneCreation}
\end{figure}
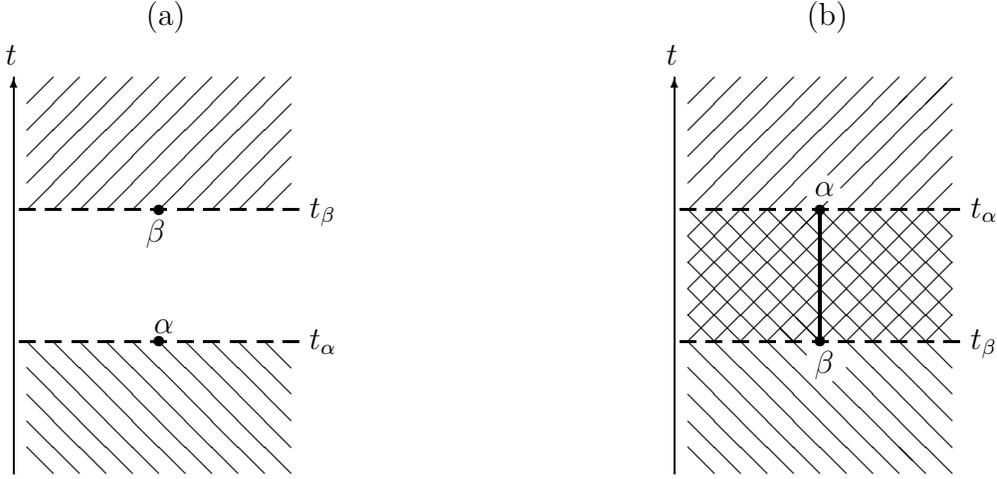
\vskip 10pt

Physically, $\a$ and $\b$ correspond to M5-brane instantons.
Let $\a$ be an instanton at time $t_\a$ and $\b$ at time $t_\b.$
Now consider switching the time order of the two instantons
from, say, $t_\a\ll t_\b$ to $t_\a\gg t_\b$ (see \figref{fig:BraneCreation}).
In this ``process'' one M5-brane passes through the other.
But this is precisely the M2-brane creation process described in \cite{Hanany:1996ie}.
After the process there is an extra M2-brane stretched along the $1^{st}, 2^{nd}$
directions and extended in time from $t_\b$ to $t_\a.$

The brane creation process has various versions for
different roots.
We will now describe a few of the versions.
In the setting that we described above, the creation of the M2-branes
can be argued as follows.
The instanton at $t_\b$ creates a jump in the flux $G_{789\,10}$
and the instanton at $t_\a$ creates a jump in the flux $G_{3456}$ so that
$$
(2\pi)^3 G_{789\,10} = 
\begin{cases}
N & \text{for $t < t_\b$}, \\
N+1 & \text{for $t > t_\b$}, \\
\end{cases}
\qquad 
(2\pi)^3 G_{3456} = 
\begin{cases}
N'-1 & \text{for $t < t_\a$}, \\
N' & \text{for $t > t_\a$}, \\
\end{cases}
$$
for some integers $N, N'.$
As we recalled in \secref{subsec:billiard}, the $\int C\wedge G\wedge G$
term of 10+1D supergravity indicates that $G\wedge G$ is a source for M2-brane
flux and there must be an equal number of anti- M2-branes to cancel that flux
\cite{Sethi:1996es}.
Therefore, together with the instanton at $t_\b,$   $N'-1$ anti- M2-branes
must also be present if $t_\b < t_\a$ and $N$ anti- M2-branes must be present
if $t_\b > t_\a.$
Setting $N=N'=0$ we see that one M2-brane is stretched between the two instantons
if $t_\b < t_\a.$

There is a U-dual process involving geometry alone 
\cite{Ganor:1996gu}\cite{Nakatsu:1997ty}.
In this case we take $\b$ to correspond to an M2-brane and $\a$ to correspond
to a Kaluza-Klein monopole such that
\be\label{apbpRoots}
\Ac_{\b'} =
2\pi e^{\inner{\b'}{\vh}} = 2\pi M_p^3 R_2 R_9 R_{10},\qquad
\Ac_{\a'} =
2\pi e^{\inner{\a'}{\vh}} = 2\pi M_p^9 R_1^2 R_2 R_3 R_4 R_5 R_6 R_7 R_8,
\ee
Then $\g=\a'+\b'$ the same as before.
This time the process of M2-brane creation can be understood entirely from
the geometry of the Kaluza-Klein monopole.
The Kaluza-Klein monopole changes by one unit
the first Chern class $c_1$ of the fibration
of the $1^{st}$ circle over the $T^2$ in the $9^{th}$ and $10^{th}$ directions.
Suppose that
$$
c_1 = 
\begin{cases}
0 & \text{for $t < t_\a.$} \\ 1 & \text{for $t > t_\a.$} \\
\end{cases}
$$
Then the M2-brane in the $2^{nd},9^{th},10^{th}$ directions cannot
pass through the Kaluza-Klein monopole.
It must get ``stuck'' at some point along 
the $9^{th}-10^{th}$ plane, and it is not hard to see that an M2-brane
that wraps the $1^{st}$ and $2^{nd}$ directions is created.

Another U-dual process involves passing a D0-brane through a D8-brane
\cite{Danielsson:1997wq} or a D4-brane through another D4-brane
\cite{Ohta:1997ir}\cite{Kitao:1998vn}. In these processes a string is created.
To relate it to our $E_{10}$ conventions, we lift type-IIA to M-theory,
taking momentum in the $2^{nd}$ direction to be related to D0-brane charge.
We then take the roots $\a''$ and $\b''$ as follows:
\be\label{appbppRoots}
\Ac_{\a''} =
2\pi e^{\inner{\a''}{\vh}} = 2\pi R_1 R_2^{-1},\qquad
\Ac_{\b''} =
2\pi e^{\inner{\b''}{\vh}} = 2\pi M_p^{12} R_1 R_2^3 R_3 R_4 R_5 R_6 R_7 R_8 R_9 R_{10}.
\ee
Again, $\g=\a''+\b''$ is the same as before and also the object that is created
is the same M2-brane stretched in the $1^{st}$ and $2^{nd}$ directions.

We conclude that the root $\g$ with
$$
\Ac_\g = 2\pi e^{\inner{\g}{\vh}} =
2\pi M_p^{12} (R_1 R_2)^2 R_3 R_4 R_5 R_6 R_7 R_8 R_9 R_{10}
$$
corresponds to a physical (temporally extended) M2-brane stretched
in the $1^{st}$ and $2^{nd}$ directions.

For another example, take $\g$ with
\be\label{gammaKK}
\Ac_\g = 2\pi e^{\inner{\g}{\vh}} =
2\pi M_p^{9} R_2 R_3 R_4 R_5 R_6 R_7 R_8 R_9 R_{10}.
\ee
We can decompose it as a sum of two real roots as $\g=\a+\b$ with
$$
\Ac_\a = 2\pi M_p^3 R_2 R_3 R_4,\qquad
\Ac_\b = 2\pi M_p^6 R_5 R_6 R_7 R_8 R_9 R_{10}.
$$
An instanton corresponding to $\a$ creates a jump by one unit in the flux
$G_{0234}$ and an instanton corresponding to $\b$ creates a jump by one unit
in the flux $G_{1234}.$
When the two fluxes $G_{0234}$ and $G_{1234}$ are present together,
we get a contribution to the field-theoretic momentum
$P^1 \equiv \int \sqrt{g} G^{0\u_1\u_2\u_3}{G^1}_{\u_1\u_2\u_3} d^{10}x.$
Since the total momentum must be zero, there must be extra Kaluza-Klein
particles with the opposite amount of momentum.
Thus, $\g$ corresponds to a Kaluza-Klein particle with momentum
in the $1^{st}$ direction. In \secref{subsec:imaction}
we will write down a mass formula for the physical objects corresponding
to the imaginary roots $\g$ that will allow us to immediately see that
$\g$ above corresponds to a Kaluza-Klein particle with mass $R_1^{-1}.$

\subsection{A mass formula}\label{subsec:imaction}
% --------------------------------------------------------------------------
There is a simple formula that relates the imaginary root $\g$
to the action of the physical brane.
Let us list the branes that we found and their ``masses,''
i.e. actions per unit time $dt.$
Let us write down the first four roots from \figref{table:ImRoots}
or equation \eqref{ImInstantonActions},
together with the masses of their corresponding branes (that we denote by $M'$),
\bear
e^{\inner{\rtB_{2\dots 10}}{\,\vh}} &=& 
   M_p^9 R_2 R_3 R_4 R_5 R_6 R_7 R_8 R_9 R_{10},
   \qquad M' = R_1^{-1},\nn\\
e^{\inner{\rtB_{1\dots 8;9\,10}}{\,\vh}} &=& 
   M_p^{12} R_1 R_2 R_3 R_4 R_5 R_6 R_7 R_8 (R_9 R_{10})^2,
   \qquad M' = M_p^3 R_9 R_{10},\nn\\
e^{\inner{\rtB_{12345;6789\,10}}{\,\vh}} &=& 
   M_p^{15} R_1 R_2 R_3 R_4 R_5 (R_6 R_7 R_8 R_9 R_{10})^2,
   \qquad M' = M_p^6 R_6 R_7 R_8 R_9 R_{10},\nn\\
e^{\inner{\rtB_{123;456789;10}}{\,\vh}} &=& 
   M_p^{18} R_1 R_2 R_3 (R_4 R_5 R_6 R_7 R_8 R_9)^2 R_{10}^3,
   \qquad M' = M_p^9 R_4 R_5 R_6 R_7 R_8 R_9 R_{10}^2.
\nn\\
&& \label{ImBraneActionsI}
\eear
The roots in equation \eqref{ImBraneActionsI} correspond
to a Kaluza-Klein particle, M2-brane, M5-brane, and Kaluza-Klein monopole,
respectively. 
The mass $M'$ can be written as
$$
M' = \frac{e^{\inner{\g}{\vh}}}{M_p^9 V_{10}},\qquad
V_{10}\equiv R_1 \cdots R_{10}.
$$
The factor $V_{10}$ might seem strange at first, but if we recall the
definition of conformal time \eqref{DefConformalTime}, we can write
the Minkowski action $\MinkAc_\g$ of the brane per unit conformal time as
\be\label{MassFormula}
\frac{d\MinkAc_\g}{d\ct} = 2\pi M_p^9 V_{10}\frac{d\Ac_\g}{dt}
 = 2\pi M_p^9 V_{10} M' = 2\pi e^{\inner{\g}{\vh}}.
\ee
We will refer to this equation as the {\it mass formula}.

The remaining roots from the table in
\figref{table:ImRoots} or equation \eqref{ImInstantonActions} are
\bear
e^{\inner{\rtB_{12;34567;89\,10}}{\,\vh}} &=& 
   M_p^{21} R_1 R_2 (R_3 \cdots R_7)^2 (R_8 R_9 R_{10})^3,
   \qquad M' = M_p^{12} R_3 \cdots R_7 (R_8 R_9 R_{10})^2,\nn\\
e^{\inner{\rtB_{12;34;56789\,10}}{\,\vh}} &=& 
   M_p^{24} R_1 R_2 (R_3 R_4)^2 (R_5 \cdots R_{10})^3,
   \qquad M' = M_p^{15} R_3 R_4 (R_5 \cdots R_{10})^2,\nn\\
e^{\inner{\rtB_{12;;3456789;10}}{\,\vh}} &=&
  M_p^{27} R_1 R_2 (R_3 \cdots R_9)^3 R_{10}^4,
   \qquad M' = M_p^{18} (R_3 \cdots R_9)^2 R_{10}^3,\nn\\
e^{\inner{\rtB_{12;3456789\,10}}{\,\vh}} &=& 
   M_p^{18} R_1 R_2 (R_3 R_4 R_5 R_6 R_7 R_8 R_9 R_{10})^2,
   \qquad M' = M_p^9 R_3 R_4 R_5 R_6 R_7 R_8 R_9 R_{10}.\nn\\
&& \label{ImBraneActionsII}
\eear
The roots in equation \eqref{ImBraneActionsII}
are unfamiliar objects, but the first three roots are Weyl reflections
(formally U-duals) of the roots of \eqref{ImBraneActionsI}.
Note that the expressions for the masses of
$\rtB_{12;34567;89\,10}, \rtB_{12;34;56789\,10}$ 
can be obtained from
the actions of the real roots
$\rtB_{2\dots 7;89\,10}, \rtB_{234;5\dots 10}$ 
(see the table in \figref{table:ReRoots}) as follows
$$
M'\left\{\rtB_{12;34567;89\,10}\right\} = 
\frac{\Ac\left\{\rtB_{2\dots 7;89\,10}\right\}}{2\pi  R_2},\qquad
M'\left\{\rtB_{12;34;56789\,10}\right\} = 
\frac{\Ac\left\{\rtB_{234;5\dots 10}\right\}}{2\pi R_2}.
$$
This is in agreement with our physical interpretation of the real roots
as instantons. The imaginary roots can be obtained by Wick rotating an
instanton back to Minkowski space.
If we replace $R_2$ with the time direction we can formally
convert the instantons to the Minkowski branes associated with the
two imaginary roots $\rtB_{12;34567;89\,10}, \rtB_{12;34;56789\,10}.$

The third imaginary root 
$\rtB_{12;;3456789;10}$ in \eqref{ImBraneActionsII} can be obtained in
a similar way from the real root 
$\rtB_{2;3456789;10}.$ The latter does not appear in the table of figure
\ref{table:ReRoots}) because its singleton count is $s=1,$ but it can
be written as $\delta + \rtB_{3\dots 9;10},$ and
$\rtB_{3\dots 9;10}$ appears in \figref{table:ReRoots} as the root
corresponding to a Kaluza-Klein monopole.

The last root satisfies $(\rtB_{12;3456789\,10})^2 = -2$ and so cannot
be a Weyl reflection of the other roots (that square to zero).
It can be obtained by a Wick rotation similar to the one
discussed above, but we have to start with $\delta\equiv\rtB_{2\dots 10}$
which is an imaginary rather than a real root, and therefore does not
correspond to an instanton. The physical interpretation of
$\rtB_{12;3456789\,10}$ is therefore different.
We will return to it in \secref{subsec:sqnegative}.

\subsection{The multiplicity}\label{subsec:immult}
% --------------------------------------------------------------------------
The imaginary roots that we studied in
 \secref{subsec:primesqzero} have a multiplicity of $m=8.$
This means that the Lie algebra $E_{10}$ has $8$ different generators for the
same root. 
The root determines the commutation relations of these generators
with the Cartan subalgebra
$\CarEX,$ and determines the mass of the brane \eqref{MassFormula}.
Thus, all $m=8$ generators with the same root yield the same mass.
In fact, from the brane creation process discussed
in \secref{subsec:primesqzero} it is
obvious that all $m=8$ generators correspond to the same object.

For example, we constructed an M2-brane stretched in the
$1^{st}$ and $2^{nd}$ directions with $\g$ given by \eqref{gCreated},
using the two instantons $\a,\b$ given by \eqref{abRoots}.
The root $\g$ was imaginary with multiplicity $m=8$ and satisfied $\g=\a+\b.$
The natural Lie algebra generator to associate with this root is
(up to a multiplicative factor)
the commutator $[J^{+\a},J^{+\b}],$ where $J^{+\a}$ and $J^{+\b}$ are the
generators associated with the roots $\a,\b.$ They are unique since $\a,\b$
are real roots with multiplicity $m=1.$
But in \eqref{apbpRoots} we decomposed $\g=\a'+\b'$ as a sum of different
real roots. It is not hard to check that 
$[J^{+\a'},J^{+\b'}]$ is linearly independent of  $[J^{+\a},J^{+\b}].$
(For this purpose,
note that a Weyl transformation in the Weyl group $\weylEX$
can be found that simultaneously maps all the roots
$\a,\b,\a',\b',\g$  to roots inside $\algEIX=E_9,$
which is tractable.)
Similarly, in \eqref{appbppRoots} we constructed yet a third decomposition
$\g=\a''+\b''$ which (as is easy to check) yields another linearly independent 
generator.

Thus, it seems that it is the root that corresponds to the brane and not
the generator.
In the following subsection we will see that the situation
is probably different for roots with negative norm.

\subsection{Roots with $\g^2 < 0$}\label{subsec:sqnegative}
% --------------------------------------------------------------------------
%
%\noindent\dotfill
%
%\noindent\dotfill
%
Take $\g=\a+\b$ with
\begin{align}
\Ac_\a &=
2\pi e^{\inner{\a}{\vh}} = 2\pi M_p^{9} R_2 R_4^2 R_5 R_6 R_7 R_8 R_9 R_{10},
\nn\\
\Ac_\b &=
2\pi e^{\inner{\b}{\vh}} = 2\pi M_p^{9} R_1 R_3^2 R_5 R_6 R_7 R_8 R_9 R_{10}.
\nn
\end{align}
Then $\g^2=-2,$ and
$$
\Ac_\g = 2\pi e^{\inner{\g}{\vh}} =
2\pi M_p^{18} R_1 R_2 (R_3 R_4 R_5 R_6 R_7 R_8 R_9 R_{10})^2.
$$
This is the root $\g=\rtB_{12;3\dots 10}$ that puzzled us at
the end of \secref{subsec:imaction}.

We need to understand what
happens when instanton $\a$ is pushed past instanton $\b.$
Instanton $\a$ creates a jump in the first Chern class $c_1$ of the fibration
of the $4^{th}$ circle over the $1^{st}$ and $3^{rd}$ directions
while $\b$ creates a jump in the first Chern class of the fibration
of the $3^{rd}$ circle over the $2^{nd}$ and $4^{th}$ directions.

We are mainly interested in the topology of the manifold.
Let $(x_1, x_2, x_3, x_4)$ be the relevant periodic coordinates with
$0\le x_1,\dots,x_4< 2\pi.$
We will describe the manifold as a $T^2$ fibration over $T^2$ with the base $B$
spanned by $x_1, x_2$ and the fiber $F$ spanned by $x_3, x_4.$
We denote a generic point of the fiber by $p\equiv (x_3, x_4).$
A point on $T^4 = B\times F$ is denoted by $(x_1, x_2, p).$

Let us first discuss the effect of a single instanton, say $\a.$
Pick an arbitrary coordinate $0< a <2\pi.$
The geometry associated with $\a$ can be described by cutting the base $B$
along the circle $x_1=a$ and gluing the part at $x_1=a-\epsilon$
(for some small $\epsilon>0$) to the part at $x_1=a+\epsilon$
by
$$
(a-\epsilon, x_2, p)\mapsto (a+\epsilon, x_2, \Monod_\a(p)),\qquad
0\le x_2<2\pi,\quad p\in F,\quad
\Monod_\a\equiv\left(\begin{array}{rr} 1 & 1 \\ 0 & 1 \\ \end{array}\right).
$$
Here $\Monod_\b\in\SL(2,\Z)$ is a linear transformation acting on the $T^2$ fiber.

\begin{figure}[t]
\begin{picture}(250,120)

\thicklines
\multiput(10,10)(100,0){2}{\line(0,1){100}}
\multiput(10,10)(0,100){2}{\line(1,0){100}}

\put(60,60){\circle*{4}}

\thinlines
\put(60,10){\line(0,1){100}}
\put(10,60){\line(1,0){100}}

\thicklines
\put(25,54){\vector(0,1){12}}\put(20,70){$\Monod_\b$}
\put(95,54){\vector(0,1){12}}\put(90,70){$\Monod_\b$}

\put(54,25){\vector(1,0){12}}\put(70,22){$\Monod_\a$}
\put(54,95){\vector(1,0){12}}\put(70,92){$\Monod_\a$}

\thinlines
\put(60,60){\circle{16}}
\put(60,52){\vector(1,0){0}}
\put(40,45){$\Monod$}

\put(10,5){\vector(1,0){100}}\put(110,1){$1$}
\put(5,10){\vector(0,1){100}}\put(3,113){$2$}
\put(62,12){$a$}
\put(12,62){$b$}

\put(150,80){
$\Monod_\a \equiv\left(\begin{array}{rr} 1 & 1 \\ 0 & 1 \\ \end{array}\right)$}
\put(150,20){
$\Monod_\b \equiv\left(\begin{array}{rr} 1 & 0 \\ 1 & 1 \\ \end{array}\right)$}
\end{picture}
\caption{
The monodromies in the $T^2$ fiber, as we pass through cuts on the $T^2$ base.
$\Monod$ is the resulting monodromy around the singular point $(a,b).$
}
\label{fig:Monodromies}
\end{figure}
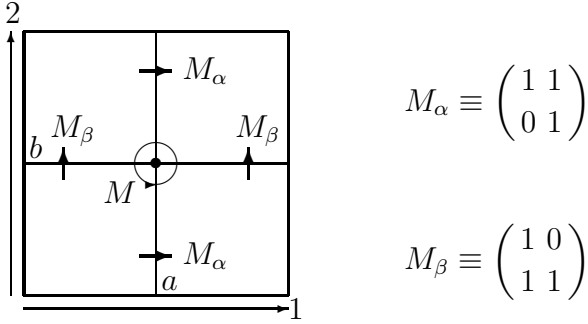

Similarly, the effect of instanton $\b$ is described by picking an arbitrary
$0\le b\le 2\pi,$ cutting the base $B$ along $x_2=b$ and gluing according to
$$
(x_1, b-\epsilon, p)\mapsto (x_1, b+\epsilon, \Monod_\b(p)),\qquad
0\le x_1<2\pi,\quad p\in F,\quad
\Monod_\b\equiv\left(\begin{array}{rr} 1 & 0 \\ 1 & 1 \\ \end{array}\right).
$$
The resulting manifold is smooth except at points that project
to $(x_1, x_2)=(a,b)$ on the base.
If we go in a circle around $(a,b)$ we discover that the fiber $F$ undergoes
a monodromy (see \figref{fig:Monodromies})
\be\label{Monodromy}
\Monod \equiv \Monod_\b \Monod_\a \Monod_\b^{-1} \Monod_\a^{-1} =
\left(\begin{array}{rr} 3 & -1 \\ 1 & 0 \\ \end{array}\right)\in SL(2,\Z).
\ee
%Note that the complex structure $\tau$ of the fiber undergoes
%$$
%\tau\rightarrow \frac{3\tau - 1}{\tau},
%$$
%which only has real fixed points 
\vskip 10pt
\vskip 10pt
So, we found out that by pushing instanton $\b$ past instanton $\a$
we create a singularity at $(x_1,x_2)=(a,b)$ that extends in directions
$5\dots 10$ and is described by a monodromy \eqref{Monodromy} in $SL(2,\Z)$
for the torus in the $3^{rd}, 4^{th}$ directions.
This is the same type of monodromies of stringy cosmic strings
\cite{Greene:1989ya} and F-theory \cite{Vafa:1996xn}.
In fact, setting $\Monod = \Monod_\b \wMonod$ with
$\wMonod\equiv \Monod_\a \Monod_\b^{-1} \Monod_\a^{-1}$ 
we see that, after reducing on the fiber $F$ to type-IIB in the spirit of F-theory,
the singularity is that of a $(0,1)$ D7-brane (associated with $\Monod_\b$)
and an anti- $(1,1)$ D7-brane (associated with $\wMonod$).

Note that a different decomposition of $\g=\a'+\b'$ with, say,
\begin{align}
\Ac_{\a'} &=
2\pi e^{\inner{\a'}{\vh}} = 2\pi M_p^{9} R_1 R_3 R_4 R_5^2 R_7 R_8 R_9 R_{10},
\nn\\
\Ac_{\b'} &=
2\pi e^{\inner{\b'}{\vh}} = 2\pi M_p^{9} R_2 R_3 R_4 R_6^2 R_7 R_8 R_9 R_{10},
\end{align}
yields an apparently different singularity.
However, the two decompositions $\g=\a+\b$ and
$\g=\a'+\b'$ define two {\it different} 1-dimensional
subspaces of the $44$-dimensional space $\algEX_\g$ as follows.
If we denote by $J^{+\a}, J^{+\b}, J^{+\a'}, J^{+\b'}\in \algEX$
nonzero Lie algebra elements in $\algEX_{\a},\dots,\algEX_{\b'}$
(unique up to a multiplicative constant) then
$[J^{+\a}, J^{+\b}]\in\algEX_\g$ and
$[J^{+\a'}, J^{+\b'}]\in\algEX_\g$ are linearly independent.
Thus, in this case it would appear that several different
objects are associated with the same root $\g$, but it might be
possible to associate them with different Lie algebra elements
in the same space $\algEX_\g.$

In any case, 
the conclusion is that the imaginary root $\g$ is associated with 
a {\it pair} of branes of different types (but perhaps not uniquely).
It would be interesting to study whether more complicated
imaginary roots can be associated with more complicated collections
of branes.
It is also interesting to note that the affine Lie algebra $E_9$
and the Kac-Moody $E_{10}$ appeared in the context of configurations
of $(p,q)$ 7-branes in the past 
\cite{DeWolfe:1998zf}\cite{DeWolfe:1998yf}\cite{DeWolfe:1998pr}
(and see also \cite{Iqbal:2001ye}).

\subsection{Nonprime roots with $\g^2 = 0$}\label{subsec:nonpsqzero}
% --------------------------------------------------------------------------
According to \propref{prop:isotropic} all imaginary roots
with $\g^2=0$ (called {\it isotropic}) are $\weylEX$-equivalent (U-dual)
to a multiple of $\delta\equiv\rtB_{2\dots 10}.$ We will now discuss
the roots $\g=n\delta$ with $n>1.$

Take the case $n=2$ and decompose $\g=\a+\b$ with
\begin{align}
\Ac_\a &=
2\pi e^{\inner{\a}{\vh}} = 2\pi M_p^{9} R_2^2  R_4 R_5 R_6 R_7 R_8 R_9 R_{10},
\nn\\
\Ac_\b &=
2\pi e^{\inner{\b}{\vh}} = 2\pi M_p^{9} R_3^2  R_4 R_5 R_6 R_7 R_8 R_9 R_{10},
\nn
\end{align}
Then $\g=2\delta,$ and
$$
\Ac_\g = 2\pi e^{\inner{\g}{\vh}} =
2\pi M_p^{18} (R_2 R_3 R_4 R_5 R_6 R_7 R_8 R_9 R_{10})^2.
$$

In this case, an interpretation of $\g$ via a brane creation process
does not work. If we try to mimic
the discussion of \secref{subsec:sqnegative}, 
we discover that the two instantons can pass through each other unharmed.
Indeed, this time
instanton $\a$ creates a jump in the first Chern class $c_1$ of the fibration
of the $2^{nd}$ circle over the $1^{st}$ and $3^{rd}$ directions
while $\b$ creates a jump in the first Chern class of the fibration
of the $3^{rd}$ circle over the $1^{st}$ and $2^{nd}$ directions.

We can create a nontrivial circle fibration of the $3^{rd}$ direction
over the $1-2$ plane by cutting a small disc around the origin of the $1-2$
plane, say of radius $\epsilon > 0,$ and gluing it back with a twist
$$
(x_1= \epsilon \cos\theta, x_2=\epsilon\sin\theta, x_3)\mapsto
(\epsilon \cos\theta, \epsilon\sin\theta, x_3 +\theta).
$$
Now let us put the two instantons together.
Start with $T^3= S^1\times S^1\times S^1$ with directions $1\dots 3.$
We can simulate the effect of $\a$ as follows.
Define
$$
\Sigma_\a\defineas \{(0, x_2, 0): \quad 0\le x_2<2\pi\}
\subset T^3.
$$
Let $\TubSig_\a$ be a small tubular neighborhood of $\Sigma_\a.$
Its topology is $\Disc\times S^1$ where $\Disc$ is the 2-dimensional disc.
The boundary of $\TubSig_\a$ has topology $T^2.$
We can pick $\TubSig_\a$ such that
$$
\partial \TubSig_\a = \{(\epsilon \cos\theta, x_2, \epsilon\sin\theta):
\quad 0\le x_2<2\pi,\quad 0\le \theta < 2\pi\},
$$
for some small $\epsilon > 0.$
Topologically, the effect of $\a$ is to cut out $\TubSig_\a$ off $T^3$
and glue it back after a Dehn twist:
$$
(\epsilon \cos\theta, x_2, \epsilon\sin\theta)\mapsto
(\epsilon \cos\theta, x_2+\theta, \epsilon\sin\theta).
$$
Similarly, $\b$ can be simulated by cutting a small tubular neighborhood
$\TubSig_\b$ around
$$
\Sigma_\a\defineas \{(0, 0, x_3): \quad 0\le x_3<2\pi\}
\subset T^3,
$$
and gluing it back with a Dehn twist.

But $\Sigma_\a$ and $\Sigma_\b$ are 1-dimensional.
We can therefore deform them so that they do not intersect inside $T^3.$
The two instantons can therefore pass through without affecting each other.
(If we had tried the same construction in \secref{subsec:sqnegative}
we would have discovered that $\Sigma_\a$ and $\Sigma_\b$ are 2-dimensional
and generically intersect at a point inside $T^4.$)

The nonprime roots must therefore have another interpretation.
We do not know what it is.

\subsection{Decomposition of level-1 imaginary roots}\label{subsec:levelone}
% --------------------------------------------------------------------------
We will now show that any level-1 imaginary root can be constructed in the
manner above, by interchanging the time order between two instantons.

\begin{claim}
Any imaginary root $\g\in\RootsEXatLevel{1}$ is $\weylEX$-dual to a sum
of two positive real roots.
\end{claim}
\begin{proof}
We have to show that there exists $w\in\weylEX$ (the Weyl group of $E_{10}$)
such that $w(\g) = \a+\b$ for $\a,\b\in\PosReRootsEX.$
Recall from \secref{subsubsec:multiplicity} that any 
$\g\in\RootsEXatLevel{1}$ can be reflected into
$\RootsEXatLevel{1}\cap\wgtEIX_{+}$ using the $E_9$ Weyl group $\weylEIX.$
Recall that $\wgtEIX_{+}$ is the set of positive dominant weights 
of $E_9\subset E_{10}$ and can be explicitly written as
$\wgtEIX_{+}\cap\RootsEXatLevel{1} = \{\a_{-1}+n\delta: n\ge 0\}.$
The latter roots are imaginary
for $n\ge 1.$ Then, for $n\ge 1,$
we can decompose $r_0(\a_{-1} + n\delta) = \a+\b$
with 
$$
\a = \a_{-1}+\a_0+\a_1,\qquad
\b = n\delta-\a_1.
$$
\end{proof}

\subsection{Decomposition of arbitrary imaginary roots}\label{subsec:gendecomp}
% --------------------------------------------------------------------------
Now let us discuss the decomposition of more general imaginary roots.
Let $\g\in\PosImRootsEX$ be an imaginary root of $E_{10}.$
Can we decompose it as a sum of two positive real roots, $\g=\a+\b$?

The problem of finding $\a$ that satisfies 
$$
\a^2 = 2,\qquad (\g-\a)^2 = 2\Longrightarrow \g^2 = 2\kform{\g}{\a},
$$
reduces to an inhomogeneous quadratic Diophantine equation in 9 integer unknowns.
(We can, for example, eliminate $n_{10}$ from the linear equation
$\g^2=2\kform{\g}{\a}$ and substitute it in $\a^2 = 2.$)
For $\g^2 < 0,$ it is not hard to see that the quadratic form is elliptic.
(Over $\R,$ the metric on $\CarEX$ is equivalent to the Lorentzian metric
on $\R^{9,1}.$ The vector $\g$ is timelike, and therefore the equations
$\a^2=2$ and $\g^2=2\kform{\g}{\a}$ define an 8-dimensional ellipsoid.)
We do not know the general answer, but we have tried to decompose several
imaginary roots by computer, and were successful each time.

For physical purposes, it is enough to address the weaker question of whether
we can find a Weyl-group element $w\in\weylEX$ such that $w(\g)=\a+\b$
for some $\a,\b\in\PosReRootsEX.$
We can actually drop the restriction of positivity for $\a,\b,$ because
of the following 
\begin{claim}
Existence of a decomposition $w(\gamma)=\a-\b$ with
$\a,\b\in\PosReRootsEX,$ implies existence of 
a  decomposition $w(\g)=\a'+\b'$
with $\a',\b'\in\PosReRootsEX.$
\end{claim}
\begin{proof}
Suppose $\g=\a-\b$.
 First assume that the minimal number of simple reflections needed to bring $\b$
to a simple root is smaller than or equal to the number required for $\a$.
Then, applying this minimal list of simple reflections, we obtain
$w'(\gamma)=\a''-\a_i,$ for some $\a''\in\PosReRootsEX,$ and
some simple root $\a_i= r_{\a_{i_s}}\circ\cdots \circ r_{\a_{i_1}}(\b).$
  To see that $\a''\in\PosReRootsEX,$ we have to use 
lemma 3.7 of \cite{KacBook} which states that the only way for a sequence of 
simple reflections $r_{\a_{i_s}} \circ r_{\a_{i_{s-1}}}\circ\cdots\circ r_{\a_{i_1}}$
to take a positive root to a negative root is by passing through a simple root 
at some stage
$\a_j = r_{\a_{i_t}}\circ\cdots \circ r_{\a_{i_1}}(\a)$ ($1\le t< s$).
But we assumed that at least $s$ simple reflections are required to turn $\a$
into a simple root, so $\a''\in\PosReRootsEX.$
We can now write
$r_{\a_i}w'(\g)=\a_i+r_{\a_i}(\a'').$
It is easy to see that $\a'\equiv r_{\a_i}(\a'')$ cannot be a simple root (otherwise
$\a'+\a_i$ would be a real root of $E_{10}$) and therefore,
by the same arguments as above, it cannot be a negative root and so must be positive.
Setting $\b'\equiv\a_i$ and $w\equiv r_{\a_i}\circ w'$
we obtain the requisite decomposition $w(\g)=\a'+\b'.$

If it is $\a$ that requires the smaller number of simple reflections to turn
it into a simple root then, following the same steps as above, we get a
decomposition $w(\g)=-\a'-\b'.$
But, according to proposition 5.2 of \cite{KacBook},
a positive imaginary root $\g$ cannot be $\weylEX$-equivalent to
a negative imaginary root $-\a'-\b'$ (in sharp contrast to real roots!).
So this case is ruled out.
\end{proof}

\subsection{Billiard cosmology with matter}\label{subsec:withmatter}
% --------------------------------------------------------------------------
To conclude this section we will apply the relation between 
physical branes and imaginary roots to billiard cosmology.
In \secref{subsec:billiard} the matter component of the universe
was provided by the fluxes. 
In this section we will add physical Kaluza-Klein particles and branes.
In the absence of fluxes, we must make
sure that the total charge of any type must be zero. We can do that
by adding an equal amount of branes and anti-branes.
As we have discussed in \secref{subsec:primesqzero}, 
the presence of fluxes can induce a brane charge.
Let $\a$ and $\b$ be two real roots such that $\g=\a+\b$
is an imaginary root that is $\weylEX$-dual to $\delta$ 
[defined in equation \eqref{DeltaRoot}].
If we turn on $N_\a$ units of flux corresponding
to $\a$ and $N_\b$ units of the $\b$-flux, then we
get effective $N_\a N_\b$ units of $\g$-charge which must be canceled
by $N_\a N_\b$ $\g$-type anti-branes.
The flux contributes a term of the form
\be\label{NaSqNbSq}
-\pi N_\a^2 e^{2\inner{\a}{\vh}}
-\pi N_\b^2 e^{2\inner{\b}{\vh}}
\ee
to the effective Lagrangian \eqref{LagrangianWithFlux}.
$N_\g$ branes (or anti-branes) 
of type corresponding to $\g$ contribute a term of the form
\be\label{gTerm}
-2\pi N_\g e^{\inner{\g}{\vh}}
\ee
to the Lagrangian.

In \cite{Damour:2000wm}\cite{Damour:2001sa}\cite{Damour:2002et}
it was argued that
each potential term $\exp\{2\inner{\a}{\vh}\}$ can be approximated
by a wall in $\vh$-space, and it was further argued that only the walls
corresponding to the simple roots $\a=\a_{-1},\dots,\a_8$ are important.
Up to a finite piece,
the other walls are generically hidden behind the walls of the simple roots.

The new terms $\exp\inner{\g}{\vh}$ that come from matter
correspond to potential terms
that are in general smaller than the terms related to the
simple roots.
If $N_\g = |N_\a N_\b|$, which is the minimal amount of branes
necessary to balance the effective charge of the fluxes, then
$$
2\pi N_\g e^{\inner{\g}{\vh}} \le \pi N_\a^2 e^{2\inner{\a}{\vh}} +\pi N_\b^2 e^{2\inner{\b}{\vh}}.
$$
In principle, however, we can let $N_\g$ be larger if we add pairs
of branes and anti-branes.
In this case, since $\g$ is lightlike, the term $\exp\inner{\g}{\vh}$
cannot be replaced by a wall because the billiard ball can 
penetrate the region where $\exp\inner{\g}{\vh}$ is large,
as can be seen after writing down the equations of motion.
[What makes this possible is the fact that the kinetic term in
the Lagrangian \eqref{LagrangianWithFlux} is not positive definite.]

In addition, the dynamics could be more complicated since the branes
could interact and annihilate.
This topic is beyond the scope of this paper.
(See \cite{StringsCosmology}
%\cite{Alexander:2003gk}\cite{Alexander:2002gj}\cite{Easther:2002mi}
%\cite{Easther:2002qk}\cite{Tseytlin:1991xk}\cite{Brandenberger:1988aj}
for a discussion on the dynamics of strings and branes in cosmology.)

It is also interesting to compare the term \eqref{gTerm} to 
the effective $\sigma$-model proposed in \cite{Damour:2002et}.
There, an effective potential which contained a sum over all positive roots
with terms of the form $\exp \{2\inner{\g}{\vh}\}$ was proposed to describe
M-theory near a spatial Kasner-like singularity.
Our term \eqref{gTerm} is different by a factor of $2$ in the exponent!
The $\sigma$-model by itself does not appear to capture this term,
as we will discuss in greater detail in \secref{sec:hamiltonian}.

\section{Interactions}\label{sec:interactions}
% ========================================================================== 
We have seen that real roots of $E_{10}$ describe fluxes and instantons,
and certain imaginary roots describe branes.
In this section we will discuss combinations of roots.
We will begin with a combination of two imaginary roots, and ask how
the features of the interactions of the corresponding branes are
related to the algebraic properties of the roots.
We will then study the effects of a flux corresponding to a real root
on a brane corresponding to an imaginary root.

\subsection{Brane interactions}\label{subsec:brinter}
% --------------------------------------------------------------------------
Take two imaginary roots $\a$ and $\b$ that correspond to physical
branes as above. 
What can we say about the interaction between the branes
from the algebraic perspective?

The inner product $\kform{\a}{\b}$ encodes the basic properties of
the interaction.
We have discussed in \secref{subsec:orthogonality} the relation
between threshold binding of instantons and the orthogonality of
their corresponding real roots.
We can now ask what is the condition on two imaginary roots $\a$ and $\b$
so that the corresponding physical branes could bind at threshold.
Let $M_\a$ and $M_\b$ be the masses (i.e. actions per
unit time) of the individual branes.
The type of interaction we are interested in
is characterized by the formation of a bound state
with mass $M_\a+M_\b,$ in the absence of $\theta$-angles.
Take for example,
$$
\Ac_\a = 2\pi M_p^{12} V_{10} R_1 R_2,\qquad
\Ac_\b = 2\pi M_p^{12} V_{10} R_3 R_4.
$$
In the absence of $\theta$-angles,
the corresponding M2-branes can bind at threshold to form an object
with mass $M_p^3 (R_1 R_2 + R_3 R_4).$
We calculate $\kform{\a}{\b} = -2.$

We conclude that the condition for binding at threshold is
\be\label{threshbind}
\kform{\a}{\b}=-2\Longrightarrow
{\mbox{binding at threshold.}}
\ee
This condition also applies
for U-dual examples, such as a Kaluza-Klein particle with
mass $R_1^{-1}$ binding to an M2-brane with mass $M_p^3 R_1 R_2,$
and so on.
In particular, the fact that an M2-brane can end on an M5-brane
\cite{Strominger:1995ac} can be traced back to the possibility
of the two objects to bind at threshold.
For this example, take
an M2-brane with mass $M_p^3 R_1 R_2$ and an M5-brane with mass
$M_p^6 R_2 R_3 R_4 R_5 R_6;$ condition \eqref{threshbind} is again satisfied.

The next type of interaction is typically characterized by forming
a bound state with mass $\sqrt{M_\a^2 +M_\b^2}.$
For example, take $\a, \b$ with
$$
\Ac_\a = 2\pi M_p^{12} V_{10} R_1 R_2,\qquad
\Ac_\b = 2\pi M_p^{12} V_{10} R_1 R_3.
$$
This corresponds to one M2-brane with mass $M_p^3 R_1 R_2$ and
a second with mass $M_p^3 R_1 R_3.$
The corresponding M2-branes can bind to form an object
with mass $M_p^3 R_1 \sqrt{R_2^2 + R_3^2},$
according to the Pythagorean theorem.

This type of interaction also occurs when a brane absorbs
a Kaluza-Klein particle and gains momentum in an orthogonal direction.
For example, take $\a, \b$ with
$$
\Ac_\a = 2\pi M_p^{9} V_{10} R_1^{-1},\qquad
\Ac_\b = 2\pi M_p^{12} V_{10} R_2 R_3.
$$
Here $\a$ corresponds to Kaluza-Klein momentum in the $1^{st}$ direction,
and $\b$ corresponds to an M2-brane in the $2^{nd}$ and $3^{rd}$ directions.
The M2-brane can absorb the momentum and get an energy of
\textit{$\sqrt{(M_p^3 R_2 R_3)^2 + (R_1^{-1})^2}$}.

A third example is furnished by an M5-brane absorbing an M2-brane which becomes
a 3-form tensor flux supported on its world-volume. In this case:
$$
\Ac_\a = 2\pi M_p^{12} V_{10} R_1 R_2,\qquad
\Ac_\b = 2\pi M_p^{15} V_{10} R_1 R_2 R_3 R_4 R_5.
$$
This is also dual to D-branes with electric or magnetic fluxes
\cite{Witten:1995im}\cite{Townsend:1995af}\cite{Douglas:1995bn}.
Inspired by the first example, we will refer to such an interaction
as {\it Pythagorean binding}.
In all these cases we have
$$
\kform{\a}{\b}=-1\Longrightarrow
{\mbox{Pythagorean binding.}}
$$
In the case of Pythagorean interaction, either $\a-\b$ or $\b-\a$ is
a positive real root.

For a third type of interaction, consider the process of brane
creation. Take, for example, the case of \cite{Hanany:1996ie} with
two M5-branes that pass through each other,
$$
\Ac_\a = 2\pi M_p^{15} V_{10} R_1 R_2 R_3 R_4 R_5,\qquad
\Ac_\b = 2\pi M_p^{15} V_{10} R_1 R_6 R_7 R_8 R_9.
$$
In this, or any of its U-dual versions, we get
$$
\kform{\a}{\b}=-4\Longrightarrow
{\mbox{Brane creation process.}}
$$
We have covered the cases $\kform{\a}{\b}=-1,-2,-4.$
It would be interesting to find the physical interpretation of other cases.

\subsection{Interactions of branes with fluxes}\label{subsec:brfluxes}
% --------------------------------------------------------------------------
In the previous section we discussed the interaction of two
branes associated to the imaginary roots $\a,\b.$
In this section we will take $\a$ to be imaginary and $\b$
to be real.

We assume that the imaginary root $\a$ corresponds to a Minkowski brane $\Brane_\a$
and the real root $\b$ corresponds to a flux.
(The instanton associated with $\b$ creates a jump in that flux.)
In this section we study the interaction of the brane $\Brane_\a$ with the flux.
We will again attempt to characterize it according to
the inner product $\kform{\a}{\b}.$

We will assume that $\vh,$ the vector of $(\log R_i)$'s, is in such an
asymptotic range that the brane $\Brane_\a$ is 
described by low-energy field theory and
that the effect of the flux can be treated perturbatively,
and we will restrict the discussion to first order.

As a first example, use the $10^{th}$ direction to reduce from M-theory
to type-IIA and take $\Brane_\a$ to be a D2-brane so that
\be\label{DtwoRoot}
e^{\inner{\vh}{\a}}  = M_p^{12} V_{10} R_8 R_9
= \frac{\Mst^{11}V_9}{\gst^3}R_8 R_9
\ee
Where we have introduced the type-IIA string scale
\textit{$\Mst\equiv M_p^{\frac{3}{2}}R_{10}^{\frac{1}{2}}$},
and coupling constant \textit{$\gst\equiv (M_p R_{10})^{\frac{3}{2}}$}.

The D2-brane is described by a $U(1)$ super-Yang-Mills theory 
with field strength $F_{\u\v}$ ($\u,\v=0\dots 2$),
$7$ scalars $\phi^I$ ($I=1\dots 7$),
and $8$ Majorana fermions $\psi^a$ ($a=1\dots 8$), with Lagrangian
\be\label{LagDtwo}
\Lag_{2+1D} = \frac{1}{4\gst}\left\lbrack
F_{\u\v} F^{\u\v} + \delta_{IJ}\px{\u}\phi^I\qx{\u}\phi^J
+ \delta_{ab}\bpsi^a\Dslash\psi^b\right\rbrack.
\ee
The index $a$ of the fermions corresponds to the spinor representation
$\rep{8}$ of the R-symmetry group $\Spin(7)$ and the index $I$ corresponds
to the vector representation $\rep{7}.$

Let us first take the flux to be an NSNS flux $H_{123}.$
The corresponding instanton is an NS5-brane in directions $4\dots 9,$ so that
$$
2\pi e^{\inner{\vh}{\b}} =
\Ac_\b = \frac{2\pi \Mst^6}{\gst^2}R_4 \cdots R_9 = 2\pi M_p^6 R_4\cdots R_9.
$$
Note that $\kform{\a}{\b} = 0.$
The effect of such a flux is to ``pin'' the brane \cite{Chakravarty:2000qd}
and add a mass term to $\Phi^1, \Phi^2,\Phi^3$ and to the fermions.
The term linear in the flux is a mass term proportional to
$H_{123}\bpsi\Gamma^{123}\psi$ where $\Gamma^I$ ($I=1\dots 7$) are Dirac matrices
of $\Spin(7).$
It is worthwhile noting that after a series of U-dualities and a Penrose limit,
the mass term above can be traced \cite{Alishahiha:2003ru}
to the mass term in the lightcone
string theory that describes pp-waves \cite{Metsaev:2001bj}\cite{Sadri:2003pr}.

For the second example, let us stay with the D2-brane but take the flux to be
an NSNS $H_{129}.$ Now the flux has one leg along the D2-brane.
The effect is 
\cite{Dasgupta:2000ry}\cite{Alishahiha:2003ru}\cite{Dasgupta:2003zx}
a nonlocal deformation of 2+1D super-Yang-Mills theory 
to a dipole 
theory \cite{Bergman:2000cw}. The first order deformation is
proportional to
$$
H_{129}F_{9\u}(\Phi^{[1}\qx{\u}\Phi^{2]}+\bpsi\Gamma^{12}\sigma^\u\psi),
$$
where $9$ is a direction on the brane,
and $\sigma^\u$ is a 2+1D Dirac matrix.

In this case
$$
2\pi e^{\inner{\vh}{\b}} =
\Ac_\b = 2\pi M_p^6 R_3 R_4\cdots R_8,
$$
and $\kform{\a}{\b} = -1.$

For a third example, we will add a Chern-Simons interaction to the 
Lagrangian \eqref{LagDtwo}.
To get the Chern-Simons interaction we will start with type-IIB
this time. Take a D3-brane in directions
$7\dots 9$ with a low energy effective action given by $N=4$ super-Yang-Mills
theory,
$\Lag_{3+1D} = \frac{1}{4\gst}
F_{\u\v} F^{\u\v} + \cdots,$ and the scalars and fermions will not concern us
this time.
To add a Chern-Simons interaction we need to recall the coupling between
the RR 0-form $\chi$ of type-IIB and the 2-form field strength $F.$
It is 
$$
\int \chi F\wedge F = -\int d\chi \wedge A\wedge F.
$$
So, if we can find an instanton that creates a constant gradient
in $\chi$ in say the $7^{th}$ direction, we can get the Chern-Simons
term $\int A\wedge F$ after dimensionally reducing to 2+1D, by forgetting
the $7^{th}$ direction. (We also get a mass term for the fermions,
as is required for supersymmetric Chern-Simons theory.)
The flux $d\chi$ is created by a D7-brane instanton.
But it is more convenient to formally T-dualize along the $7^{th}$ direction.
The D3-brane becomes a D2-brane corresponding to the 
root $\a$ as before \eqref{DtwoRoot}.
The D7-brane instanton becomes a D8-brane with formal action,
$$
2\pi e^{\inner{\vh}{\b}} =
\Ac_\b = \frac{2\pi \Mst^9}{\gst}R_1 \cdots R_9 = 2\pi M_p^{12} R_1\cdots R_9 R_{10}^3.
$$
Note that $\kform{\a}{\b} =-2.$
The D8-brane instanton turns type-IIA into a massive type-IIA 
\cite{Romans:tz}\cite{Polchinski:1995df}\cite{Hull:1998vy},
and we can arrive at the same Chern-Simons term
by studying D-branes in massive type-IIA theory \cite{Lowe:2003qy}.
\vskip 10pt
\begin{figure}[t]
\begin{picture}(420,200)

% ==========
% Mass term
% ==========
\thicklines
\put(10,20){\begin{picture}(100,150)
\put(40,135){(a) Mass term}
\put(80,115){$\kform{\a}{\b}=0$}
\multiput(0,0)(80,0){2}{\line(1,1){40}}
\multiput(0,0)(40,40){2}{\line(1,0){80}}
\put(60,60){\vector(0,1){25}}
\put(47,90){$1,2,3$}
\put(65,70){$H$}
\put(47,70){$\b$}
\put(40,-10){$8,9$}
\put(20,-10){$\a$}
\put(40,20){$\sim \bpsi\psi$}
\end{picture}}

% ===========
% Dipole term
% ===========
\thicklines
\put(160,20){\begin{picture}(100,150)
\put(40,135){(b) Dipole term}
\put(80,115){$\kform{\a}{\b}=-1$}
\multiput(0,0)(80,0){2}{\line(1,1){40}}
\multiput(0,0)(40,40){2}{\line(1,0){80}}
\put(60,60){\vector(0,1){25}}
\put(60,60){\vector(1,0){25}}
\put(53,90){$1,2$}
\put(88,58){$9$}
\put(65,70){$H$}
\put(47,70){$\b$}
\put(40,-10){$8,9$}
\put(20,-10){$\a$}
\put(35,25){$\sim F \phi\partial\phi$}
\put(45,10){$+ F\bpsi\psi$}
\end{picture}}

% =================
% Chern-Simons term
% =================
\thicklines
\put(310,20){\begin{picture}(100,150)
\put(40,135){(c) Chern-Simons term}
\put(80,115){$\kform{\a}{\b}=-2$}
\multiput(0,0)(80,0){2}{\line(1,1){40}}
\multiput(0,0)(40,40){2}{\line(1,0){80}}
\put(60,60){\vector(0,1){25}}
\put(60,60){\vector(1,0){25}}
\put(50,90){$1\dots 7$}
\put(88,58){$8,9$}
%\put(65,70){$$}
\put(47,70){$\b$}
\put(40,-10){$8,9$}
\put(20,-10){$\a$}
\put(40,20){$\sim A\wedge F$}
\end{picture}}

\end{picture}
\caption{Three types of interactions of a D2-brane with flux.
The D2-brane is in the plane of the $8^{th},9^{th}$ directions (and time).
The imaginary root associated with it is $\a.$ The flux is associated with the
real root $\b.$ The arrows indicate the directions of the flux:
(a)~A mass term appears as a result of an NSNS flux orthogonal to the brane;
(b)~A dipole interaction appears as a result of an NSNS flux with two legs 
orthogonal to the brane and one leg parallel to the brane;
(c)~A Chern-Simons term appears in massive type-IIA theory 
    (the flux permeates throughout space);
}
\label{fig:BranesAndFlux}
\end{figure}
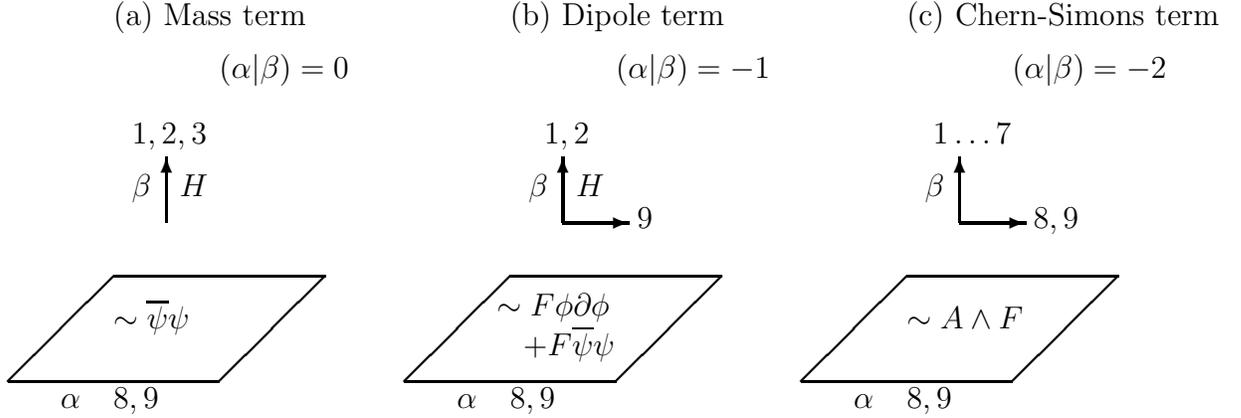
\vskip 10pt

To conclude, we have found the following interactions of fluxes with branes
(see \figref{fig:BranesAndFlux}),
\bear
\kform{\a}{\b}=0 &\Longrightarrow& {\mbox{Mass term,}}\nn\\
\kform{\a}{\b}=-1 &\Longrightarrow& {\mbox{Dipole interaction,}}\nn\\
\kform{\a}{\b}=-2 &\Longrightarrow& {\mbox{Chern-Simons.}}\nn\\
&& \label{BraneFluxInner}
\eear

\subsection{Interaction potentials}\label{subsec:potentials}
% --------------------------------------------------------------------------
How can we connect the interactions on the righthand column of
\eqref{BraneFluxInner} with the algebraic properties of the roots?

In this subsection we will write down formulas for the potential
energies of the interactions. The formulas are in the spirit of
the mass formula \eqref{MassFormula} and relate the derivative of the 
action with respect to conformal time $\ct$ to the roots.

Let us start with the case of a D2-brane in the $8^{th}, 9^{th}$ 
directions that is immersed in $H_{123}$ NSNS flux,
as in \figref{fig:BranesAndFlux}-a.
We need to calculate the mass term that 
is generated on the D2-brane world-volume.
The magnitude of the flux is \textit{$H_{123}/R_1 R_2 R_3$},
and it therefore follows that the mass term is proportional 
to $\exp\inner{\a}{\vh},$ in units dual to conformal time.

However, this formula does not tell us which degrees of freedom on the
brane (i.e. which components of the fermions) receive a mass term.
In order to distinguish the components,
it will be more convenient to work with a mass term that preserves some
supersymmetry.  This can be achieved by adding $H_{145}$ so that we now
have both $H_{123}$ and $H_{145}$ fluxes perpendicular to the brane.
The magnitudes of the fluxes $H_{123}/R_1 R_2 R_3$ and $H_{145}/R_1 R_4 R_5$
must be equal for some supersymmetry to be preserved.
Since the mass term is supersymmetric
(both fermions and bosons
get the same mass), the ground state energy will not change.
It will be simply the mass of the D2-brane. We need to find some way
to coax the mass term to show itself as a change in energy.

We have at our disposal the option to add more branes and fluxes,
and this is what we will do. We will adopt the same method used
in \cite{Ganor:1999ui}-\cite{Ganor:1999uy}.
We first note that the details of the $H_{123}$-related mass term are such 
that every state on the brane with angular momentum in the $2-3$ plane
(which corresponds to an R-symmetry generator in the field theory
on the D2-brane) gets an additional energy proportional to the angular momentum.
Similarly, the $H_{145}$ term is related to angular momentum in the $4-5$ plane.
The ground state, having zero angular momentum, is not lifted.

Thus, to test the mass term we need to add angular momentum in the $2-3$ plane, say,
and check the extra term in the energy of the state.
But from the $E_{10}$ perspective we can only easily add Kaluza-Klein momentum
in some direction $1\dots 7$ perpendicular to the brane, not angular momentum.
We need to find a trick to convert angular momentum to ordinary Kaluza-Klein momentum.

The trick is to add a ``spectator'' Kaluza-Klein monopole. 
Consider a type-IIA Kaluza-Klein monopole in $\R^{8,1}\times S^1$ space,
with $S^1$ corresponding to the $5^{th}$ direction and let the monopole be
extended in the $1^{st},6^{th},\dots, 9^{th}$ directions.
We will ignore the $1^{st}, 6^{th},\dots, 9^{th}$ directions, for the moment.
The Taub-NUT solution, corresponding to the Kaluza-Klein monopole, is
$$
ds^2 = R_5^2 U(dx_5 - \sum_{i=2}^4 A_i dx_i)^2 
+ U^{-1}\sum_{i=2}^4 dx_i^2,\quad
0\le x_5\le 2\pi,\quad
U \equiv \left(1 + {{R_5}\over {\sqrt{\sum_{i=2}^4 x_i^2 }}}\right)^{-1},
$$
where $A_i$ is the gauge field of a monopole centered at the origin.
The Taub-NUT solution is a fibration of a circle (the $5^{th}$ direction)
over $\R^3$ such that at $\infty$ the circle 
has a constant radius $R_5$.
The Taub-NUT solution is smooth at the origin.
The relevant point for us is that there is an isometry that looks like a
translation in the $5^{th}$ (the $S^1$'s) direction at $\infty$ 
and as a rotation in $SO(4)$ of the $\R^4$ tangent space 
at the origin. If we place a D-brane at the origin (and allow it to extend
in some of the other directions $1,6,\dots,9$) 
we can convert angular
momentum in directions $2,3,4,5$ perpendicular to the brane to Kaluza-Klein
momentum in the $5^{th}$ direction at $\infty$ far from the brane.
The upshot is that together with the Kaluza-Klein monopole, states with 
Kaluza-Klein momentum in the $5^{th}$ direction should get extra energy.

Now let us rephrase the story in $E_{10}$ language.
First, it will be convenient to generate the NSNS fluxes $H_{123}$ and
$H_{145}$ not via an instanton, as we did in \secref{subsec:brfluxes},
but via a $\theta$-angle.
The spectator Kaluza-Klein monopole helps us with that too
\cite{Dasgupta:2000ry}.
Suppose that far away from the origin we try to set up a constant
NSNS $B_{15}$-field. The Taub-NUT geometry looks locally like $\R^3\times S^1$
with a constant $S^1,$ so unless we get very close to the origin there
is no problem in setting up the constant $B_{15}$-field.
But if try to extend $B_{15}$ to the full Taub-NUT geometry we run into an obstacle.
We have to set the 
$B$-field to be proportional to the global angular 1-form
of the fibration, but that form is not closed, so there has to be an
$H=dB$ flux. In fact, at the origin the value of the flux turns out to be
$|H|^2\sim (B_{15}/R_1 R_5^2)^2,$ where $0\le B_{15}<2\pi$
is the asymptotic value of
the $B$-field at infinity \cite{Dasgupta:2000ry}.

Now we are ready to translate to $E_{10}$-roots.
Let us lift back from type-IIA to M-theory along the $10^{th}$ direction.
We have the imaginary root $\a$ that corresponds to the D2-brane, the 
imaginary root $\sigma$ that corresponds to the spectator Kaluza-Klein monopole,
the imaginary root $\g$ that corresponds to Kaluza-Klein momentum in
the $5^{th}$ direction (to be converted to angular momentum by the 
Kaluza-Klein monopole) and finally, we have the real root $\eta$ that corresponds
to the $\theta$-angle that is the NSNS 2-form flux $\Fl_\eta\equiv B_{15}.$
The corresponding actions are listed in the following table:
\vskip 10pt
\begin{tabular}{lll}
Object & Root & Action$/2\pi$ \\ \hline\hline
D2-brane & $\a$ & $e^{\inner{\a}{\vh}} = M_p^{12}V_{10} R_8 R_9$ \\ 
Spectator & $\sigma$ & 
    $e^{\inner{\sigma}{\vh}} = M_p^{18}V_{10}^2 R_5/R_2 R_3 R_4$ \\
Momentum & $\g$ & 
    $e^{\inner{\g}{\vh}} = M_p^{9}V_{10}/R_5$ \\
$B$-flux & $\eta$ & 
    $e^{\inner{\eta}{\vh}} = M_p^3 R_1 R_5 R_{10}$ \\ \hline
\end{tabular}
\vskip 10pt
The extra energy due to the interaction of the flux with the brane that we
expect is 
$$
\Delta\Ptl\sim \frac{\Fl_\eta}{\Mst^2 R_1 R_5^2} 
= \frac{\Fl_\eta}{M_p^3 R_1 R_5^2 R_{10}}.
$$
In the spirit of the mass formula \eqref{MassFormula}, we write it as
\be\label{InteractionFormula}
\frac{d\MinkAc_I}{d\ct} = 2\pi M_p^9 V_{10}\Delta\Ptl
\sim e^{\inner{\g-\eta}{\vh}}\Fl_\eta,\qquad
\kform{\g}{\eta}=-1.
\ee
where $\MinkAc_I$ is the extra term in the action due to the interaction.
We see that the spectator root $\sigma$ does not enter into the interaction
formula. Note also that
$$
\kform{\a}{\g}=-1,\qquad\kform{\a}{\eta}=-1.
$$

We can similarly study the case depicted in \figref{fig:BranesAndFlux}-b.
In this case, states with $2-3$ or $4-5$ angular momentum
have an effective electric dipole on the D2-brane worldvolume.
The dipole vector is proportional to the angular momentum and is directed
along the $9^{th}$ direction.
To probe it we need to add an additional electric field on the D2-brane,
as was done in \cite{Dasgupta:2000ry}. We can do it by adding an extra
fundamental string charge to the setting, but we will not do that here.

\section{A note on supersymmetry}\label{sec:fermions}
% ========================================================================== 
For M-theory on $T^8$, each instanton that corresponds to a (real) 
positive root of $E_{8(8)}$ breaks half the supersymmetry.
The supersymmetry generators transform in the vector representation
of the double cover of the
maximal compact subgroup $K_8=\Spin(16)/\Z_2$ of $E_{8(8)}.$\footnote{We
are grateful to R. Borcherds for pointing out to us that $K_8$ should not
be denoted by $SO(16).$}
($\Spin(16)$ has a $\Z_2\times\Z_2$ center. The first $\Z_2$ factor
is trivial in the spinor representation, while the second factor is
trivial in the vector representation. The $\Z_2$ factor in $K_8$
is the second one, since the adjoint representation of $E_8$ decomposes
as the $\rep{120}$ adjoint of $so(16)$ plus the $\rep{128}$ spinor
of $so(16),$ but does not contain the vector $\rep{16}.$)
 
Physically, each real positive root therefore defines a subalgebra of
the Lie algebra $so(16).$ This is the subalgebra that preserves the
unbroken supersymmetry.
We will now study this relation from the group theoretic point of view.
For a related discussion see 
\cite{Duff:2003ec}\cite{Hull:2003mf}\cite{West:2003fc}\cite{Keurentjes:2003yu}\cite{Keurentjes:2003hc}.

First, let us explain what we mean by the action of $K.$
Classically, the low-energy limit of M-theory on $T^8$ is described by
a supersymmetric $\sigma$-model with target space $G/K$ where $G=E_{8(8)}$
and $K=\Spin(16)/\Z_2.$ A point in the target space can be parameterized as
a coset $g K$ with $g\in G.$
The $\sigma$-model can be formulated as a gauged $\sigma$-model
with target space $G$ and gauge group $K$ acting on $G$ from the right.
The fermions $\psi$ are in the vector representation of $K.$
But this action of $K$ from the right is not physically interesting, because
it is merely a gauge symmetry. We are interested in the {\it gauge invariant}
combinations $g\psi$ 
on which $K$ acts from the {\it left} as
$g\psi\mapsto x g\psi,$ for $x\in K.$
(Note that $g\psi$ is defined only up to a sign ambiguity because
of the $\Z_2$ factor in $K$, but bilinears in $\psi$ are well defined.)
This $K$-symmetry, being broken by instantons, is not a good quantum symmetry.
But this is precisely the point here -- each instanton term breaks
a part of $K$ and defines an unbroken subgroup.

Consider 10+1D uncompactified M-theory.
The supersymmetry generators are Majorana spinors of $so(10,1).$
Under $so(8)\oplus so(2,1)\subset so(10,1)$ they decompose as
$\rep{32} = (\rep{8}_c\oplus \rep{8}_s)\otimes \rep{2},$
where $\rep{8}_c$ and $\rep{8}_s$ are the two real spin representations of
$so(8)$ and $\rep{2}$ is the real spin representation of $so(2,1).$
Let $V$ be the 16-dimensional real vector space $\rep{8}_c\oplus \rep{8}_s.$
Both $\rep{8}_c$ and $\rep{8}_s$ have an $so(8)$-invariant
bilinear form that we denote by $(\cdot|\cdot)_c$ and $(\cdot|\cdot)_s,$
respectively.
The supersymmetry generators are
2+1D spinors which take values in $V.$
The R-symmetry algebra $so(16)$ acts on $V$ as the subset of $gl(V,\R)$ that
preserves the bilinear form $(\cdot|\cdot)_c + (\cdot|\cdot)_s.$
Choose a Majorana representation so that the Dirac $\Gamma$-matrices
are purely imaginary.
A massless particle with momentum $\vp=(p_0,p_1,\dots,p_{10})$
 preserves the supersymmetry
generators that satisfy $(p_0\Gamma^0+\sum_1^{10} p_i\Gamma^i)x = 0,$
for $x\in V.$
Now consider a Kaluza-Klein instanton with momentum in the $j^{th}$
direction and world-line in the $k^{th}$ direction, so that
the action is $R_k R_j^{-1}.$ After Wick rotating the massless particle
we find that the instanton preserves $x\in V$ that commute with $\Gamma^{kj}.$
It defines the subspace
$$
W_{k; j} \defineas \{x\in V: i\Gamma^{k j}x = x\}
\subset V.
$$
We denote the subalgebra of $so(16)$ that preserves
$W_{k;j}$ by
$$
U_{k;j} \defineas \{g\in so(16):
\Gamma^{kj}g = g\Gamma^{kj}\}
\subset so(16).
$$
$U_{k;j}$ is isomorphic to $u(8).$

Similarly,
an M2-brane instanton stretched in directions $j_1, j_2, j_3$ defines a subspace
$$
W_{j_1 j_2 j_3} \defineas \{x\in V: i\Gamma^{j_1 j_2 j_3}x = x\}
\subset V
$$
We denote the subalgebra of $so(16)$ that preserves
$W_{j_1 j_2 j_3}$ by
$$
U_{j_1 j_2 j_3} \defineas \{g\in so(16):
\Gamma^{j_1 j_2 j_3}g = g\Gamma^{j_1 j_2 j_3}\}
\subset so(16)
$$
$U_{j_1 j_2 j_3}$ is also isomorphic to $u(8)$.

To see this,
let us take, without loss of generality, $(j_1, j_2, j_3)=(1,2,3),$
and let us decompose the representation $V$ of $so(8)$ under the Lie algebra
$so(3)\oplus so(5)\subset so(8).$
We find that $V$ decomposes as the complex representation 
$(\rep{2},\rep{4})$ of $so(3)\oplus so(5).$
The components of a vector $z\in V$ can be written as $z^{\a a}$
where $\a=1,2$ and $a=1,\dots,4.$
The bilinear form $(\cdot|\cdot)\equiv(\cdot|\cdot)_c + (\cdot|\cdot)_s$
can be written as $(z|z) = \sum_{\a,a}|z^{\a a}|^2.$
We can decompose each component into its real and imaginary parts as
$z^{\a a} = u^{(\a a)} +i v^{(\a a)}.$
The elements of $so(3)$ and $so(5)$ mix the components $u^{(\a a)}$
with $v^{(\a a)}$ and $i\Gamma^{123}$ acts as
$z^{\a a}\rightarrow i z^{\a a}$
and therefore as
$u^{(\a a)}\rightarrow -v^{(\a a)},$ and
$v^{(\a a)}\rightarrow u^{(\a a)}.$
The subalgebra $U_{j_1 j_2 j_3}\subset so(16)$
is therefore the subalgebra that commutes with the transformation above
and is isomorphic to $u(8),$ as we claimed.

On the other hand, as we have reviewed in \secref{subsec:instantons},
the Kaluza-Klein and 
the M2-brane instantons correspond to positive roots of $E_8.$
Thus, in the same way as above, every positive root $\a$ of $E_8$ defines
a subalgebra $U_\a$ of $so(16).$

The subalgebra $so(16)\subset E_8$ is generated by $e_i-f_i$ ($i=1\dots 8$),
where $e_i, f_i$ are Chevalley generators as in \eqref{subsubsec:construction}.
Let $u$ be a generator of the 1-dimensional root space $\algEVIII_\a\subset E_8.$
The compact involution on $E_8$ is defined by the generating relations
\be\label{eqn:CompactInv}
\omega(e_i) = -f_i,\qquad \omega(f_i) = -e_i,\qquad
\omega(h_i) = -h_i.
\ee
Define $q_\a\defineas u+\omega(u).$ Then $U_\a$ is the subalgebra
of $so(16)$ that commutes with $q_\a.$

To see this consider the subsets
$$
U_{j_1;j_2} \defineas \{g\in so(16):
\Gamma^{j_1 j_2}g = g\Gamma^{j_1 j_2}\}
\subset so(16),\qquad
1\le j_1 < j_2 \le  8.
$$
corresponding to Kaluza-Klein particles.
Note that if we drop the $8^{th}$ node of the Dynkin diagram of $E_8,$
we get the subalgebra $sl(8)\subset E_8.$ This subalgebra is generated
by $e_i, f_i, h_i$ for $i=1\dots 7.$ The combinations $e_i-f_i$ for
$i=1\dots 7$ generate $so(8)\subset sl(8).$
The matrix $\Gamma^{j_1 j_2}$ can be identified with an $so(8)$ generator
on the spinor representation $V.$ It is easy to see that for $j_1 =i,$
and $j_2 = i+1,$ this generator can be identified with $e_i -f_i.$
Thus $U_{j_1 j_2}\cap so(8)\subset so(16)$ is the subgroup that
commutes with $e_i - f_i.$
It is also not hard to see that the $8^{th}$ generator $e_8 - f_8$
can be identified with $i\Gamma^{678}$ acting on $V.$
The statement that $U_{i;i+1}$ is the subspace of $so(16)$ that commutes
with $e_i - f_i$ (for $i=1\dots 7$) follows.

These constructions can be extended to the infinite dimensional Lie algebras
$\algEIX=E_9$ and $\algEX=E_{10}.$
For $\algEIX,$ the algebra $\komEIX$ is defined as the $\omega$-invariant
subalgebra of $\algEIX.$ It  is denoted by
$so(16)^\infty$ \cite{Nicolai:1997hi}
and is {\it not} to be confused with the affine $\widehat{so}(16)$
Lie algebra.\footnote{
We are grateful to A. Keurentjes for pointing out
an incorrect statement we had made in a previous version.
\label{KeurentjesCorrection}}
Any root $\a$ of $E_9$ defines the subalgebra
$$
U_\a\defineas\{v\in\komEIX: [v,u+\omega(u)]=0\quad\forall
u\in \algEIX_\a\}
\subset\komEIX
$$
where
$\algEIX_\a$ is the root space of $\a,$ which could now be of dimension
higher than $1$ if $\a$ is an imaginary root.

For a real root $\a$, one can argue that
$U_\a \sim su(8)^\infty\oplus u(1),$ where $su(8)^\infty$ is constructed
from the affine Lie algebra $\hat{E}_7$ in a similar way to 
the construction of $so(16)^\infty$ from the affine Lie algebra
$\algEIX =\hat{E}_8,$ that is, by considering the generators
that are invariant under an involution.${}^{{\tiny \ref{KeurentjesCorrection}}}$
(To see this,
take $\a=\a_{-1}$ without loss of generality, and note that the only
elements of the form $e_\b-f_\b$ that commute with $e_\a-f_\a$ are such that
$\b=n\delta +\b'$ with $\b'$ a root of $E_7\subset E_8\subset E_9.$)

For imaginary roots $\a=n\delta$ it is also not hard
to see that $U_{n\delta}$ is trivial.

It would be interesting to find a nontrivial extension of the definition
of $U_\a$ for imaginary roots and to explore its relationship with
its associated brane. Perhaps, one needs to find an element $u$
of the root space $\algEIX_\a,$ such that the centralizer 
of $u+\omega(u)$ (i.e. all $v\in\komEIX$ such that $[u+\omega(u),v]=0$)
is maximal in some sense. We will leave this for future work.

\section{Constructing a Hamiltonian}\label{sec:hamiltonian}
% ========================================================================== 
It is time to collect all the pieces into one framework.
In this section we will construct a Hamiltonian, based on $E_{10}$,
that describes 
some of the features of M-theory on $T^{10},$ that we discussed above.
As we do not know the full details,  we will only 
present a few simple observations.

The Lie algebra $E_{10}$ is integrable, which means that the Lie group
$G_{10}\equiv \exp E_{10}$ can be defined. We can also define 
$\Kom_{10}\equiv\exp\komEX.$
(Here $\komEX$ is defined to be the $\omega$-invariant subalgebra of $E_{10},$
where $\omega$ is the compact involution, defined similarly to \eqref{eqn:CompactInv}.)
$\Kom_{10}$ is not actually compact but seems to be what we need 
(see \cite{Nicolai:1997hi}).
A natural starting point is 
a 0+1D (quantum mechanics) $\sigma$-model on the coset space $G_{10}/\Kom_{10}.$
The $G_{10}$ invariant 
metric on $G_{10}/\Kom_{10}$ is ``almost'' positive definite.
To see what this means, consider the metric on the Lie algebra $E_{10}.$
The Lie algebra has an invariant bilinear form \cite{KacBook}, but it is not
positive definite.
This form turns out to be negative definite in the directions 
of $\komEX.$ 
In addition, the Cartan subalgebra $\CarEX$ has signature
$(9,1)$ which means that it has a negative-norm element $x$.
However, when restricted to the subspace orthogonal to $x$ and $\komEX,$
the invariant bilinear form is positive definite.
(See theorem 11.7 of \cite{KacBook} for more details.)
Thus, after modding out $G_{10}$ by $\Kom_{10}$ we get
rid of all the negative-norm directions except the one in
the Cartan subalgebra.\footnote{We are grateful to Edward Witten for
raising this issue.}

We now take the effective Hamiltonian to be 
proportional to the Laplacian $\Ham=-\triangle$
on the infinite dimensional space $G_{10}/\Kom_{10}.$
The wave functions are required to satisfy a generalized Wheeler-DeWitt 
equation
\be\label{wdw}
\Ham\Psi \equiv -\triangle\Psi = 0.
\ee
The manifest $G_{10}$ invariance could be spontaneously
broken to the U-duality subgroup $E_{10}(\Z)$ by requiring
$\Psi$ to be only $E_{10}(\Z)$ invariant.
This can be implemented by defining the target space,
on which the Laplacian $\triangle$ acts, as the coset
$E_{10}(\Z)\backslash G_{10}/\Kom_{10}.$

Before we proceed,
we have to mention that equation \eqref{wdw} appeared in similar contexts before.
A $\sigma$-model on the same coset space $G_{10}/\Kom_{10}$
was presented in \cite{Damour:2002cu}
and an extension to $G_{11}/\Kom_{11}$ was presented in 
\cite{West:2000ga}\cite{West:2001as}\cite{Englert:2003py}.
Furthermore, equation \eqref{wdw} was also proposed in \cite{Ganor:1999ui}.

The new point of this paper is that we can identify a mechanism to
go beyond dimensionally reduced classical supergravity and to
test the approximation \eqref{wdw} quantum mechanically.
We will attempt a quantum mechanical treatment
of the variables of $G_{10}/\Kom_{10}$ associated
with imaginary roots, and we will compare the resulting energy levels
to the energies of branes and
Kaluza-Klein particles that can be introduced into the evolving universe.

Specifically, ``excited states'' of the universe, with branes or Kaluza-Klein particles,
appear to be related to excited Landau levels of a certain effective magnetic field
that is naturally generated inside $E_{10}(\Z)\backslash G_{10}/\Kom_{10}$
when the canonical momenta dual to the
variables associated with imaginary roots are nonzero.
The separation between Landau levels roughly matches the expected
energies of the branes, but unfortunately there is a mismatch by a factor of $2\pi.$
%This is a quantum test of the conjecture, since branes are quantized objects.
%In particular, the ability to excite Kaluza-Klein states means that
%we can go beyond dimensional reduction and allow the supergravity fields to vary
%along the spatial directions.
There are also a few other puzzles, related to the zero-point energy and to total neutrality.
We present the ideas here anyway, in the hope that there might be some way to ``fix''
the problems.
Let us now construct the model.

\subsection{The variables}\label{subsec:variables}
% --------------------------------------------------------------------------
The model is a 0+1D quantum mechanics. ``Time'' is taken to be
M-theory's conformal time defined in \eqref{DefConformalTime}.

Skipping the proof, which can be found elsewhere 
(see \cite{Damour:2002et} and also
\cite{Obers:1998fb} for the finite dimensional case),
the variables of the coset 
$G_{10}/\Kom_{10}$ can be described as follows.
We have $10$ real variables, each taking values in $\R$,
given by the components of $\vh$
that are related to the physical radii as in \eqref{vhDef}.
In addition, we have an infinite tower of periodic variables with period $2\pi$;
there is one variable $\Fl_\a$ associated with every {\it positive real root}
$\a$ of $E_{10}$, and there are $\mult(\g)$ variables
$\Fl_{\g,j}$ ($j=1\dots \mult(\g)$)
associated with any positive imaginary root $\g$ of $E_{10}.$
Here $\mult(\g)$ is the multiplicity of the root $\g$.
Occasionally, it will be convenient to suppress the index $j$.
In that case, it will be understood that $\Fl_\g$ denotes some linear combination
of the $\Fl_{\g,j}$'s.

We will now construct the Hamiltonian.
It is going to be convenient to identify the charges of the variables
under the $\R^{10}$ Cartan subalgebra $\CarEX$ of $E_{10}$ that acts as
$$
\vh\mapsto\vh+\veps,\qquad\veps\in\CarEX.
$$
Under this symmetry
\be\label{CarCharge}
\Fl_\a\mapsto e^{\inner{\a}{\veps}}\Fl_\a,\qquad
\Fl_{\g,j}\mapsto e^{\inner{\g}{\veps}}\Fl_{\g,j}.
\ee
This symmetry does not preserve the periodicity of $\Fl_\a, \Fl_{\g,j},$
but it is a symmetry of the Hamiltonian.

The Hamiltonian is constructed from functions of $\vh,\Fl_\a,\Fl_{\g,j}$
and their first derivatives 
$\partial/\partial h_i,$ $\partial/\partial \Fl_\a,$ 
$\partial/\partial \Fl_{\g,j}.$
It is probable that we also need to include fermionic degrees of freedom,
but we will completely ignore the fermions in this section, for simplicity.

\subsection{The Hamiltonian}\label{subsec:hamiltonian}
% --------------------------------------------------------------------------
The Hamiltonian $\Ham$ preserves $G_{10},$ and hence all the terms
appearing in it conserve the $\R^{10}$ charges of \eqref{CarCharge}.
Up to a factor of $-1,$ it is the
$E_{10}$-invariant Laplacian $\Ham =-\triangle.$
Explicitly, it contains the following terms.

First, there is a term that contains only $\vh$ and is given by
\be\label{HamZeroH}
\Ham_{h}\defineas -\frac{1}{8\pi}\left[\sum_{k=1}^{10} \pspxs{h_k} 
-\frac{1}{9}\left(\sum_{k=1}^{10}\ppx{h_k}\right)^2
+\sum_{k=1}^{10}\bigl(2k-\tfrac{56}{3}\bigr)\ppx{h_k}\right].
\ee
The linear term might appear strange, but it can be deduced by extrapolation
from the $E_8$ case. [It is also required in order for the instanton terms
$\exp(-\Ac_{\a_k}+i\Fl_{\a_k})$ to be harmonic functions for the simple roots
$\a_k.$ See \secref{subsec:instantons}.] The apparent \textit{$S_{10}$}-asymmetric
form of \eqref{HamZeroH} is also not a problem since the decomposition into
positive and negative roots already breaks this \textit{$S_{10}$} permutation symmetry.

Then we have the terms
\be\label{HamZero}
\Ham_{0}\defineas
-\pi\sum_{\a^2=2}e^{2\inner{\a}{\vh}}\pspxs{\Fl_\a}
-\pi\sum_{\a^2\le 0}\sum_{j=1}^{\mult(\a)} e^{2\inner{\a}{\vh}}\pspxs{\Fl_{\a,j}}
\ee
Note that this term is invariant under \eqref{CarCharge} as
$\partial/\partial\Fl_\a$ has charge $-\a.$
(The factors of \textit{$\pi$} appear because of our choice
of periodicity of \textit{$\Fl_\a$}.)

In addition to $\Ham_{h}$ and $\Ham_{0}$ we have an infinite series 
of ever more complex terms, so that
$$
\Ham = \Ham_{h}+\Ham_{0} + \Ham_{1}+ \Ham_{2}+\cdots
$$
where $\Ham_{n}$ is quadratic in $\partial/\partial \Fl_\a$
(or $\partial/\partial \Fl_{\g,j}$) but is a polynomial
of degree $n$ in $\Fl_\a$ (or $\partial/\partial \Fl_{\g,j}$).
The first terms look schematically like
\be\label{HamZeroOne}
\Ham_{1}\sim \sum_{\g=\a+\b} 
e^{\inner{\b+\g-\a}{\vh}}\Fl_\a\pspxpx{\Fl_\b}{\Fl_\g}
= \sum_{\g=\a+\b} 
e^{2\inner{\b}{\vh}}\Fl_\a\pspxpx{\Fl_\b}{\Fl_\g}.
\ee
Note that the dependence on $\vh$ is entirely determined by conservation
of $\R^{10}$-charge. The expression $\Ham_{1}$ can be deduced from the $E_{10}$
transformation properties of $\Ham_{0}$ and the invariance of
the total expression $\Ham.$ Similarly, each consecutive $\Ham_{n}$
can be deduced from the $E_{10}$-transformation properties of its predecessors.
We will not do the explicit computation here.
It can be found in \cite{Damour:2002et} (see also \cite{Ganor:1999uy}).

% potential walls.
% nontrivial bundles
The kinetic term $\Ham_{0}$ already contains all the ``wall'' potential
terms required for billiard cosmology without branes 
(see \secref{subsec:billiard}).
A state with $N_\a$ units of the flux associated with the real root $\a$
has a wave function that behaves as $\sim \exp{i N_\a \Fl_\a}.$
It is an eigenstate of $\partial/\partial\Fl_\a.$ 
Setting $\partial/\partial\Fl_\a\rightarrow i N_\a$
in $\Ham_{0}$ we obtain the potential
\textit{$\pi N_\a^2\exp(2\inner{\a}{\vh})$}.
That leads to the
expressions discussed in \secref{subsec:billiard}
for the wall potentials of billiard cosmology \cite{Damour:2002cu}.

The term $\Ham_{1}$ is also interesting in that it tells us that
the target space of the $\sigma$-model is not just a product
$\R^{10}\times S^1\times S^1\times\cdots$ -- with each $S^1$ corresponding
to a different $\Fl_\a$ -- but is a nontrivial circle bundle.
For example, take two positive real roots $\a, \b,$ such that
$\g=\a+\b$ is also real.
Being periodic,
the associated variables $\Fl_\a, \Fl_\b$ parameterize a $T^2.$
If $\g=\a+\b$ and the commutator of 
the Lie algebra generators $[J^{+\a}, J^{+\b}]$ is proportional 
to $J^{+\g}$ then the circle associated with the periodic variable
$\Fl_\g$ is nontrivially fibered over the $T^2.$ The first
Chern class of the fibration is $c_1=1.$
This is easily seen by noting that the infinitesimal $E_{10}$
transformation $\exp (\epsilon J^{+\a})$ acts as
$$
\Fl_\a\rightarrow \Fl_\a+\epsilon,\qquad
\Fl_{\g}\rightarrow\Fl_{\g}+\frac{\epsilon}{2\pi}\Fl_\b.
$$
For real roots, this geometrical
fact has some interesting physical consequences
such as Wess-Zumino terms, but we will not discuss this here
(see for instance \cite{Ganor:1999ui}\cite{Damour:2002et}).
If $\g$ is an imaginary root, we have to be careful because of its
multiplicity.
The statement is that if the commutator of the Lie algebra generators
$[J^{+\a}, J^{+\b}]$ is proportional to a generator
$J^{+\g,j}$ in the root space $\algEX_\g,$ then
$\Fl_{\g,j}$ is nontrivially fibered over $T^2$ with first Chern class
$c_1 = 1.$ This fact will be {\it crucial} in \secref{subsec:ll}.

To summarize, $\Ham$ is a quadratic differential operator
(which also contains the linear term in $\Ham_{h}$).
It is essentially determined by $E_{10}$-invariance.

\subsection{Instanton effects}\label{subsec:insteffects}
% --------------------------------------------------------------------------
Universes with a flux $\Fl_\a$ turned on must have a wave-function of
the form
\be\label{UniverseWithFlux}
\Psi_\a = e^{-\Ac_\a +i \Fl_\a}\left(\cdots\right).
\ee
where $(\cdots)$ is independent of $\Fl_\a.$
The prefactor expresses the tunneling amplitude from a state without
flux to a state with flux.
A state with $N_\a$ units of flux is an eigenstate of
$-i\partial/\partial\Fl_\a.$
The action $\Ac_\a$ is in general a complicated expression of the 
fluxes and of $\vh,$ but when all the fluxes (except $\Fl_\a$ of course)
are set to zero, $\Ac_\a$ reduces to $2\pi\exp\inner{\a}{\vh}$ -- the 
simplified expression that we have been using throughout this paper.
The fact that the prefactor 
$\exp\{-\Ac_\a + i\Fl_\a\}$ is a harmonic function on $G_{10}/\Kom_{10}$
if $\a$ is a simple root (see \secref{subsec:instantons})
is intriguing, but it seems that extra terms must be added to $\Ham$
in order for \eqref{UniverseWithFlux} to be an eigenfunction.

\subsection{Branes and Landau levels}\label{subsec:ll}
% --------------------------------------------------------------------------
In \secref{sec:imroots} we argued that
a prime imaginary root $\g$ with $\g^2=0$ corresponds to a Minkowski brane.
We found a mass formula \eqref{MassFormula} that expresses the mass
(defined with respect to conformal time) in terms of $\g,$ as
$2\pi\exp\inner{\vh}{\g}.$
If there are $n$ branes, we expect a contribution to the Hamiltonian
of the form $2\pi n\exp\inner{\vh}{\g}.$
We will now suggest 
a way in which such a term could come from quantizing the variables
$\Fl_{\g,j}.$
Our result will reproduce the correct $n\exp\inner{\vh}{\g}$ factor,
but will be off by a factor of $2\pi$ as well as an $n$-independent term.

Decompose $\g=\a+\b$ as a sum of two positive real roots, and
suppose that the Lie algebra element corresponding to $\Fl_{\g,j}$
is proportional to the commutator $[J^{+\a}, J^{+\b}].$
Then, $\Fl_{\g,j}$ is a local coordinate on a circle bundle
over the $\Fl_\a,\Fl_\b$ torus, as explained in \secref{subsec:hamiltonian}.
Since the $\Fl_{\g,j}$-circle is nontrivially fibered over
the $\Fl_\a,\Fl_\b$ torus,
it follows that the negative of the Laplacian $\triangle$ contains terms of the form
\be\label{Habg}
\Ham' = 
-\pi e^{2\inner{\vh}{\a}}\pspxs{\Fl_{\a}}
-\pi e^{2\inner{\vh}{\b}}\left(\ppx{\Fl_\b} -\frac{\Fl_{\a}}{2\pi}\ppx{\Fl_{\g,j}}\right)^2
-\pi e^{2\inner{\vh}{\g}}\pspxs{\Fl_{\g,j}}
\ee
[The $\vh$-dependent coefficients are determined by the $\R^{10}$-symmetry
\eqref{CarCharge}.]

Suppose we have a state $\Psi$
for which $-i\partial/\partial\Fl_{\g,j}=N_{\g,j}.$
Plugging that into \eqref{Habg}
we find that $\Ham'$ describes
the Hamiltonian of an abstract charged particle
on a torus (parameterized by the coordinates $\Fl_\a, \Fl_\b$)
with $N_{\g,j}$ units of magnetic flux.
The ``cyclotron'' frequency is
\textit{$\omega = \exp\inner{\vh}{\a+\b}\abs{N_{\g,j}} =
\exp\inner{\vh}{\g}\abs{N_{\g,j}}$}.
Eliminating $\Fl_{\a}$ we get ``Landau levels'' with energy
$$
\Ham'' = e^{\inner{\vh}{\g}}\abs{N_{\g,j}} (n+\tfrac{1}{2}) 
+ \pi e^{2\inner{\vh}{\g}}\abs{N_{\g,j}}^2
$$
It is now tempting to
compare these states with states of the universe that contain $n$ bound states of \textit{$N_{\g,j}$}
branes.
The $n$-dependent part of the energy is
\textit{$n\abs{N_{\g,j}}\exp\inner{\vh}{\g}$}. According to the mass formula
\eqref{MassFormula} this is similar to the contribution of $n$ bound states
of \textit{$N_{\g,j}$} branes
to the energy, but unfortunately there is a $2\pi$ mismatch.

The remaining terms in $\Ham'',$ which include the zero-point
energy and the $\Fl_{\g,j}$-flux contribution, are independent of $n.$
We do not know how to interpret them, but perhaps
they behave like a cosmological constant.
Perhaps they can be cancelled if 
supersymmetry is properly taken into account.
It might also be possible to consistently leave the problematic terms
\textit{$\exp(2\inner{\vh}{\g})\partial^2/\partial\Fl_{\g,j}^2$} out of $\Ham.$

We also have to mention, however, that 
when all of space is compact we cannot add branes at will, because
the total charge has to cancel. But we can add pairs of branes and anti-branes.
If there are $N_p$ pairs we expect a contribution to the Hamiltonian
of the form $2N_p\times 2\pi\exp\inner{\vh}{\g}.$
Furthermore, in \secref{sec:imroots} we constructed branes from pairs of instantons
corresponding to real roots $\a$ and $\b$ with $\g=\a+\b$ such that $\g^2 = 0.$
We argued that if there are $N_\a$ units of flux associated with the
real root $\a$ and $N_\b$ units of flux associated with the real root $\b$
then, in the setting of \secref{subsec:primesqzero}, there
must also be a net number of $N_\a N_\b$ branes, for charge neutrality.
At the moment, we do not know how charge neutrality appears in
the $E_{10}$ formalism.
In fact, it appears that the $\sigma$-model by itself cannot capture
the term \eqref{gTerm}. It seems consistent to set to zero all the operators
$\partial/\partial\Fl_{\g,j}$ for $\g>\a,\b,$
but we will then get only the term \eqref{NaSqNbSq}.

Finally, let us show that the mismatch factor of $2\pi$ between the Landau levels
and the expected masses of branes is not an artifact of the conventions.
To see this compare the energy levels of a (nonrelativistic) free 
particle on $T^2=S^1\times S^1$ with
one unit of magnetic flux, to the energy levels of the same particle on the
same $T^2$ without any magnetic flux.
In the second case, the energy levels are of the form \textit{$C_1 n_1^2 + C_2 n_2^2$},
where $C_1, C_2$ are constants and $n_1, n_2\in\Z.$
In the first case, the energy levels are \textit{$(2n+1)\sqrt{C_1 C_2}/2\pi$}.
The factor of $2\pi$ in the last formula is the source of the mismatch.

\subsection{Comparison with the ``small tension'' expansion}
\label{subsec:comparison}
% --------------------------------------------------------------------------
In \cite{Damour:2002cu} a different interpretation for $E_{10}$ roots,
including imaginary ones, was proposed.
The analysis of \cite{Damour:2002cu}
was done for the case of uncompactified M-theory, at the level of 
the supergravity equations of motion.
We will now briefly compare that proposal to ours by discussing
a particular example of an imaginary root -- the prime isotropic
root $\g$ from \eqref{gammaKK}.
According to \cite{Damour:2002cu},
the root $\g$ labels certain fluxes.
These fluxes, $8=\mult\g$ in number, were
encoded together with $442$ other fluxes, corresponding
to other roots, in the variable that was denoted by $DA^{b|a_1 a_2\cdots a_8},$
where all indices $b, a_1,\dots, a_8$ are spacelike (from $1\dots 10$),
and the variable was antisymmetric with respect to $a_1\dots a_8.$
Our imaginary root $\g$ is related to this flux,
up to factors of $R_1,\dots, R_{10},$ by
$$
-i\frac{\partial}{\partial\Fl_{\g,j}}
\rightarrow 
{\mbox{linear combination of 
$DA^{2|3\dots 10}, DA^{3|2 4\cdots 10},\dots, DA^{10|2\dots 9},$}}
$$
where on the left we used the notation of \secref{subsec:variables}.
One of the main points of \cite{Damour:2002cu} is that
$DA^{b|a_1 a_2\cdots a_8}$ can be written in terms of 
11D supergravity fields as
$$
DA^{b|a_1\cdots a_8} = 
\frac{3}{2}\epsilon^{a_1\dots a_8 b_1 b_2}
\left(
{C^b}_{b_1 b_2}
+\frac{2}{9}\delta^{b}_{[b_1}{C^c}_{b_2]c}
\right),
$$
where ${C^c}_{ab}$ is the connection that is related to
the zehnbein $\theta^a$ by
$$
d\theta^c = \frac{1}{2}{C^c}_{ab}\theta^a\wedge\theta^b.
$$
For other values of the indices ($b, a_1,\dots,a_8$), 
the flux $DA^{b|a_1\cdots a_8}$ corresponds to:
(i)~an isotropic imaginary root that can be obtained from $\g$ by
an $S_{10}$ permutation of the indices, which is the case
if $b\notin\{a_1,\dots,a_8\},$ or
(ii)~a real root $\a$ that corresponds to a ``gravitational wall''
of the form \eqref{GrWall},
if $b\in\{a_1,\dots, a_8\}.$
The square $DA^{b|a_1\cdots a_8}DA_{b|a_1\cdots a_8}$
appears as a term in the Einstein-Hilbert action.
This term is directly related to
the quadratic term $-(\partial/\partial\Fl_{\g,j})^2$
from \eqref{HamZero}, which is the quantized version of the
classical $\sigma$-model that was used in \cite{Damour:2002cu}.
When compactified on $T^{10},$
these terms are proportional to $\exp(2\inner{\g}{\vh}),$
if we assume that $\Fl_{\g,j}$ is periodic, as implied by 
$E_{10}(\Z)$ U-duality.

The point of our paper is that in addition to such terms,
there have to be terms proportional to $\exp\inner{\g}{\vh}.$
In particular, this is the case in the presence of fluxes
associated with real roots $\a,\b$ such that $\a+\b=\g,$
as we discussed in \secref{subsec:withmatter}.
Such terms cannot be deduced purely from classical supergravity,
since they describe quantized objects such as particles and branes.
Furthermore, it would be interesting to understand why
$DA^{b|a_1 a_2\cdots a_8}$ is quantized on $T^{10}.$
This is clear for $b\in\{a_1,\dots,a_8\},$
since the flux is then related to the ``gravitational wall''
\eqref{GrWall}, and it
would be interesting to study the quantization condition for
$b\notin\{a_1,\dots,a_8\}.$
(The quantization requirement of course follows from $E_{10}(\Z)$
U-duality.)
If indeed the flux is quantized then, as we have argued in
\secref{subsec:ll}, terms that are proportional to particle masses
naturally arise from the $\sigma$-model formalism.

\subsection{Summary}\label{subsec:FinalConjecture}
% --------------------------------------------------------------------------
Some features of M-theory on $T^{10}$ 
are effectively described by a harmonic function
on the target space \textit{$E_{10}(\Z)\backslash G_{10}/\Kom_{10}$} satisfying
$$
\triangle\Psi = 0.
$$
The Wheeler-DeWitt wave-function $\Psi$ is a sum of terms
with different eigenvalues of the various fluxes 
$N_\a=-i\partial/\partial\Fl_\a.$
The behavior of $\Psi$ as a function of the radii (encoded in $\vh$)
crucially depends on $N_\a.$
Different pieces of $\Psi$ therefore describe completely different
evolutions of the universe and can thus be separated.

The term without fluxes (all $N_\a=0$) describes possible Kasner
evolutions with \textit{$\norm{\vp}^2=0$}
(in the notation of \secref{subsec:billiard}) and,
according to \cite{Banks:1998vs}, can never describe a classical
universe in the far future.
Terms in $\Psi$ for which only the fluxes $N_\a$ that are associated to
simple roots are nonzero describe a chaotic evolution as in
\cite{Damour:2000wm} and are also never classical in the far future.

But terms with nonzero quantum numbers $N_\g$ associated to imaginary roots can
describe, as we suggested, universes with an ordinary matter component
composed of Kaluza-Klein states, or branes.
These universes can have a classical evolution in the far future
(a ``safe'' region of moduli space, in the terminology of \cite{Banks:1998vs}).
Unfortunately, the brane masses that we obtain are smaller than
the correct masses by a factor of $2\pi.$

\section{Conclusions and discussion}\label{sec:disc}
% ==========================================================================
The infinite dimensional Lie algebra $E_{10}$ is likely to play an important
role in a fundamental formulation of M-theory.
Its roots encode the kinematic properties of branes.
Real roots encode instanton actions, and, as we have proposed in this paper,
certain imaginary roots correspond to branes.
We have also seen that the inner product of two imaginary roots $\a,\b$ encodes 
basic properties of the interaction between the two corresponding branes.
We have interpreted the values $\kform{\a}{\b}=-1,-2,-4.$
Similarly, we have seen that the inner product of an imaginary root
and a real root encodes the basic properties of the interactions of the 
corresponding brane with the corresponding flux. 
We have interpreted the values $\kform{\a}{\b}=0,-1,-2.$

We have begun to construct a Hilbert space and 
an effective Hamiltonian that can describe some features of M-theory
in this setting. 
We have argued that this Hilbert space has states that describe
branes and Kaluza-Klein particles.
The variables associated to imaginary roots play an
important role in the reproduction of the mass of these branes
and particles.
Including branes corresponds, in this Hilbert space,
to exciting a certain subset of the variables to higher Landau levels
of an abstract particle in a magnetic field.
Unfortunately, the masses of the branes are off by a factor of $2\pi,$
although their dependence on the metric is correct.

Many open questions remain:
\begin{enumerate}
\item
What is the physical significance of the multiplicities of imaginary roots?
The imaginary roots that we studied all have a multiplicity of $m=8,$
but, as we have seen in \secref{subsec:immult}, all $8$ generators
that correspond to the same root also correspond to the same brane.
Could this multiplicity be related to the multiplicity of the supersymmetric
multiplets?
\item
In \secref{sec:interactions} we classified some interactions between
branes and fluxes and between pairs of branes according to the inner product
of the corresponding roots. We covered the cases $\kform{\a}{\b}=0,-1,-2$
for branes and fluxes, and the cases $\kform{\a}{\b}=-1,-2,-4$
for pairs of branes. It would be interesting to study other values
of the inner products.
\item
We have argued in \secref{subsec:sqnegative} that a certain imaginary
root is associated to a pair of Minkowski branes of different types.
It would be interesting to relate the properties of the individual branes
to the properties of the root.
\item
It would be interesting to prove or disprove our general decomposition conjecture
that every positive imaginary root can be decomposed as a sum of 
positive real roots. (See \secref{subsec:gendecomp}.)
\item
Can the process used in \secref{subsec:primesqzero}
be generalized to other imaginary roots?
\item
Can the process of passing one instanton through another,
used in \secref{subsec:primesqzero},
be generalized to triple
commutators such as $[J^{+\a_1},[J^{+\a_2},J^{+\a_3}]]$?
\item
In \secref{sec:fermions} we explored various
definitions for the subalgebra 
$U_\g\subset \komEX\subset E_{10}$ associated with the root $\g.$
It would be interesting to 
study the physical interpretation of $U_\g$ and its relation to
the supersymmetry that is preserved by the brane.
\item
It would be interesting to extend the discussion to heterotic string theory
where the Kac-Moody algebra $DE_{18}$ plays a role \cite{Motl:1999cy}.
It is intriguing that D7-branes can actually be created entirely with 
$E_{10}$, since after the lift from type-IIB to M-theory they correspond to 
imaginary roots, as in \secref{sec:imroots}.
Therefore, it might be possible to rephrase the F-theory construction
\cite{Vafa:1996xn} entirely in terms of $E_{10}$ variables.
It would be interesting to find out how this works!
\item
In the Hamiltonian formulation discussed in \secref{sec:hamiltonian},
can the zero-point energy be cancelled?
Can the condition of total charge neutrality be incorporated?
\item
Perhaps the most intriguing question is whether we can create an arbitrary 
collection of branes via a process as in \secref{subsec:sqnegative},
or using the formalism of \secref{sec:hamiltonian}.
If true, it would mean that an arbitrary state of the universe can be
described with the variables associated to imaginary roots of $E_{10}$!

\end{enumerate}

% ==========================================================================
%\section*{Acknowledgments}
% ==========================================================================
\acknowledgments
It is a pleasure to thank Stephon Alexander,
Michel Bauer, Richard Borcherds, Shyamoli Chaudhuri, Alexandre Givental,
Marty Halpern, Arjan Keurentjes, Bom-Soo Kim, Indrajit Mitra,
Lubos Motl, Boris Pioline,
Mukund Rangamani, Igor Schnakenburg and Edward Witten for helpful discussions
and correspondence.
We are also greatful to Nobuyoshi Ohta and Vladimir Ivashchuk for bringing
a few references to our attention,
and we are grateful to the (anonymous) JHEP referee for helpful comments.
The work of OJG was supported in part by the Director, Office of Science,
Office of High Energy and Nuclear Physics, of the U.S. Department of
Energy under Contract DE-AC03-76SF00098,
and in part by the NSF under grant PHY-0098840. 

\newpage
\appendix
\setcounter{equation}{0}
\renewcommand{\theequation}{A-\arabic{equation}}

\section{The singleton count of real and imaginary roots}
\label{app:proofs}
% ========================================================================= 
In this appendix we will prove some of the theorems from \secref{sec:comb}.
We start with \clmref{thm:nonegni}:
a positive imaginary root has no negative $n_i$'s.
A positive real root has negative $n_i$'s only if it is a permutation of
$(1,-1,0,\dots,0).$
\begin{proof}
If at least one $n_i=0,$ say for $i=1,$ then the root is in an $E_9$ subalgebra
for which the roots are completely classified. They are the roots of $E_8$ plus
an integer multiple of $(0,1,1,\dots,1).$ It is easy to verify that the theorem
holds in that case. So we assume that all $n_i\neq 0.$
Suppose without loss of generality that $n_k \le n_{k+1} \le \cdots\le n_{10} < 0$
and $0 < n_1 \le n_2 \le \cdots\le n_{k-1}$
for some $2\le k\le 9.$
Then
\bear
\a^2 &=&
\frac{1}{9}\sum_{1\le i<j\le 10} (n_i - n_j)^2
-\frac{1}{9}\sum_{i=1}^{10} n_i^2
\nn\\ &=&
\frac{1}{9}\sum_{1\le i<j\le k-1} (n_i - n_j)^2
+\frac{1}{9}\sum_{k\le i<j\le 10} (n_i - n_j)^2
+\frac{1}{9}\sum_{i\le k-1}\sum_{j\ge k} (n_i + |n_j|)^2
-\frac{1}{9}\sum_{i=1}^{10} n_i^2
\nn\\ &=&
\frac{1}{9}\sum_{1\le i<j\le k-1} (n_i - n_j)^2
+\frac{1}{9}\sum_{k\le i<j\le 10} (n_i - n_j)^2
+\frac{2}{9}\sum_{i\le k-1}\sum_{j\ge k} n_i|n_j|
\nn\\ &&
+\frac{1}{9}\sum_{i=1}^{k-1} (10-k)n_i^2
+\frac{1}{9}\sum_{j=k}^{10} (k-2)n_i^2
\ge
\frac{1}{9}[4(k-1)(11-k)-10] > 2.
\nn
\eear
\end{proof}

Next, we prove \thmref{thm:imaginaryroots}:
the only imaginary roots with a singleton count $s\ge 2$ are permutations of
\bear
\a &=& (0,1,1,1,1,1,1,1,1,1),\qquad \a^2 = 0,\qquad s=9\nn\\
\a &=& (1,1,1,1,1,1,1,1,2,2),\qquad \a^2 = 0,\qquad s=8\nn\\
\a &=& (1,1,1,1,1,2,2,2,2,2),\qquad \a^2 = 0,\qquad s=5\nn\\
\a &=& (1,1,1,2,2,2,2,2,2,3),\qquad \a^2 = 0,\qquad s=3\nn\\
\a &=& (1,1,2,2,2,2,2,3,3,3),\qquad \a^2 = 0,\qquad s=2\nn\\
\a &=& (1,1,2,2,3,3,3,3,3,3),\qquad \a^2 = 0,\qquad s=2\nn\\
\a &=& (1,1,3,3,3,3,3,3,3,4),\qquad \a^2 = 0,\qquad s=2\nn\\
\a &=& (1,1,2,2,2,2,2,2,2,2),\qquad \a^2 = -2,\qquad s=2\nn
\eear
and there is an infinite number of imaginary roots with singleton count $s=1.$
\begin{proof}
According to the previous theorem, $n_i\ge 0$ for $i=1\dots 10.$
If $n_i=0,$ for some $n_i,$ then $\a$ is a root of an $E_9$ subalgebra.
But the only imaginary roots of $E_9$ are given by
$$
\a = (n,n,n,n,n,n,n,n,n),\qquad \a^2 = 0.
$$
This has a singleton count $s=9$ for $n=1$ and singleton count $s=0$ for $n>1.$
So, suppose without loss of generality that 
$n_1 = n_2 = \cdots = n_s = 1$ and that $2\le n_{s+1}\le n_2\le \cdots\le n_{10}$
for $s\ge 1.$
In order for $\a$ to be a root we need $s + \sum_{i=s+1}^{10} n_i \in 3\Z.$
Then
\bear
\a^2 &=&
\frac{1}{9}\sum_{1\le i<j\le 10} (n_i - n_j)^2
-\frac{1}{9}\sum_{i=1}^{10} n_i^2
\nn\\ &=&
\frac{1}{9}\sum_{s+1\le i<j\le 10} (n_i - n_j)^2
+\frac{s}{9}\sum_{i=s+1}^{10} (n_i -1)^2
-\frac{1}{9}\sum_{i=s+1}^{10} n_i^2-\frac{s}{9}
\nn
\eear
If $s>1$ we can write 
\bear
\a^2 &=&
\frac{1}{9}\sum_{s+1\le i<j\le 10} (n_i - n_j)^2
+\frac{s-1}{9}\sum_{i=s+1}^{10} \left(n_i -\frac{s}{s-1}\right)^2
-\frac{s}{s-1}
\nn
\eear
There is only a finite number of sequences $2\le n_{s+1}\le \cdots\le n_{10}$
for which the righthand side is not positive.
A quick exhaustive computer search
yielded the 8 imaginary roots stated above.
For $s=1$ we get
\be\label{sEqualOne}
\a^2 =
\frac{1}{9}\sum_{2\le i<j\le 10} (n_i - n_j)^2
-\frac{2}{9}\sum_{i=2}^{10} n_i+\frac{8}{9}
\ee
and there is an infinite number of imaginary roots with $s=1$ because
for any given imaginary root $\a$ we can always change $n_i\rightarrow n_i+1$
for all $i=1\dots 9$ and get a root with a smaller $\a^2.$
\end{proof}

Finally, we prove \thmref{thm:realroots}:
the only real roots ($\a^2 = 2$) with a singleton count $s\ge 2$ are permutations of
\bear
\a &=& (0,0,0,0,0,0,0,1,1,1),\qquad s=3\nn\\
\a &=& (0,0,0,0,1,1,1,1,1,1),\qquad s=6\nn\\
\a &=& (0,0,1,1,1,1,1,1,1,2),\qquad s=7\nn\\
\a &=& (0,1,1,1,1,1,1,2,2,2),\qquad s=6\nn\\
\a &=& (0,1,1,1,2,2,2,2,2,2),\qquad s=3\nn\\
\a &=& (1,1,1,1,1,1,1,1,1,3),\qquad s=9\nn\\
\a &=& (1,1,1,1,1,1,2,2,2,3),\qquad s=6\nn\\
\a &=& (1,1,1,1,2,2,2,2,3,3),\qquad s=4\nn\\
\a &=& (1,1,1,2,2,2,3,3,3,3),\qquad s=3\nn\\
\a &=& (1,1,1,3,3,3,3,3,3,3),\qquad s=3\nn\\
\a &=& (1,1,2,2,2,2,2,2,3,4),\qquad s=2\nn\\
\a &=& (1,1,2,2,2,3,3,3,3,4),\qquad s=2\nn\\
\a &=& (1,1,2,3,3,3,3,3,4,4),\qquad s=2\nn\\
\a &=& (1,1,3,3,3,3,4,4,4,4),\qquad s=2\nn\\
\a &=& (1,1,3,4,4,4,4,4,4,4),\qquad s=2\nn
\eear
and there is an infinite number of imaginary roots with singleton count $s=1.$
\begin{proof}
The proof is very similar to that of \thmref{thm:imaginaryroots}.
Note that to satisfy $\a^2 = 2$ in equation \eqref{sEqualOne},
we can start by fixing some difference,
say $n_3-n_2,$ and pick an otherwise arbitrary sequence
$2\le n_2\le \cdots\le n_{10}$
such that $1+\sum_2^{10}n_i$ is divisible by $3$ and
the righthand side of \eqref{sEqualOne} is positive.
(It is not hard to see that there
are an infinite number of such sequences for any value of $n_3-n_2$.)
It is also easy to see that the righthand side is then an even integer
(as it must, being an element of the $E_{10}$ root lattice $\latEX$).
If we now change $n_i\rightarrow n_i + k$ for $i=2,\dots 10$ we see
that we decrease the righthand side of \eqref{sEqualOne} by $2k.$
we can therefore find the appropriate $k$ for which $\a^2=2,$
and there is an infinite number of roots like that.
\end{proof}

\end{document}